\documentclass[12pt, onecolumn,a4paper]{quantumarticle}
\pdfoutput=1 
\usepackage[T1]{fontenc}

\usepackage{tikz}
\usetikzlibrary{positioning, calc}
\usetikzlibrary{calc}
\usetikzlibrary{arrows}
\usetikzlibrary{intersections,through}
\usepackage{tikz-3dplot}
\usepackage{graphicx}
\usepackage{mdwlist}
\usepackage{color}
\usepackage{thmtools}
\usepackage{amssymb}
\usepackage{physics}
\usepackage{amsmath}
\usepackage{array}
\usepackage{doi}
\usepackage{url,hyperref}
\usepackage{cancel}

\usepackage[numbers,sort&compress]{natbib}
\usetikzlibrary{fadings}
\usetikzlibrary{decorations.pathmorphing}
\usetikzlibrary{patterns.meta}
\usepackage{mathrsfs}
\usepackage{authblk}
\usepackage[mathscr]{euscript}
\usepackage{enumitem}
\usetikzlibrary{quantikz2}

\tikzfading[name=fade left,
            left color=transparent!100,
            right color=transparent!30]

\tikzfading[name=fade right,
            left color=transparent!30,
            right color=transparent!100]

\DeclareMathAlphabet\mathbfcal{OMS}{cmsy}{b}{n}

\usepackage{cleveref}

\usetikzlibrary{decorations.pathreplacing}
\tikzset{snake it/.style={decorate, decoration=snake}}

\usetikzlibrary{decorations.pathreplacing,decorations.markings}

\usepackage{bm}

\tikzset{
    >=stealth',
    punkt/.style={
           rectangle,
           rounded corners,
           draw=black, very thick,
           text width=6.5em,
           minimum height=2em,
           text centered},
    pil/.style={
           ->,
           thick,
           shorten <=2pt,
           shorten >=2pt,},
  on each segment/.style={
    decorate,
    decoration={
      show path construction,
      moveto code={},
      lineto code={
        \path [#1]
        (\tikzinputsegmentfirst) -- (\tikzinputsegmentlast);
      },
      curveto code={
        \path [#1] (\tikzinputsegmentfirst)
        .. controls
        (\tikzinputsegmentsupporta) and (\tikzinputsegmentsupportb)
        ..
        (\tikzinputsegmentlast);
      },
      closepath code={
        \path [#1]
        (\tikzinputsegmentfirst) -- (\tikzinputsegmentlast);
      },
    },
  },
  mid arrow/.style={postaction={decorate,decoration={
        markings,
        mark=at position .5 with {\arrow[#1]{stealth'}}
      }}}
}

\usepackage{hyperref}
\usepackage{slashed}
\usepackage{float} 
\usepackage{graphicx} 
\usepackage{subcaption}

\usepackage{comment}

\newcommand{\p}{\partial}

\usepackage{bbold}

\mathchardef\mhyphen="2D

 \newcommand{\badat}{\begin{alignedat}}
 \newcommand{\eadat}{\end{alignedat}}
  \def\be{\begin{equation}}
\def\ee{\end{equation}}

\newcommand{\lads}{L_{\textnormal{AdS}}}
\newcommand{\V}{\hat{\mathcal{V}}}
\newcommand{\R}{\hat{\mathcal{R}}}
\renewcommand{\S}{\hat{\mathcal{S}}}

\newtheorem{theorem}{Theorem}
\newtheorem{conjecture}[theorem]{Conjecture}

\newtheorem{definition}[theorem]{Definition}

\newenvironment{proof}[1][Proof]{\noindent\textbf{#1.}}{\ \rule{0.5em}{0.5em}}
\begin{document}

\title{Cryptographic tests of the python's lunch conjecture}

\author[1,2]{Alex May}
\email{amay@perimeterinstitute.ca}
\orcid{0000-0002-4030-5410}

\author[1]{Sabrina Pasterski}
\email{spasterski@perimeterinstitute.ca}
\orcid{0000-0003-3672-4169}

\author[1]{Chris Waddell}
\email{cwaddell@perimeterinstitute.ca}
\orcid{0000-0002-5972-6151}

\author[3]{Michelle Xu}
\email{mdx@stanford.edu}
\orcid{0000-0002-0948-5222}

\affiliation[1]{Perimeter Institute for Theoretical Physics, Waterloo, Ontario}
\affiliation[2]{Institute for Quantum Computing, Waterloo, Ontario}
\affiliation[3]{Stanford Institute for Theoretical Physics, Stanford University}

\abstract{
In the AdS/CFT correspondence, a subregion of the CFT allows for the recovery of a corresponding subregion of the bulk known as its entanglement wedge. 
In some cases, an entanglement wedge contains a locally but not globally minimal surface homologous to the CFT subregion, in which case it is said to contain a python's lunch. 
It has been proposed that python's lunch geometries should be modelled by tensor networks that feature projective operations where the wedge narrows. 
This model leads to the python's lunch (PL) conjecture, which asserts that reconstructing information from past the locally minimal surface is computationally difficult.
In this work, we use cryptographic tools related to a primitive known as the Conditional Disclosure of Secrets (CDS) to develop consequences of the projective tensor network model that can be checked directly in AdS/CFT. 
We argue from the tensor network picture that the mutual information between appropriate CFT subregions is lower bounded linearly by an area difference associated with the geometry of the lunch. 
Recalling that the mutual information is also computed by bulk extremal surfaces, this gives a checkable geometrical consequence of the tensor network model.
We prove weakened versions of this geometrical statement in asymptotically AdS$_{2+1}$ spacetimes satisfying the null energy condition, and confirm it in some example geometries, supporting the tensor network model and by proxy the PL conjecture. 
}
\maketitle

\pagebreak

\tableofcontents

\flushbottom

\section{Introduction}\label{sec:introduction}

In the context of the AdS/CFT correspondence, the python's lunch (PL) conjecture \cite{brown2020python} gives a geometrical characterization of when large complexity appears in the dictionary relating bulk and boundary degrees of freedom.\footnote{Specifically, it concerns the complexity of operator reconstruction, rather than e.g. the complexity of geometry reconstruction discussed in \cite{bouland2019computational, Akers:2024wre}.}
Concretely, the appearance of locally but not globally minimal extremal surfaces indicates the presence of large complexity, with the bulk region between the local and global minimum being highly complex to recover from the boundary degrees of freedom. 
The appearance of large complexities has been argued to be necessary for the consistency of semi-classical physics: a developing perspective argues that while severe breakdowns of semi-classical physics can occur in quantum gravity even at low energy, they are not witnessed by computationally limited observers \cite{harlow2013quantum, susskind2016computational, susskind2016typical, kim2020ghost, brown2020python, akers2022black}. 

\begin{figure}
    \centering
    \includegraphics[width=0.75\linewidth]{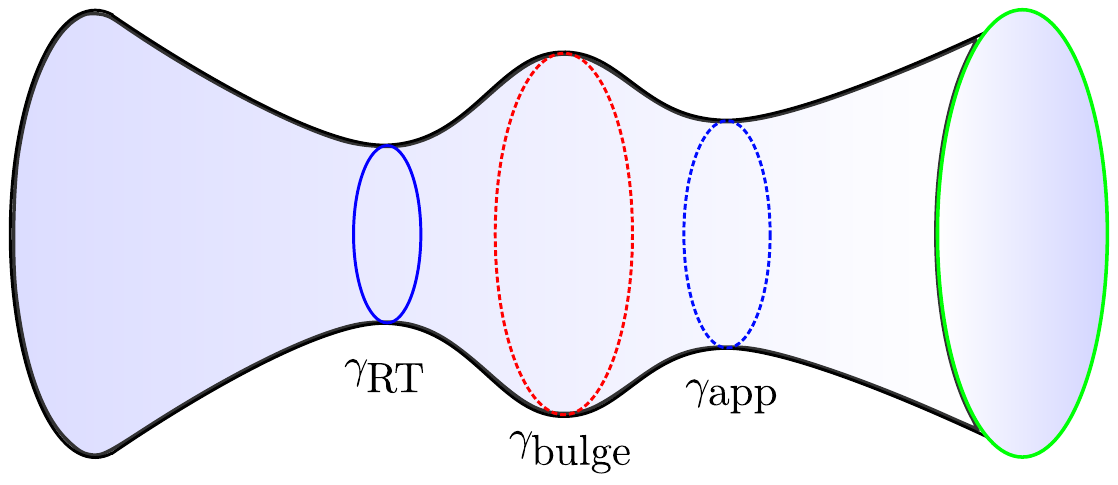}
    \caption{Spatial slice of a spacetime with a python's lunch. The light green curve denotes a boundary region $\hat{\mathcal{R}}$, the solid blue curve denotes the corresponding Ryu-Takayanagi surface, the dashed blue curve denotes a locally minimal ``appetizer'' surface, and the dashed red curve denotes the ``bulge'' surface. The region between the Ryu-Takayanagi surface and the appetizer surface is the ``python's lunch" for the region $\hat{\mathcal{R}}$.}
    \label{fig:intro_PL_geom}
\end{figure}

\begin{figure}
    \centering
    \includegraphics[width=0.75\linewidth]{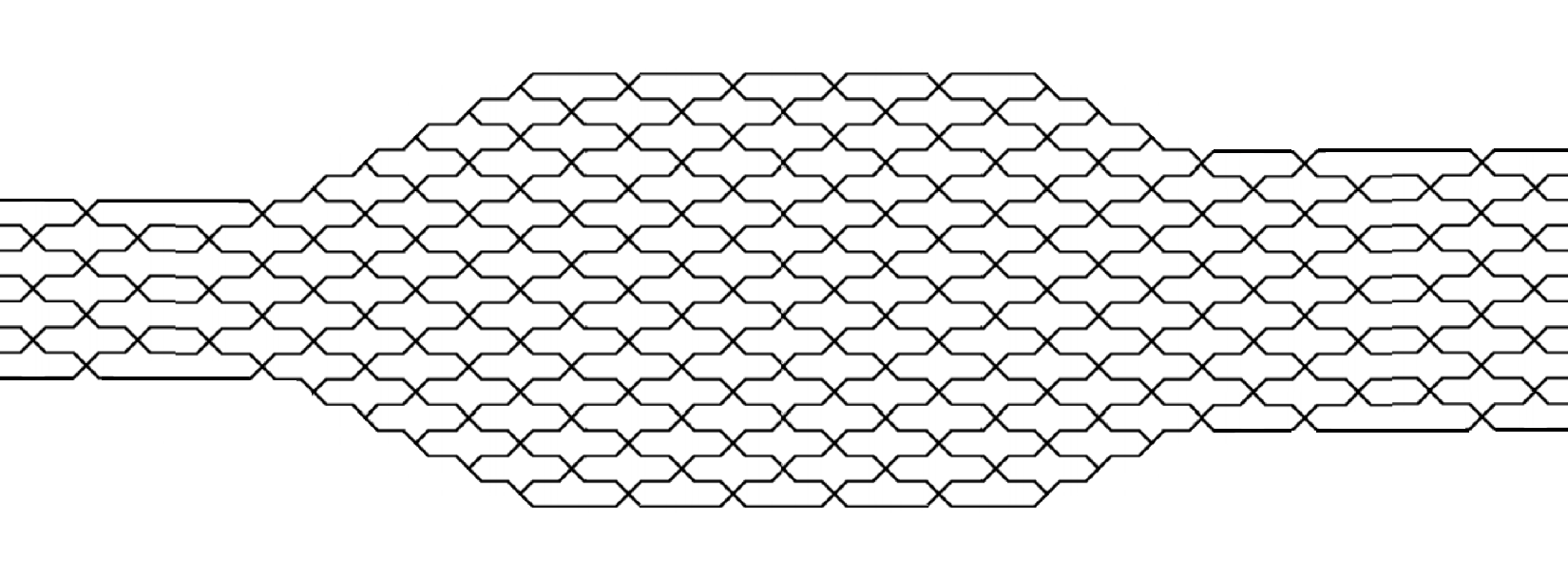}
    \caption{A tensor network model of a python's lunch geometry. Each vertex corresponds to a tensor, with the edges corresponding to tensor indices. Vertices are connected by edges according to the pattern in which tensors are contracted. 
    Figure from \cite{brown2020python}.}
    \label{fig:TNlunch}
\end{figure}

The PL conjecture is motivated by a tensor network model of holographic AdS spacetimes. 
The spacetime picture is shown in figure \ref{fig:intro_PL_geom}.\footnote{Our illustration depicts a wormhole spacetime with $\hat{\mathcal{R}}$ the entire right boundary, but similar considerations apply to any CFT subregion.} 
We consider a subregion $\R$ of the CFT and the globally minimal spacelike extremal surface which encloses it, known as the Ryu-Takayanagi surface \cite{Ryu:2006bv}. 
A well developed line of work \cite{czech2012gravity, headrick2014causality, wall2014maximin, dong2016reconstruction, cotler2019entanglement} has revealed that information in the entanglement wedge of $\R$, which extends from $\R$ to it's Ryu-Takayanagi surface, can be recovered from $\R$. 
The tensor network is a simple model of the geometry of the entanglement wedge; we show an example in figure \ref{fig:TNlunch}.  
We can read the tensor network as mapping from the globally minimal surface (the open legs on the left) to the boundary legs representing region $\R$ (the open legs on the right). 
In some cases there is a surface within the entanglement wedge of $\R$ which is locally but not globally minimal with respect to spacelike deformations. 
In this case the tensor network model has a constriction, as shown. 
We can view the tensors in the network as unitaries.
Then, where the network becomes wider, ancilla systems are being introduced, and where it becomes narrower, projections are occurring. 

In the tensor network model the number of these projections controls the complexity of recovering from the lunch. 
In particular, to recover information past the constriction in the tensor network model, the complexity is argued in \cite{brown2020python} to be $\mathcal{C}_0 2^{m_R/2}$, where $m_R$ is the number of qubits that are projected into on the right side of the lunch, and $\mathcal{C}_0$ is the complexity of the unitary defined by the tensor network.
Assuming this tensor network picture is a good model of the AdS/CFT setting, this becomes a complexity of
\begin{align}
    \mathcal{C} = \mathcal{C}_0 e^{\frac{\Delta A_{\text{PL}}}{8G_N}}
\end{align}
to recover from inside the lunch in the geometric picture.
Here we've defined the area difference
\begin{equation}
    \Delta A_{\text{PL}} = A[\gamma_{\text{bulge}}] - A[\gamma_{\text{app}}] \: ,
\end{equation}
with $\gamma_{\text{bulge}}$ (the ``bulge'') a maximal area surface sitting in the interior of the lunch, and $\gamma_{\text{app}}$ (the ``appetizer'') the outermost locally minimal surface.\footnote{These objects are defined more carefully in the main text.}

The python's lunch conjecture is remarkable in that it gives a geometrical characterization of complexity, a quantity which is typically difficult to compute and understand. 
This aspect of the conjecture also means it is difficult to verify: while the argument within the tensor network picture for the large complexity is compelling, it is more difficult to justify the claim that this tensor network picture accurately models this aspect of the AdS/CFT correspondence. 
Similar models have been effective in capturing information theoretic properties of AdS/CFT, for instance in elucidating the role of the entanglement wedge as the region reconstructable from a boundary subregion \cite{Pastawski:2015qua, Hayden:2016cfa}.  
In the information-theoretic case properties expected on the basis of tensor network models can be verified directly from the perspective of the gravitational path integral, where for example entropies can be computed \cite{Lewkowycz:2013nqa, Faulkner:2013ana}. 
However, the situation could be much different for computational properties.
For instance no such verification at the level of the gravitational path integral appears possible in the computational case. 

In this work we find implications of these projections in the tensor network model for boundary correlation. 
This bridges computational and information-theoretic properties of the network, and in particular produces verifiable predictions of the projective tensor network model for holographic CFTs. 
To develop consequences of this model, we consider quantum information processing tasks which rely on the existence of a hard-to-reconstruct region of spacetime. 
Specifically, we employ the conditional disclosure of secrets (CDS) cryptographic setting \cite{GERTNER2000592, gay2015communication, applebaum2017conditional, applebaum2020power, applebaum2021placing,allerstorfer2024relating, asadi2024conditional}. 
In CDS two parties, Alice and Bob, receive classical inputs $x$ and $y$ respectively. 
Alice and Bob cannot communicate with each other, but wish to reveal a secret to a third party, the referee, if and only if a condition on their joint inputs is met, say $f(x,y)=1$ for an appropriately chosen function $f$. 
We require that when $f(x,y)=0$ the referee cannot recover the secret using low complexity operations, but when $f(x,y)=1$ they should be able to recover the secret easily. 
We prove that accomplishing this requires the correlation shared by Alice and Bob to be lower bounded in a way that grows with how secure and how correct the CDS implementation is. 

\begin{figure}
    \centering
    \tdplotsetmaincoords{15}{0}
    \begin{tikzpicture}[scale=1.6,tdplot_main_coords]
    \tdplotsetrotatedcoords{0}{30}{0}
    
    \draw[gray] (-2,1,0) -- (-2,4.5,0);
    \draw[gray] (2,1,0) -- (2,4.5,0);
    
    \begin{scope}[tdplot_rotated_coords]
    
    \draw[domain=0:45,variable=\x,smooth, fill=black!60!,opacity=0.8] plot ({-2*sin(\x)}, {1+\x/45}, {2*cos(\x)}) -- plot ({-2*sin((45-\x))}, {3-(45-\x)/45}, {2*cos(45-\x)}) --  plot ({2*sin(\x)}, {3-\x/45}, {2*cos(\x)}) -- plot ({2*sin(45-\x)}, {1+(45-\x)/45}, {2*cos(45-\x)});
    
    \draw[domain=0:45,variable=\x,smooth,thick] plot ({-2*sin(\x)}, {1+\x/45}, {2*cos(\x)});
    \draw[domain=0:45,variable=\x,smooth,thick] plot ({2*sin(\x)}, {1+\x/45}, {2*cos(\x)});
    \draw[domain=0:45,variable=\x,smooth,thick] plot ({-2*sin(\x)}, {3-\x/45}, {2*cos(\x)});
    \draw[domain=0:45,variable=\x,smooth,thick] plot ({2*sin(\x)}, {3-\x/45}, {2*cos(\x)});
    
    \begin{scope}[canvas is xz plane at y=1]
    \draw[gray] (0,0) circle[radius=2] ;
    \end{scope}
    
    \begin{scope}[canvas is xz plane at y=4.5]
    \draw[gray] (0,0) circle[radius=2] ;
    \end{scope}
    
    \draw[red] (0,1,-2) -- (-0.25,1.75,-1);
    \draw[red] (0,1,2) -- (-0.25,1.75,1);
    
    \begin{scope}[canvas is xz plane at y=2]
    
    \draw[gray] (0,0) circle (2);
    
    \draw [domain=-45:45,fill=lightgray,opacity=0.8] plot ({2*cos(\x+90)}, {2*sin(\x+90)}) -- (-1.41,1.41) to [out=-45,in=45] (-1.41,-1.41) -- plot ({2*cos(\x-90)}, {2*sin(\x-90)}) -- (1.41,-1.41) to [out=135,in=-135] (1.41,1.41);
    
    \draw [green,ultra thick,domain=-45:45] plot ({2*cos(\x+90)}, {2*sin(\x+90)});
    
    \draw[blue,thick] (1.41,1.41) to [out=-135,in=+135] (1.41,-1.41);
    \draw[blue,thick] (-1.41,1.41) to [out=-45,in=45] (-1.41,-1.41);
    
    \end{scope}

    \begin{scope}[canvas is xz plane at y=3]
    \draw[red, dashed] (0,2) -- (0,-2);
    \draw[blue] (0,2) to[out=-60,in=60] (0,-2);
    \draw[dashed, blue] (0,2) to[out=-120,in=120] (0,-2);
    \draw (-0.9,0) -- (-1.1,0);
    \draw (-1,-0.1) -- (-1,0.1);
    \draw (-0.15,0) -- (-0.35,0);
    \draw (-0.25,-0.1) -- (-0.25,0.1);
    \draw[domain=0:180,variable=\x,smooth, green, ultra thick] plot ({2*cos(\x+90)},{2*sin(\x+90)});
    \end{scope}
    
    \draw[thick, red] (-0.5,2.5,0) -> (-1,3,0);
    \draw[thick, red] (-0.5,2.5,0) -> (-0.25,3,0);
    
    \draw[thick,red,-triangle 45] (-0.25,1.75,-1) -- (-0.5,2.5,0);
    \draw[thick,red,-triangle 45] (-0.25,1.75,1) -- (-0.5,2.5,0);
    
    \draw[domain=0:45,variable=\x,smooth, fill=black!50!,opacity=0.8] plot ({-2*sin(\x+180)}, {1+\x/45}, {2*cos(\x+180)}) -- plot ({-2*sin((45-\x)+180)}, {3-(45-\x)/45}, {2*cos(45-\x+180)}) --  plot ({2*sin(\x+180)}, {3-\x/45}, {2*cos(\x+180)}) -- plot ({2*sin(45-\x+180)}, {1+(45-\x)/45}, {2*cos(45-\x+180)});
    
    \begin{scope}[canvas is xz plane at y=2]
    \draw [green,ultra thick,domain=-45:45] plot ({2*cos(\x-90)}, {2*sin(\x-90)});
    \end{scope}
    
    \draw[domain=0:45,variable=\x,smooth,thick] plot ({-2*sin(\x+180)}, {1+\x/45}, {2*cos(\x+180)});
    \draw[domain=0:45,variable=\x,smooth,thick] plot ({2*sin(\x+180)}, {1+\x/45}, {2*cos(\x+180)});
    \draw[domain=0:45,variable=\x,smooth,thick] plot ({-2*sin(\x+180)}, {3-\x/45}, {2*cos(\x+180)});
    \draw[domain=0:45,variable=\x,smooth,thick] plot ({2*sin(\x+180)}, {3-\x/45}, {2*cos(\x+180)});
    
    \draw plot [mark=*, mark size=1.5] coordinates{(0,1,-2)};
    \node[below] at (0,1,-2) {$c_1$};
    \draw plot [mark=*, mark size=1.5] coordinates{(0,1,2)};
    \node[below] at (0,1,2) {$c_2$};
    \draw plot [mark=*, mark size=1.5] coordinates{(-0.5,2.5,0)};

    \node[left] at (0,4,-4) {$\R$};
    
    \end{scope}
    \end{tikzpicture}
    \caption{An asymptotically AdS$_{2+1}$ geometry. The boundary region $\R$ features a lunch, bounded by an extremal surface called the appetizer (dashed blue) which is locally minimal with respect to spacelike deformations, and an extremal surface called the RT surface (solid blue) which is globally minimal with respect to such deformations. Between them is a bulge surface (dashed red), which is extremal but not locally minimal. We argue that, assuming the python's lunch conjecture, when signals from $c_1$ and $c_2$ can meet in the bulk and then travel to either side of the appetizer surface, associated boundary regions $\V_1$ and $\V_2$ (shaded grey) must have mutual information satisfying a lower bound of the form \eqref{eq:lowerboundintro}, and in particular they must have a connected entanglement wedge. 
    }
    \label{fig:phasetransition3d}
\end{figure}
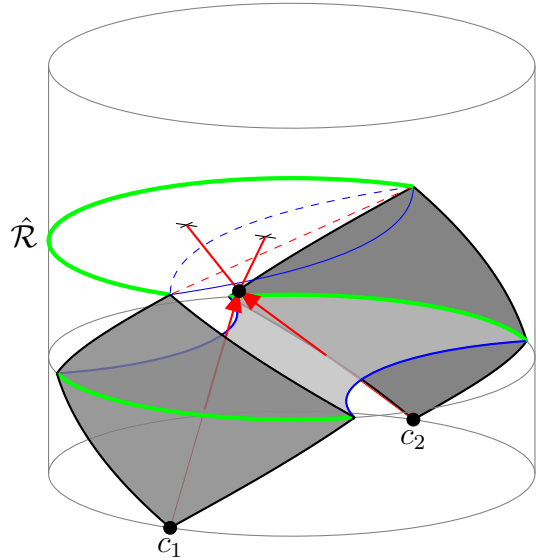

To relate the CDS setting to AdS/CFT, we give Alice, Bob and the referee control of carefully chosen subregions of the CFT. 
See figure \ref{fig:phasetransition3d} for an illustration. 
For appropriate geometries, Alice and Bob can exploit the fact that the correlations between them are described by a geometrical connection to complete the CDS task. 
Consider settings where Alice and Bob can meet in their dual bulk geometry, compute $f(x,y)$, and then choose to send the secret into the portion of the bulk spacetime that can be easily reconstructed from $\R$ or into the python's lunch of $\R$, contingent on the value of $f(x,y)$.  
According to the python's lunch conjecture, 
this ensures the secret is easy to reconstruct when $f(x,y)=1$ and hard to reconstruct when $f(x,y)=0$. 
Our lower bound on CDS then leads to the claim that Alice and Bob must share substantial correlation, and in particular a connected entanglement wedge. 

In fact, our bound implies more.
As we argue in more detail in the main text, in the tensor network model the success probability of learning information inside the lunch with low complexity operations is small, and determined by the number of projections appearing in the network between the secret and the boundary. 
Thus we expect the security parameter for the CDS protocol to be lower bounded by a geometrization of this number of projections. 
Combined with our lower bound on correlation in CDS in terms of the security parameter, we are led to propose a lower bound of the form
\begin{align}\label{eq:lowerboundintro}
    I(\V_1:\V_2) \stackrel{?}{\geq} \alpha_0 \left( \frac{ A[\gamma_{\text{bulge}}^X]-A[\gamma_{\text{app}}]}{4G_N} \right) \: .
\end{align}
Here $\gamma_{\text{bulge}}^X$ is a generalization of the bulge surface usually appearing in the python's lunch conjecture, which is intended to capture the number of projections referenced above. 
It's definition depends on a region, denoted $X$, which is the portion of the python's lunch into which the secret can be hidden.
In some cases, where $X$ is large and the secret can be hidden deep inside of the python's lunch, we expect $\gamma_{\text{bulge}}^X=\gamma_{\text{bulge}}$, and in general we know $A[\gamma_{\text{bulge}}^X]\leq A[\gamma_{\text{bulge}}]$.\footnote{We provide the definition of the region $X$ and the surface $\gamma_{\text{bulge}}^{X}$ in section \ref{sec:AdSlowerbound}.} 
The parameter $\alpha_0$ is a constant that is not known but is universal, by which we mean that $\alpha_0$ is the same value for all choices of regions for Alice, Bob, and the referee respecting the necessary boundary causal structure, and all choices of geometry.

We then study geometrically if the lower bound \eqref{eq:lowerboundintro} holds. 
As a first step we note that if $X$ is non-empty, so that the secret can be hidden somewhere in the python's lunch, the lower bound involves $A[\gamma_{\text{bulge}}^X]-A[\gamma_{\text{app}}] > 0$. 
This means in particular that $I(\V_1:\V_2)=\Theta(1/G_N)$.
We prove this for all asymptotically AdS$_{2+1}$ spacetimes satisfying the null energy condition. 
Going further, we also prove that 
\begin{align}\label{eq:weakenedlowerboundintro}
    I(\V_1:\V_2) \geq \frac{1}{2} \left( \frac{ A[\gamma_{\text{bulge}}^Y]-A[\gamma_{\text{app}}]}{4G_N} \right)
\end{align}
for a subset $Y \subset X$ that we construct, so that $A[\gamma_{\text{app}}] \leq A[\gamma_{\text{bulge}}^Y] \leq A[\gamma_{\text{bulge}}^X]$. 
This constitutes a weakened version of \eqref{eq:lowerboundintro}. 
These facts represent progress toward the lower bound \eqref{eq:lowerboundintro}, 
and can be viewed as circumstantial evidence validating the tensor network model and the role of projections in describing bulk geometry. 
Proving the full lower bound geometrically, or showing it is violated, remains an open problem. 

Ultimately, we would like to be able to apply this line of argument to provide evidence for or against the python's lunch conjecture directly, rather than for the projective tensor network model. 
To do this requires a lower bound on mutual information in CDS where the complexity of recovery appears.\footnote{The bound we currently use involves a parameter $\delta$ which is small whenever the secret is well hidden from low complexity operations. We relate this to the number of projections in the tensor network model to obtain \ref{eq:lowerboundintro}.} 
We explore a new lower bound on CDS of the form
\begin{align} \label{eq:lb_comp}
    I(A:B) \stackrel{?}{\geq} \min \{\beta_0 \ln t, I_{IT} \}
\end{align}
where $t$ is the complexity needed to recover the secret with high probability, $\beta_0$ is a constant, and $I_{IT}$ is the cost to perform information-theoretically secure CDS. 
This bound is plausible: a lower bound of this form is easily established for the \textit{dimensions} of the quantum systems to which Alice and Bob have access rather than the correlation they share, suggesting a similar bound on correlation is likely unless local randomness plays a significant role in an optimal CDS protocol. 
We leave proving this lower bound as an open problem; proving it seems likely to allow more direct probes of complexity in AdS/CFT.\footnote{There is a complication here in that one must find regimes where the $\ln t$ term is dominant over the information-theoretic term.}

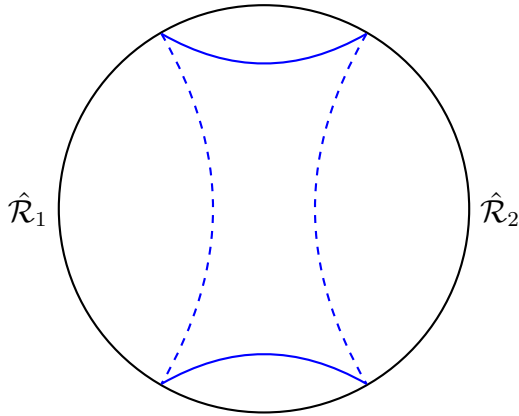
\begin{figure}
    \centering
    \begin{tikzpicture}[scale=0.9]
    
    \draw[thick] (0,0) circle (3);
    
    \draw[blue,dashed, thick] (1.5, 2.58) to [out=-120,in=120] (1.5, -2.58);
    \draw[blue,dashed, thick] (-1.5, 2.58) to [out=-60,in=60] (-1.5, -2.58);
    
    \draw[blue, thick] (-1.5, 2.58) to [out=-30,in=-150] (1.5, 2.58);
    \draw[blue, thick] (-1.5, -2.58) to [out=30,in=150] (1.5, -2.58);
    
    \node[right] at (3,0) {$\R_2$};
    \node[left] at (-3,0) {$\R_1$};
    
    \end{tikzpicture}
    \caption{Two intervals $\R_1$ and $\R_2$ in the boundary of vacuum AdS$_{2+1}$. The RT surface $\gamma_{\R_1\cup \R_2}$ defining the entanglement wedge of $\R_1\cup \R_2$ is shown in solid blue; the appetizer surface $\gamma_{\R_1}\cup \gamma_{\R_2}$ is in dashed blue. The lunch region sits between the appetizer and RT surfaces. We observe that situations in which the region $\hat{\mathcal{R}}$ consists of two intervals can furnish counter-examples to the bound \eqref{eq:conjlowerbound}.
    }
    \label{fig:counterexample}
\end{figure}

Another direction we explore is a natural strengthening of our conjectured lower bound \eqref{eq:lowerboundintro} to a bound
\begin{align}\label{eq:conjlowerbound}
    I(\V_1:\V_2) \stackrel{\times}{\geq} \alpha_0 \left( \frac{ A[\gamma_{\text{bulge}}]-A[\gamma_{\text{app}}]}{4G_N} \right) \: .
\end{align}
Such a lower bound is simpler than equation \eqref{eq:lowerboundintro}, and hence natural to consider from a geometrical standpoint. 
As well, it would follow from our CDS conjecture \eqref{eq:lb_comp} in settings where $\ln t \geq I_{IT}$,  assuming the python's lunch conjecture, further motivating its consideration. 
We study this bound in various families of asymptotically AdS$_{2+1}$ spacetimes, where we can compute both sides of the relation \eqref{eq:conjlowerbound}.
In the AdS$_{2+1}$ defect, the BTZ black hole, and vacuum AdS with a static end-of-the-world brane, we find agreement with \eqref{eq:conjlowerbound} for any $\alpha_0 \leq 1$. 
However, we construct an example in pure AdS$_{2+1}$, shown in figure \ref{fig:counterexample}, where \eqref{eq:conjlowerbound} is violated for all constants $\alpha_0$;  hence, we rule out \eqref{eq:conjlowerbound}. 
Consistency with both the python's lunch conjecture and the lower bound \eqref{eq:lb_comp} then would require that $I_{IT}$ is smaller than $\ln t$, which we argue should be the case in that example.  

The outline of our paper is as follows.
In section \ref{sec:toolsandnotation}, we introduce our notation and some quantum information tools required for later sections. 
Section \ref{sec:PLreview} then reviews the statement of the python's lunch conjecture and its motivation from tensor network models, while section \ref{sec:CDSintro} introduces the conditional disclosure of secrets setting from cryptography. 

In section \ref{sec:CDSlowerbound}, we develop a lower bound on the correlation needed to complete CDS; section \ref{sec:AdSlowerbound} develops the holographic setting and applies this CDS lower bound to AdS/CFT, culminating in the conjectured lower bound \eqref{eq:lowerboundintro}.
In section \ref{sec:lb_comp}, we provide additional context pertaining to the more speculative lower bound \eqref{eq:conjlowerbound}, which we show later can be violated. 

Section \ref{sec:geometryproof} proves a geometrical theorem that follows as a consequence of our lower bound \eqref{eq:lowerboundintro}. 
In particular, we show the mutual information is $\Theta(1/G_N)$ whenever the bulk geometry allows the computationally secure CDS task to be completed. 
Section \ref{sec:weak_bound} then strengthens this observation from a parametric statement to a more direct geometrical statement by establishing \eqref{eq:weakenedlowerboundintro}. 

Section \ref{sec:examples} gives the detailed calculations for the checks of the bound \eqref{eq:conjlowerbound} performed in example geometries, where we can study the dependence of the mutual information on $\Delta A_{\text{PL}}$.  
We begin in section~\ref{sec:vacuumexample} with the two interval example discussed in the introduction. 
Section \ref{sec:ETWbrane} gives our results for an ETW brane geometry. 
Section \ref{sec:defectandBTZ} reports the results of studying our lower bound in the AdS$_{2+1}$ defect and the BTZ black hole.
Appendix \ref{app:examples} supports these results with our detailed calculations. 

We conclude in section \ref{sec:discussion} with a discussion of our results and some comments on possible refinements of our work that may lead to a deeper understanding of the lunch conjecture and the role of complexity in gravity.  

\subsection{Tools and notation}\label{sec:toolsandnotation}

We summarize the notation used in this article below. 

\vspace{0.2cm}
\noindent \textbf{Geometrical notation:}
\begin{itemize}
    \item Bulk spacetime regions are denoted with capital script Latin letters, $\mathcal{V}, \: \mathcal{U}, \: \ldots$ We add a hat to take the restriction to the conformal boundary, $\V, \: \hat{\mathcal{U}}, \: \ldots$
    \item The causal future of a spacetime region $\mathcal{A}$ is denoted by $J^+(\mathcal{A})$; the causal past of a spacetime region $\mathcal{A}$ is denoted by $J^-(\mathcal{A})$.
    \item The domain of dependence of a region $\mathcal{A}$ is defined as the set of points $p$ such that every inextendible causal curve through $p$ intersects $\mathcal{A}$. We denote the domain of dependence by $D(\mathcal{A})$.
    \item The spacelike complement of a region $\mathcal{A}$ is denoted by $\mathcal{A}'$. 
    The spacelike boundary of a region $\mathcal{A}$ is denoted by $\partial \mathcal{A}$.
\end{itemize}
\noindent \textbf{Quantum notation:}
\begin{itemize}
    \item Quantum systems are labelled with capital Latin letters, $A, B, C,...$
    \item Bold, capital, script letters denote quantum channels, $\mathbfcal{N}, \mathbfcal{M}, ...$
    \item Bold, capital letters denote unitaries or isometries, $\mathbf{U}$, $\mathbf{V}$,...
\end{itemize}
We use big-O and related notations in the sense used by computer scientists. 

Next we introduce some tools from quantum information theory. 
We mostly follow the conventions of \cite{wilde2013quantum}. 
Recall the von Neumann entropy, 
\begin{align}
    S(A)_\rho = - \tr(\rho_A\ln \rho_A)
\end{align}
where we use a natural logarithm. 
The mutual information $I(A:B)_\rho$ is defined as
\begin{align}
    I(A:B)_\rho = S(A)_\rho + S(B)_\rho - S(AB)_\rho.
\end{align}
The relative entropy $D(\rho||\sigma)$ is defined as
\begin{align}
    D(\rho||\sigma) = \tr(\rho \ln \rho) - \tr(\rho \ln \sigma).
\end{align}
The fidelity is defined as
\begin{align}
    F(\rho,\sigma)=\tr\sqrt{\sqrt{\rho}\,\sigma \sqrt{\rho}}.
\end{align}
The mutual information can be equivalently expressed in terms of the relative entropy $D(\rho||\sigma)$ as
\begin{align}\label{eq:IandD}
    I(A:B)_\rho = D(\rho_{AB}||\rho_A\otimes \rho_B).
\end{align}
The relative entropy is related to the fidelity $F(\rho,\sigma)$ by,\footnote{This inequality follows for example from the observation that the relative entropy and log-fidelity are special cases of the $\alpha$-$z$ divergences \cite{audenaert2013alpha}, and these have a monotonicity property that relates these two objects.} 
\begin{align}\label{eq:DanfF}
    D(\rho||\sigma) \geq - 2\ln F(\rho,\sigma).
\end{align}
The diamond norm distance of two quantum channels $\mathbfcal{N}$, $\mathbfcal{M}$ is defined by
\begin{align}
    ||\mathbfcal{N}_{A\rightarrow B}-\mathbfcal{M}_{A\rightarrow B}||_\diamond  \equiv \sup_{n}\max_{\psi_{AR_n}} ||\mathbfcal{N}_{A\rightarrow B}(\psi_{AR_n})-\mathbfcal{M}_{A\rightarrow B}(\psi_{AR_n}) ||_1
\end{align}
where $\log \dim R_n = n$. 
It is always possible to restrict the dimension of the reference system $R_n$ in the above definition to at most $\dim A$.

By a \emph{quantum algorithm}, we mean a quantum channel $\mathbfcal{A}_{Q\rightarrow Z}$ which takes in quantum system $Q$ and outputs a single classical bit $Z$. 
We will denote the complexity  of an algorithm $\mathbfcal{A}_{Q\rightarrow Z}$ by $\mathcal{C}(\mathbfcal{A})$. 
By complexity we mean the number of quantum gates chosen from a fixed gate set needed to implement the channel within constant error (measured by the diamond norm). 

Given two density matrices $\rho$, $\sigma$, we define the computationally constrained distance by
\begin{align}
    D_{t}(\rho,\sigma) = \max_{\mathcal{C}(\mathcal{A})\leq t} |\text{Prob}[\mathcal{A}(\rho)=1]-\text{Prob}[\mathcal{A}(\sigma)=1]|.
\end{align}
This is related to the probability of distinguishing $\rho$ and $\sigma$ using an algorithm of complexity at most $t$ by
\begin{align}
    p_{\text{dist},t}(\rho,\sigma) =\frac{1}{2} + \frac{1}{2}D_t(\rho,\sigma).
\end{align}
It is straightforward to prove $D_t(\rho,\sigma)$ satisfies the triangle inequality, is symmetric, and is positive semi-definite. 
It need not be the case, however, that $D_t(\rho,\sigma)=0$ implies $\rho=\sigma$. 
To see why, notice that we can easily construct two states that can't be distinguished under a single qubit measurement (e.g. two different Bell states). 
Then taking $t=1$ gives that the $D_t$ distance between these is zero. 
This means $D_t(\rho,\sigma)$ does not define a metric; objects lacking only this property are sometimes called pseudo-metrics. 

We can also define a computational version of the diamond norm distance as follows, 
\begin{align}
    ||\mathbfcal{N}_{A\rightarrow B}-\mathbfcal{M}_{A\rightarrow B}||_{t} = \sup_d \max_{\psi_{AR_d}} D_t(\mathbfcal{N}_{A\rightarrow B}(\psi_{AR}), \mathbfcal{M}_{A\rightarrow B}(\psi_{AR}))
\end{align}
Similarly to the diamond norm, we can also notice that it suffices to restrict attention to reference systems with $n\leq t$ qubits, since if there are more qubits they cannot all be acted on by a gate, so those qubits can be removed without affecting the maximum.

\section{Background}

\subsection{Review of the python's lunch conjecture}\label{sec:PLreview}

Starting with Harlow and Hayden \cite{harlow2013quantum}, a number of works have explored the role of complexity in decoding the radiation of black holes \cite{hayden2007black,yoshida2017efficient, aaronson2016lecture, brakerski2023black}. 
A key idea that has emerged from this literature is that the computational difficulty of the black hole radiation decoding task may need to be high for the low-energy effective description of black holes to be consistent.

In \cite{brown2020python}, the authors conjectured a geometric origin for the computational hardness of decoding black hole radiation, or more broadly, of entanglement wedge reconstruction. 
To understand their claim, it is helpful to first consider a somewhat different setting pertaining to when decoding is possible in an information-theoretic sense. 
Consider a subregion $A$ of a holographic CFT. 
To calculate the entropy of $A$, we can use the Ryu-Takayanagi (RT) formula \cite{Ryu:2006bv, Hubeny:2007xt, Lewkowycz:2013nqa, Faulkner:2013ana, Engelhardt:2014gca}
\begin{align}\label{eq:RT}
    S(A) = \min_{\gamma_{\text{ext}}\in \text{Hom}(A)} S_{\text{gen}}(\gamma_{\text{ext}}) \: , \qquad S_{\text{gen}}(\gamma_{\text{ext}}) \equiv \frac{A[\gamma_{\text{ext}}]}{4G_N} + S_{\Sigma [\gamma_{\text{ext}}]} \: .
\end{align}
Here the minimization is over all spacelike, codimension-2 surfaces $\gamma_{\text{ext}}$ that are extrema of $S_{\text{gen}}$ and that are homologous to $A$; the surface which achieves the minimum in \eqref{eq:RT} is known as the RT surface $\gamma_{\text{RT}}$. We say that $\gamma$ is homologous to $A$ if there exists a codimension-1 surface $\Sigma [\gamma]$, called a homology slice, such that
\begin{align}
    \partial \Sigma[\gamma] = A \cup \gamma \: .
\end{align}
The quantity $S_{\Sigma[\gamma]}$ then denotes the von Neumann entropy of the bulk state on $\Sigma[\gamma]$. 
We are often interested in situations where the area term dominates the bulk entropy term in $S_{\text{gen}}$; in this case, the surfaces $\gamma_{\text{ext}}$ are extremal area surfaces, which are minimal when deformed in spacelike directions, and maximal when deformed in a timelike direction.

The domain of dependence of $\Sigma[\gamma_{\text{ext}}]$ is known as the entanglement wedge of $A$, which we denote $E_{A}$. 
Importantly, bulk operators in $E_{A}$ are exactly those operators that can be represented in the boundary as operators with support restricted to $A$ \cite{czech2012gravity, headrick2014causality, wall2014maximin, dong2016reconstruction, cotler2019entanglement}. 
Equivalently, the state on bulk degrees of freedom localized to $E_{A}$ can be recovered given access to $A$. 
More briefly, we say that $E_{A}$ is the bulk region that can be reconstructed from $A$. 

Returning to the computational setting, we can ask which portion of the bulk spacetime can be reconstructed \emph{efficiently}, in the sense of requiring only polynomial complexity quantum operations. 
The work \cite{brown2020python} proposed the following:
if there is only a single extremal surface candidate appearing in the minimization in the RT formula \eqref{eq:RT}, the full entanglement wedge can be efficiently reconstructed. 
If there is more than one extremal surface candidate, a portion of the bulk will be information-theoretically reconstructable but computationally hard to access. 

More specifically, when there exist multiple quantum extremal surfaces $\{ \gamma_{i}\}_{i}$ which are spacelike separated from one another, we can define an ordering on these surfaces via $W_{O}[\gamma_{i}] \subseteq W_{O}[\gamma_{j}]$ for $i > j$, where the ``outer wedge'' $W_{O}[\gamma]$ of $\gamma$ is defined to be the domain of dependence of the homology slice $\Sigma[\gamma]$. Intuitively, this ordering proceeds from the innermost extremal surface to the outermost extremal surface. 
A surface $\gamma$ which is a local minimum of $S_{\text{gen}}$ with respect to spacelike deformations is said to be ``outer minimal'' if there is no other extremum $\gamma' \in W_{O}[\gamma]$ with $S_{\text{gen}}(\gamma') < S_{\text{gen}}(\gamma)$ \cite{Engelhardt:2023bpv}. 
Then the region of the bulk in $W_{O}[\gamma_{\text{RT}}]$ but not in $W_{O}[\gamma_{i}]$ for any other outer minimal extremal surface, referred to as the \emph{python's lunch}, is claimed to have reconstruction complexity \cite{brown2020python, Engelhardt:2021qjs, Engelhardt:2023bpv}
\begin{equation} \label{eq:gen_PL}
    \ln \mathcal{C} = \max_{j > i} \left[ \frac{S_{\text{gen}}(\gamma_{i}) - S_{\text{gen}}(\gamma_{j})}{2} \right] + O(1) \: .
\end{equation}
On the other hand, the region within the entanglement wedge but not within the python's lunch, sometimes referred to as the ``simple wedge'', is claimed to have reconstruction complexity which is polynomial in $1 / G_{\text{N}}$. 
We refer to the $\gamma_{i}, \gamma_{j}$ which realize the maximum in \eqref{eq:gen_PL} as the ``bulge'' and ``appetizer'' surfaces respectively; by construction, the appetizer surfaces lies outside of the bulge surface, which itself lies outside of the RT surface. Given a homology surface $\Sigma[\gamma_{\text{RT}}]$ containing $\gamma_{\text{app}}$, the high complexity region of $\Sigma[\gamma_{\text{RT}}]$ is the region bounded by $\gamma_{\text{RT}} \cup \gamma_{\text{app}}$. 

A useful observation is that the bulge surface appearing in \eqref{eq:gen_PL} can be obtained via an extremization procedure, analogous to the ``maximin'' prescription for the Ryu-Takayanagi surface \cite{wall2014maximin}. 
Suppose we are given the RT surface $\gamma_{\text{RT}}$ and the appetizer surface $\gamma_{\text{app}}$, and let $\mathcal{W}_{\text{PL}}$ denote the domain of dependence of any partial Cauchy slice whose boundary is $\gamma_{\text{RT}} \cup \gamma_{\text{app}}$. Let $\Sigma_{\text{PL}}$ denote a partial Cauchy slice for $\mathcal{W}_{\text{PL}}$. We define a ``sweep-out'' of $\Sigma_{\text{PL}}$ as a smooth function $f : \Sigma_{\text{PL}} \rightarrow [0, 1]$ satisfying $f(\gamma_{\text{RT}}) = 0$ and $f(\gamma_{\text{app}}) = 1$, which one should think of as inducing a smooth foliation of $\Sigma_{\text{PL}}$ via level sets.\footnote{A definition of the bulge surface utilizing an alternate definition of sweep-outs, thought of more generally as continuous paths between $\gamma_{\text{RT}}$ and $\gamma_{\text{app}}$ through the space of homologous surfaces with a suitable topology, is provided in \cite{Arora:2024edk}. This definition allows for sweep-outs $f$ for which $\{ f(\eta) \}_{0 \leq \eta \leq 1}$ need not be a foliation. }
Then we can consider a ``maximinimax'' procedure
\begin{equation}
    \max_{\{\Sigma\}} \min_{\{ f_{\Sigma}\}} \max_{0 \leq \eta \leq 1} S_{\text{gen}}(f_{\Sigma}^{-1}(\eta)) \: ,
\end{equation}
which identifies the surface within a sweep-out of a partial Cauchy slice which maximizes $S_{\text{gen}}$, then minimizes over all such sweep-outs, and finally maximizes over all partial Cauchy slices. The bulge surface is precisely the surface which realizes this maximinimax procedure.  

The claim that the lunch is hard to reconstruct is motivated by a tensor network model of the spatial geometry of the entanglement wedge. 
We illustrate this in figure \ref{fig:TNlunch}.
The tensor network prepares the state on the boundary by a series of contractions. 
In this model, most of the tensors can be viewed as unitary gates, but tensors in the regions where the network is becoming thinner need to be viewed as unitaries with an additional projection on some of the legs. 
For a network composed of Haar random tensors, operators supported in the full entanglement wedge can be represented on the boundary legs. 
However, past the appetizer surface, where projections appear, these operators may be represented only by high complexity operators in $A$. 

In more detail, consider the reconstruction problem of mapping from the boundary Hilbert space to the bulk Hilbert space describing the interior of the lunch. 
Let $\ket{\alpha}$ be the state on the bulk legs of the tensor network. 
Then the network prepares a boundary state according to 
\begin{align}
    \ket{s} \propto \bra{0}^{\otimes m_R}\mathbf{U}_{TN} \ket{\alpha}\ket{0}^{\otimes m_L}.
\end{align}
Here $\mathbf{U}_{TN}$ is the action of the tensor network where we view each tensor as a unitary, $\ket{0}^{\otimes m_L}$ are the ancilla qubits inserted on the left side of the network as it widens, and $\bra{0}^{\otimes m_R}$ are the ancilla qubits projected onto on the right side of the network as it thins again. 
To invert this and recover $\ket{\alpha}$ from $\ket{s}$ is non-trivial because of the projections. 
The work \cite{brown2020python} gives an algorithm, similar to Grover's search algorithm, to do this inversion.
The algorithm involves applying a low complexity unitary $\mathbf{V}$ repeatedly, which has the following effect on the boundary state $\ket{s}$, 
\begin{align}\label{eq:lunchrecoveryalg}
    \mathbf{V}^\ell (\ket{s}\ket{0}^{\otimes n}) = \sin\left((2\ell+1)\theta\right) \ket{\alpha}\ket{0}^{\otimes m_R} + \cos\left((2\ell+1)\theta\right)\ket{\beta}\ket{0}^{\otimes m_R}.
\end{align}
The value of $\theta$ is set by the square root of the probability of projecting the $m_R$ ancilla qubits onto $\ket{0}^{\otimes m_{R}}$.
For a sufficiently long tensor network composed of random tensors, we will have $\theta \sim 2^{-m_R}$.
This means that, using the proposed algorithm, we need to apply  $\mathbf{V}$ a number $\Theta(2^{m_R/2})$ times to recover $\ket{\alpha}$ with order $1$ fidelity. 

The work \cite{brown2020python} argues that the above Grover-like algorithm is optimal, and hence the complexity of recovering from the lunch with fidelity near $1$ is
\begin{align}
    \mathcal{C} = \mathcal{C}_0 2^{\frac{m_R}{2}} \: ,
\end{align}
with $C_0$ polynomial in $m_R$ and in the number of qubits describing the bulk. 
To relate this to the geometry of the lunch, rather than to the geometry of a tensor network, they further argue that $m_R$ should be associated with the area difference between the bulge surface and appetizer surface of the lunch, measured in Planck units. 
This gives a complexity of reconstruction from the lunch of
\begin{align}
    \mathcal{C} = \mathcal{C}_0 e^{\frac{1}{2}\frac{\Delta A_{\text{PL}}}{4G_N}} \: ,
\end{align}
where $\Delta A_{\text{PL}} = A[\gamma_{\text{bulge}}] - A[\gamma_{ \text{app}}]$ and $\mathcal{C}_0$ is polynomial in $1/G_N$, so we arrive at the python's lunch conjecture.\footnote{More generally, the exponent should be one half of the difference in generalized entropies defined by the bulge and appetizer surfaces. The generalized entropy is defined with natural logarithms, which accounts for the change from $2$ to $e$ in our exponential compared to the qubit case.} 

One existing consistency check on the python's lunch proposal comes from the quantum focusing conjecture \cite{bousso2016quantum}.
The work \cite{engelhardt2021world} points out that, given quantum focusing, the interior of the lunch for a boundary region $\R$ is always outside the causal wedge\footnote{The causal wedge of a boundary spacetime region $\R$ is defined as $C(\R)=J^+(\R)\cap J^-(\R)$.} of $\R$.
This is necessary for the lunch to be high complexity to decode, since there are known efficient procedures for reconstructing the operators in the causal wedge of $\R$.
Further, they give a simple prescription for reconstructing the bulk up to the appetizer surface. 
This verifies the claim in the lunch conjecture that this region is easy to reconstruct. 

Note that there exists an extension of the python's lunch conjecture as we have formulated it here applicable to situations where there exist multiple locally minimal surfaces homologous to $A$ which may not be spacelike separated from one another. An account of this case can be found in \cite{Engelhardt:2023bpv}.  

\subsection{Computationally secure CDS}\label{sec:CDSintro}

\begin{figure}[ht]
\begin{subfigure}{0.45\textwidth}
\begin{center}
\begin{tikzpicture}[scale=0.4]
    
    \draw[thick] (-5,-5) -- (-5,-3) -- (-3,-3) -- (-3,-5) -- (-5,-5);
    
    \draw[thick] (5,-5) -- (5,-3) -- (3,-3) -- (3,-5) -- (5,-5);
    
    \draw[thick] (5,5) -- (5,3) -- (3,3) -- (3,5) -- (5,5);
    
    \draw[thick, mid arrow] (4,-3) -- (4.5,3);
    
    \draw[thick, mid arrow] (-4,-3) to [out=90,in=-90] (3.5,3);
    
    \draw[thick] (-3.5,-5) to [out=-90,in=-90] (3.5,-5);
    \draw[black] plot [mark=*, mark size=3] coordinates{(0,-7.05)};

    \node[left] at (0,1) {$M_a$};
    \node[right] at (4.5,0) {$M_b$};
    
    \draw[thick] (-4.5,-6) -- (-4.5,-5);
    \node[below] at (-4.5,-6) {$x, s$};
    
    \draw[thick] (4.5,-6) -- (4.5,-5);
    \node[below] at (4.5,-6) {$y$};
    
    \draw[thick] (4,5) -- (4,6);
    \node[above] at (4,6) {s iff $f(x,y)=1$};
    
    \end{tikzpicture}
\end{center}
\caption{}
\label{fig:CDSprotocolstructure}
\end{subfigure}
\begin{subfigure}{0.5\textwidth}
\begin{center}
    \begin{tikzpicture}[scale=0.4]
    
    \draw[thick] (-5,-5) -- (-5,-3) -- (-3,-3) -- (-3,-5) -- (-5,-5);
    
    \draw[thick] (5,-5) -- (5,-3) -- (3,-3) -- (3,-5) -- (5,-5);
    
    \draw[thick] (5,5) -- (5,3) -- (3,3) -- (3,5) -- (5,5);
    
    \draw[thick] (4,-3) -- (4.5,3);
    
    \draw[thick] (-4,-3) to [out=90,in=-90] (3.5,3);
    
    \draw[thick,dashed] (-3,-5.5) -- (3,-5.5);
    \draw[black] plot [mark=*, mark size=3] coordinates{(-3,-5.5)};
    \draw[black] plot [mark=*, mark size=3] coordinates{(3,-5.5)};
    \node[above] at (0,-5.5) {$r_0$};

    \draw[thick,dashed] (-3,-6.5) -- (3,-6.5);
    \draw[black] plot [mark=*, mark size=3] coordinates{(-3,-6.5)};
    \draw[black] plot [mark=*, mark size=3] coordinates{(3,-6.5)};
    \node[below] at (0,-6.5) {$r_1$};
    
    \draw[thick] (-4.5,-6) -- (-4.5,-5);
    \node[below] at (-4.5,-6) {$x,s$};
    
    \draw[thick] (4.5,-6) -- (4.5,-5);
    \node[below] at (4.5,-6) {$y$};

    \node[left] at (0,1) {$m_a=s\oplus r_x$};
    \node[right] at (4.5,0) {$m_b=r_y$};
    
    \draw[thick] (4,5) -- (4,6);
    
    \end{tikzpicture}
\end{center}
\caption{}
\label{fig:CDSexample}
\end{subfigure}
\caption{(a) General form of a CDQS protocol. Alice, on the bottom left, holds input $x$ and a secret $s$. Bob, on the bottom right, holds input $y$. Alice and Bob may share entanglement, represented as the lower curved wire. Alice and Bob perform quantum channels on their locally held systems and send the resulting outputs to the referee. The referee knows $x,y$ and should be able to determine $s$ if and only if $f(x,y)=1$, where $f$ is a function known to all parties. (b) A (classical) CDS protocol for the equality function on single bit inputs. Alice and Bob share two random bits, $r_0$ and $r_1$.  Alice sends the XOR of the secret $z$ with $r_x$. Bob sends $r_y$. If $x=y$, then the referee can compute $m_a\oplus m_b=s$, while if $x\neq y$ then $m_a$ and $m_b$ are independent random bits which reveal nothing about $s$.}
\end{figure}
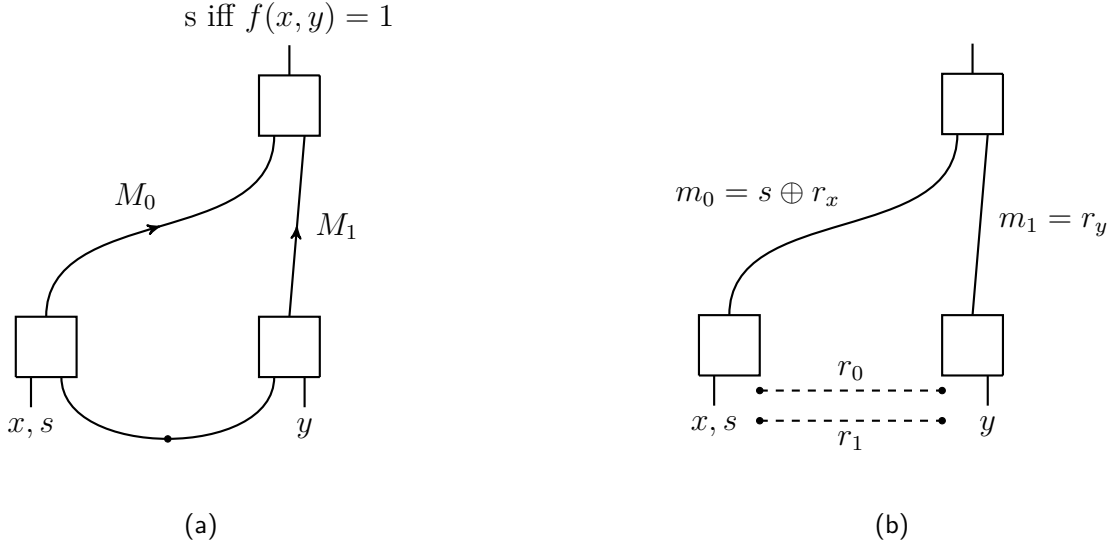

To obtain our constraints on boundary correlation from complexity, a key tool is the notion of \emph{conditional disclosure of secrets} (CDS) studied in cryptography \cite{GERTNER2000592, gay2015communication, applebaum2017conditional, applebaum2020power, applebaum2021placing,allerstorfer2024relating, asadi2024conditional}. 
The CDS setting involves three parties, labelled Alice, Bob and the referee. 
Alice holds an input string $x$ of length $n$ as well as a secret $s$, while Bob holds an input string $y$ of length $n$. 
Alice and Bob do not communicate but share a random string or hold subsystems of a joint quantum state, which may be entangled. 
Alice and Bob each compute a message from the data they respectively hold, and each send this message to the referee. 
The goal is for the referee to learn the secret $s$ if and only if a Boolean function on the inputs satisfies $f(x,y)=1$.
The general form of a CDS protocol is illustrated in figure \ref{fig:CDSprotocolstructure}. 

Naively, performing CDS might seem impossible. 
In fact, it turns out that CDS can be implemented for any function $f(x,y)$ given sufficient correlation \cite{GERTNER2000592}. 
For intuition, we give an example protocol for performing CDS in the classical setting on the function $f(x,y)=EQ(x,y)$, the equality function. 
See figure \ref{fig:CDSexample} for the protocol. 
We consider single bit inputs for concreteness. 
A similar strategy to the one given for equality can be generalized to perform CDS for any function. 

The protocol in figure \ref{fig:CDSexample} is information-theoretically secure, meaning that the message received on $f(x,y)=0$ inputs is uncorrelated with the secret. 
This means there is no computation acting on the message which guesses the secret with high probability. 
We can also ask instead for computational security, which is weaker, in that it requires only that a referee restricted to low complexity algorithms is unable to determine the secret when $f(x,y)=0$. 
In addition to relaxing to computational security, we can also allow the players access to quantum resources. 
It is computationally secure quantum CDS that we focus on in this work. 
We define it formally below. 

\begin{definition}\label{def:CDQS}
    A \textbf{conditional disclosure of quantum secrets (CDQS) with computational security} task is defined by a choice of function $f:\{0,1\}^{2n}\rightarrow \{0,1\}$, and a $d_Q$ dimensional Hilbert space $\mathcal{H}_Q$ which holds the secret.
    The task involves inputs $x\in \{0,1\}^{n}$ and secret $Q$ given to Alice, and input $y\in \{0,1\}^{n}$ given to Bob.
    Alice sends message system $M_a$ to the referee, and Bob sends message system $M_b$. 
    Label the combined message systems as $M=M_aM_b$.
    Label the quantum channel defined by Alice and Bob's combined actions $\mathbfcal{N}_{Q\rightarrow M}^{x,y}$. 
    We put the following two conditions on a CDQS protocol. 
    \begin{itemize}
        \item $\epsilon$\textbf{-correct:} There exists a channel $\mathbfcal{D}^{x,y}_{M\rightarrow Q}$, called the decoder, implementable by a polynomial-time computable circuit such that
        \begin{align}
            \forall (x,y)\in X\times Y \,\,\, s.t. \,\, f(x,y)=1,\,\,\, ||\mathbfcal{D}^{x,y}_{M\rightarrow Q}\circ \mathbfcal{N}^{x,y}_{Q\rightarrow M} - \mathbfcal{I}_{Q\rightarrow Q}||_\diamond \leq \epsilon.
        \end{align}
        \item $\delta$\textbf{-secure:} There exists a quantum channel $\mathbfcal{S}_{\varnothing \rightarrow M}^{x,y}$, called the simulator, such that
        \begin{align}
            \forall (x,y)\in X \times Y \,\,\, s.t. \,\, f(x,y)=0,\,\,\, ||\mathbfcal{S}^{x,y}_{\varnothing\rightarrow M}\circ \tr_Q- \mathbfcal{N}^{x,y}_{Q\rightarrow M}||_t \leq \delta(t,\lambda). \nonumber 
        \end{align}
    \end{itemize}
    Here $\delta(t,\lambda)$ is a real valued function with $\lambda$ a security parameter; we require $\delta \rightarrow 0$ as $\lambda \rightarrow \infty$. 
\end{definition}
The definition of security in the second condition makes use of the computational variant of the diamond norm distance introduced in section \ref{sec:toolsandnotation}.
To the best of our knowledge, computationally secure CDS has not been studied before, in either the classical or quantum settings. 

Before continuing, we should provide some intuition for the parameter $\delta$.
Let's suppose that the secret is a state $\ket{i}$ drawn uniformly from some fixed basis $\{\ket{j}\}_{j=1}^{d_{Q}}$, and that the referee will try to determine $j$ using operations of complexity at most $t$, despite being in an instance $f(x, y) = 0$. 
To do this, the referee will take the message system $\rho_{M}^{(i)} = \mathcal{N}_{Q \rightarrow M}^{x, y}(|i \rangle \langle i|)$ and perform some POVM $\{ E_{j} \}_{j=1}^{d_{Q}}$ (using operations of complexity at most $t$), obtaining outcome $j$ with probability
\begin{equation}
    p(j | i) = \text{tr} \left( \rho_{M}^{(i)} E_{j} \right) \: .
\end{equation}
We can consider one of $d_{Q}$ algorithms $\{\mathcal{A}_{j}\}_{j=1}^{d_{Q}}$ that output $1$ if this measurement returns $j$, and $0$ otherwise. 
Then our security definition says that the probability such an algorithm returns $1$ must be $\delta$ close to the probability that it would output $1$ if $\mathbfcal{N}^{x,y}_{Q\rightarrow M}$ was replaced with the simulator channel, so
\begin{equation}
    p(j|i) \leq \tr \left( \sigma_{M} E_{j} \right) + \delta \: , \qquad \sigma_{M} \equiv (\mathbfcal{S}^{x,y}_{\varnothing \rightarrow M} \circ \tr_Q) (\ketbra{i}{i}_Q) \: .
\end{equation}
Importantly, $\sigma_M$ is independent of $i$. 
It follows that the probability of guessing correctly, averaged over uniformly distributed inputs, satisfies
\begin{equation}
\begin{split}
    p_{\text{correct}} & \equiv \frac{1}{d_{Q}} \sum_{i=1}^{d_{Q}} p(i|i) \\
    & \leq \delta + \frac{1}{d_{Q}} \sum_{i=1}^{d_{Q}} \text{tr} \left( \rho_{\mathcal{S}} E_{i} \right) = \delta + \frac{1}{d_{Q}} \: .
\end{split}
\end{equation}
Thus, $\delta$ corresponds to the bias with which one can guess the secret when it is a uniformly distributed classical string $i$. 
Our definition is somewhat more general than this requirement alone; in particular, it extends the security requirement to the case where the secret is a general quantum state.  

A comment is that in the classical setting an information-theoretically secure CDS which has a low-complexity decoder can be adapted into a computationally secure CDS that uses less shared randomness. 
To do this, we simply replace the shared randomness by a shared, short key, that each party uses to compute a pseudo-random string.
This pseudo-random string is then used to replace the randomness in the original (information-theoretic) CDS protocol.
To a low complexity referee, the pseudo-random string is indistinguishable from a truly random string, so the protocol must behave the same as the information theoretic one in the low complexity regime. 
In particular, since the decoder is assumed to be low complexity, it must still correctly recover the secret. 
Further, any low-complexity referee cannot learn the secret when $f(x,y)=0$, since they couldn't do so in the original information-theoretic protocol.

A key result that we will prove and make use of about computationally secure CDQS is the following theorem. 
\begin{theorem}\label{thm:CDSlowerbound}
Suppose we have an $\epsilon$-correct and $\delta$-secure CDQS for a function $f$, and which hides a $d_Q$ dimensional secret. 
Label the shared resource state held by Alice and Bob as $\Psi_{AB}$. Assume that $f(x,y)$ has the property that there exists an $x=x_*$ such that $f(x_*,\cdot)$ is non-constant.
Then, the resource state must satisfy
\begin{align}
    I(A:B)_\Psi \geq -\ln\left(\frac{1}{\sqrt{d_Q}}+\delta+\epsilon \right) - 1
\end{align}
\end{theorem}
This is saying that to achieve a highly correct and highly secure CDQS protocol that hides a large secret, we need large correlation.
There is also a requirement that the function $f(x,y)$ not be too trivial.

\section{Lower bound on CDS and application to AdS/CFT}\label{sec:lowerbound}

\subsection{Lower bound on correlation in CDS}\label{sec:CDSlowerbound}

In this section, we prove theorem \ref{thm:CDSlowerbound}. 
Intuitively, the proof starts with the observation that when the resource system is uncorrelated between Alice and Bob, all protocols must be insecure.
The reason is that the referee can, when $f(x,y)=0$, throw away Bob's system and locally produce the message Bob would have sent for a different value of $y$. 
Choosing the new value of $y$, call it $y'$, to be such that $f(x,y')=1$, the referee can then recover the secret by correctness of the protocol. 
If the resource state is close to product this strategy will work well, so to achieve high security and correctness we need a resource system that is far from product.
Being far from product then imposes a lower bound on the mutual information. \\

\begin{proof}\, \textbf{(Of theorem \ref{thm:CDSlowerbound})}
    We first describe a recovery procedure that works well when the resource system is close to
    product.
    Let the shared state used in the CDQS protocol be $\Psi_{AB}$. 
    By assumption, there exists a value $x_*\in X$ such that $f(x_*,\cdot):Y\rightarrow Z$ is non-constant.
    Consider inputs $(x_*,y_*)$ such that $f(x_*,y_*)=0$, and a second choice of $y$, call it $y_*'$ such that $f(x_*,y_*')=1$.
    Let Alice's operation on the left given input $x$ be denoted $\mathbfcal{N}^x_{QA\rightarrow M_a}$ and Bob's operation on the right given input $y$ be denoted $\mathbfcal{N}^y_{B\rightarrow M_b}$. 
    Label system $M_aM_b$ as $M$.
    For inputs $(x_*,y_*)$, the referee receives the system
    \begin{align}
        \rho_M(x_*,y_*) = \mathbfcal{N}^{x_*}_{QA\rightarrow M_a}\otimes \mathbfcal{N}_{B\rightarrow M_b}^{y_*} (\psi_{RQ}\otimes \Psi_{AB}).
    \end{align}
    We take the input $\psi_{RQ}$ to be the maximally entangled state between the secret system $Q$ and a reference system $R$. 
    
    The recovery procedure works as follows: the referee traces out $M_b$ and replaces it with $\mathbfcal{N}^{y_*'}_{B\rightarrow M_b}(\Psi_B)$, obtaining the state
    \begin{align}
        \sigma_M(x_*,y_*') = \mathbfcal{N}^{x_*}_{QA\rightarrow M_a}\otimes \mathbfcal{N}_{B\rightarrow M_b}^{y_*'} (\psi_{RQ}\otimes \Psi_A\otimes \Psi_B).
    \end{align}
    Then, the referee applies the decoder $\mathbfcal{D}^{x_*,y_*'}_{M\rightarrow Q}$. 
    A short calculation gives a bound on how well this procedure recovers the secret,
    \begin{align}
        ||\mathbfcal{D}^{x_*,y'_*}(\sigma_M(x_*,y_*')) -\psi_{RQ}||_1 &\leq ||\mathbfcal{D}^{x_*,y'_*}(\sigma_{M}(x_*,y_*'))-\mathbfcal{D}^{x_*,y'_*}(\rho_{M}(x_*,y_*'))||_1\,+ \nonumber \\
        & \qquad \qquad \qquad ||\mathbfcal{D}^{x_*,y'_*}(\rho_{M}(x_*,y_*'))-\psi_{RQ}||_1 \nonumber \\ 
        &\leq ||\sigma_{M}(x_*,y_*')-\rho_{M}(x_*,y_*')||_1 + \epsilon \nonumber \\
        &\leq ||\Psi_A \otimes \Psi_B - \Psi_{AB}||_1 + \epsilon \nonumber \\
        & \leq 2\sqrt{1-F(\Psi_{AB},\Psi_A \otimes \Psi_B)} + \epsilon.
    \end{align}
    To obtain the second line, we used that the trace distance is decreasing under the action of a quantum channel to bound the first term, and used correctness on the $(x_*,y_*')$ input to bound the second term. 
    To obtain the third line we again used that the trace distance is decreasing under the action of a quantum channel, where now the channel considered is the encoding procedure of Alice and Bob. 
    The final line follows from the Fuchs-van de Graaf inequality.
    The final line expresses that if the resource state is close to product, then the referee recovers well. 
    
    We now compare this to the $\delta$-security requirement of CDQS that the referee should not be able to recover well. 
    By $\delta$-security, and taking $\sigma_Q$ to be the output of the simulator channel appearing in the definition, we see that there exists a state $\sigma_Q$ such that
    \begin{align}
        ||\mathbfcal{D}^{x_*,y'_*}(\sigma_{M}(x_*,y_*')) - \pi_{R}\otimes \sigma_Q ||_1 \leq \delta
    \end{align}
    where $\pi_R$ is the maximally mixed state.
    This says the recovered state is close to product. 
    Now, we see that we can get an upper bound on the fidelity $F(\Psi_{AB},\Psi_A\otimes \Psi_B)$: this can't be too small because we can't have the recovered state be both near product and near the maximally entangled state $\psi_{QR}$. 
    In particular, a calculation gives how far in trace distance a product state is from the maximally entangled state:
    \begin{align}
        ||\psi_{RQ}- \pi_{R} \otimes \sigma_Q|| \geq 2\left(1-\frac{1}{\sqrt{d_Q}}\right).
    \end{align}
    Now, we combine the last three statements.
    First, use the reverse triangle inequality to write,
    \begin{align}
        ||\mathbfcal{D}^{x_*,y'_*}(\sigma_M(x_*,y_*')) -\psi_{RQ}||_1 \geq ||\psi_{RQ}- \pi_{R} \otimes \sigma_Q|| - ||\mathbfcal{D}^{x_*,y'_*}(\sigma_{M}(x_*,y_*')) - \pi_{R}\otimes \sigma_Q ||_1 \nonumber 
    \end{align}
    Then we substitute the last three numbered statements. 
    Doing so we find
    \begin{align}
        F(\Psi_{AB},\Psi_A\otimes \Psi_B) \leq 2\left( \frac{1}{\sqrt{d_Q}} + \epsilon + \delta \right).
    \end{align}
    Now use \eqref{eq:IandD} giving the mutual information in terms of the relative entropy and the lower bound \eqref{eq:DanfF} on the relative entropy from fidelity to obtain
    \begin{align}
        I(A:B)_{\Psi} \geq -2\ln F(\Psi_{AB}, \Psi_A \otimes \Psi_B)
    \end{align}
    which gives the claimed lower bound. 
\end{proof}

In our model of CDQS, the referee obtains the message systems but holds no other quantum systems. 
We can consider extending this to allow the referee to hold an arbitrary additional `advice' system, which they might use to help decode the message.
This system can be high complexity to prepare, and can depend on $(x,y)$, but must be prepared before receiving the messages.
Our theorem above continues to apply to that setting so long as the advice is uncorrelated with the resource system. 

\subsection{Holographic correlation and CDS lower bounds}\label{sec:AdSlowerbound}

In this section, we give our argument relating mutual information and the geometry of the python's lunch. 
The main idea is to realize a CDS protocol using AdS/CFT. 
In the boundary picture, we arrange things such that the communication pattern is limited to the general form of a CDS protocol (figure \ref{fig:CDSprotocolstructure}). 
Considering the bulk perspective, we use the bulk geometry to allow Alice and Bob to meet and decide to reveal the secret or not, then send the secret towards the referee or away from them as appropriate. 
In particular we assume the python's lunch conjecture and use the bulk region it designates as ``hard to reconstruct'' to hide the secret when $f(x,y)=0$. 
We then return to the boundary picture and apply theorem \ref{thm:CDSlowerbound} to place a constraint on boundary correlation.
Our reasoning is in the spirit of an operational viewpoint on holography developed in \cite{may2019quantum, may2020holographic, may2021holographic, may2022connected, may2022complexity}.

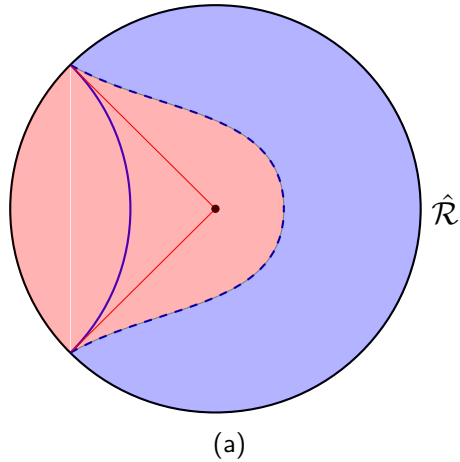
\begin{figure}
    \centering
    \subfloat[\label{fig:withoutmatter}]{
    \begin{tikzpicture}[scale=0.9]
    
    \draw[thick] (0,0) circle (3);
    
    \draw[blue, thick] (-2.12, 2.12) to [out=-45,in=45] (-2.12, -2.12);
    \draw[blue,dashed,thick] (-2.12, 2.12) to [out=-30,in=90] (1,0);
    \draw[blue, dashed, thick] (-2.12, -2.12) to [out=30,in=-90] (1,0);
    
    \draw[red] (-2.12,2.12) -- (0,0) -- (-2.12,-2.12);
    
    \draw[black] plot [mark=*, mark size=1.5] coordinates{(0,0)};
    
    \node[right] at (3,0) {$\R$};
    \draw[domain=225:-45,variable=\x,smooth,fill=blue,opacity=0.3] (-2.12, 2.12) to [out=-30,in=90] (1,0) to [out=-90,in=30] (-2.12, -2.12) plot ({3*sin(\x)}, {3*cos(\x)});

    \draw[domain=-45:-135,variable=\x,smooth,fill=red,opacity=0.3] (-2.12, 2.12) to [out=-30,in=90] (1,0) to [out=-90,in=30] (-2.12, -2.12) plot ({3*sin(\x)}, {3*cos(\x)});
    
    \end{tikzpicture}
    }
    \caption{An example of a python's lunch geometry, with a single time slice depicted. The blue dashed surface is the appetizer, the red surface is the bulge, and the solid blue surface is the Ryu-Takayanagi surface. The easy wedge $E_{\R}^{ \text{easy}}$ is shown in blue and the hard wedge $(E^{\text{easy}}_{\R})'$ is shown in red.}
    \label{fig:simpleandhard}
\end{figure}

\vspace{0.2cm}
\noindent \textbf{Geometric set-up:} It will be helpful to introduce definitions for the bulk subregions defined by the python's lunch construction. 
Consider a boundary subregion $\R$. 
We will focus for simplicity on settings where there are exactly two locally minimal extremal surfaces homologous to $\R$, and a single bulge surface between them. 
We refer to the locally but not globally minimal surface as the appetizer surface $\gamma_{\text{app}}$, the globally minimal surface as the RT surface $\gamma_{\text{RT}}$, and the locally maximal surface as the bulge surface $\gamma_{\text{bulge}}$. 
Then we make the following definitions:
\begin{itemize}
    \item The \emph{easy wedge} of $\R$ is the domain of dependence of the codimension $1$ region $r_s$ such that $\partial r_s = \R \cup \gamma_{\text{app}}$. We label this region $E^{\text{easy}}_{\R}$.
    \item The \emph{hard wedge} of $\R$ is the spacelike complement of $E^{\text{easy}}_{\R}$. We denote this as $(E^{\text{easy}}_{\R})'$. 
\end{itemize}
See figure \ref{fig:simpleandhard} for an illustration of these regions in a simple example. 
Notice that the hard wedge, in our definition, includes the spacelike complement of the entanglement wedge.  
Thus, the hard wedge also includes the bulk region that cannot be reconstructed on $\R$ even information-theoretically.\footnote{This is somewhat awkward, but will simplify some remarks later. The reader may wish to think of this region as the `at least hard, but maybe impossible' region.} 

We now begin constructing our CDS protocol within AdS/CFT. 
Choose two spacelike separated points $c_1$ and $c_2$ in the conformal boundary of an asymptotically AdS$_{2+1}$ spacetime, and a boundary region $\R$. 
Consider introducing inputs $x,Q$ at $c_1$, and the input $y$ at $c_2$.
The region $\R$ corresponds to the data available to the referee in the CDS protocol.
Define boundary spacetime regions
\begin{align}
    \V_1 &= \hat{J}^+(c_1) \cap \hat{J}^-(\R)\cap \hat{J}^-(\hat{\mathcal{R}'}), \nonumber \\
    \V_2 &= \hat{J}^+(c_2) \cap \hat{J}^-(\R)\cap \hat{J}^-(\hat{\mathcal{R}'}).
\end{align}
We refer to these as the \emph{decision regions}; they are the boundary spacetime regions where one of the inputs is available, and where signals can be sent to $\R$ or its complement. 
The dynamics happening in the decision regions correspond to the lower left and right operations in the CDS protocol. 
Note that it is crucial that the $\V_i$ be restricted to the past of the spacelike complement of $\R$; requiring this corresponds in the circuit picture to allowing Alice and Bob to perform partial traces (since sending a system to $\R'$ keeps it away from the referee, and hence traces it out). 
Identifying the decision regions $\V_1$ and $\V_2$ with Alice and Bob's systems, and $\R$ as the referee's systems, our boundary set-up shares the structure of a CDQS protocol.

Now consider the bulk picture. 
We are interested in the case where the bulk spacetime region
\begin{align} \label{eq:scatt_reg_def}
    J_{2 \rightarrow 2} \equiv J^+(E_{\mathcal{V}_1})\cap J^+(E_{\mathcal{V}_2}^{\text{easy}}) \cap J^-(E^{\text{easy}}_{\mathcal{R}}) \cap J^-((E^{\text{easy}}_{\R})')
\end{align}
is non-empty. 
This allows for the following bulk process: Alice and Bob take their respective inputs and send them into the above bulk spacetime region.\footnote{Notice we take the future of the easy wedge of $\V_{2}$ rather than the future of its entanglement wedge; this will be needed later, as we will need that Bob's encoding procedure is low complexity.} 
There, the inputs meet and the function $f(x,y)$ is computed locally. 
If $f(x,y)=1$, the secret $Q$ is sent into $E^{\text{easy}}_{\R}$. 
If $f(x,y)=0$, $Q$ is sent into $(E^{\text{easy}}_{\R})'$.
From this bulk picture, and assuming the python's lunch conjecture, we see that the boundary CFT implements computationally secure CDQS: the referee who has access to $\R$ can reconstruct $Q$ with low complexity operations if $f(x,y)=1$ and cannot reconstruct $Q$ with low complexity operations if $f(x,y)=0$, completing the CDQS protocol. 
We will call the set of points into which the secret can be hidden $X$. 
Formally, $X$ is the intersection of the future of the scattering region with the hard wedge,
\begin{equation} \label{eq:X_def}
    X = J^{+}[J_{2 \rightarrow 2}] \cap (E^{\text{easy}}_{\R})' \: .
\end{equation}

Next, recall theorem \ref{thm:CDSlowerbound} states
\begin{align}
    I(A:B)_\Psi \geq -\ln\left(\frac{1}{\sqrt{d_Q}}+\delta+\epsilon \right) - 1 \: .
\end{align}
This applies whenever the whenever the referee can apply a decoder with complexity equal to Bob's encoder.
We consider a low complexity referee who can access the easy wedge, and will understand the $\delta$, $\epsilon$ parameters relevant for that referee.
Recall also that this bound applies to every density matrix $\Psi_{AB}$ that suffices to complete CDQS with a $d_Q$ dimensional secret, security error $\delta$, and correctness error $\epsilon$.
This means we can phrase the above as a lower bound on $I_{\text{min}}=I_{\text{min}}(\delta, \epsilon, n_Q)$, the minimal mutual information needed to complete the task with parameters $\delta, \epsilon, n_Q$, 
\begin{align}\label{eq:Imin}
    I_{\text{min}} \geq -\ln\left(\frac{1}{\sqrt{d_Q}}+\delta+\epsilon \right) - 1 \: .
\end{align} 
To apply this in the holographic setting, we need to understand the relevant values of $\delta$, $\epsilon$, and $d_Q$ in AdS/CFT. 

\begin{figure}[h!]
    \centering
    \begin{tikzpicture}[scale=0.35]

\draw[thick] (19,3) -- (16,3) -- (14,5) -- (12,5) --( 11,6)--(9,6) -- (8,7) -- (6,7)--(1,2);
\draw[thick] (19,-3) -- (16,-3) -- (14,-5) -- (12,-5) --( 11,-6)--(9,-6) -- (8,-7) -- (6,-7)--(1,-2);

\draw[thick] (0.5,2.5) -- (1,2);
\draw[thick] (0.5,1.5) -- (1,2);
\draw[thick] (0.5,0.5) -- (1,0);
\draw[thick] (0.5,-0.5) -- (1,0);
\draw[thick] (0.5,-2.5) -- (1,-2);
\draw[thick] (0.5,-1.5) -- (1,-2);

\draw[thick] (1,2) -- (9,-6);
\draw[thick] (1,-2) -- (9,6);
\draw[thick] (1,0) -- (8,7);
\draw[thick] (1,0) -- (8,-7);
\draw[thick] (2,3) -- (11,-6);
\draw[thick] (2,-3) -- (11,6);
\draw[thick] (3,4) -- (12,-5);
\draw[thick] (3,-4) -- (12,5);
\draw[thick] (4,5) -- (14,-5);
\draw[thick] (4,-5) -- (14,5);
\draw[thick] (5,6) -- (15,-4);
\draw[thick] (5,-6) -- (15,4);
\draw[thick] (6,7) -- (16,-3);
\draw[thick] (6,-7) -- (16,3);
\draw[thick] (9,6) -- (18,-3);
\draw[thick] (9,-6) -- (18,3);
\draw[thick] (12,5) -- (19,-2);
\draw[thick] (12,-5) -- (19,2);
\draw[thick] (16,3) -- (19,0);
\draw[thick] (16,-3) -- (19,0);
\draw[thick] (18,3) -- (19,2);
\draw[thick] (18,-3) -- (19,-2);

\draw[thick,red] (12.5,-6) -- (12.5,6);

\draw[blue] plot [mark=*, mark size=5] coordinates{(13,0)};
  
\end{tikzpicture}
\caption{Tensor network representing an entanglement wedge. We consider reconstructing a quantum state located at the site marked with the blue dot. The number of projections appearing in the map from the red curve to the boundary may be less than the full area difference between the bulge and the appetizer.}
\label{fig:tikztensor}
\end{figure}
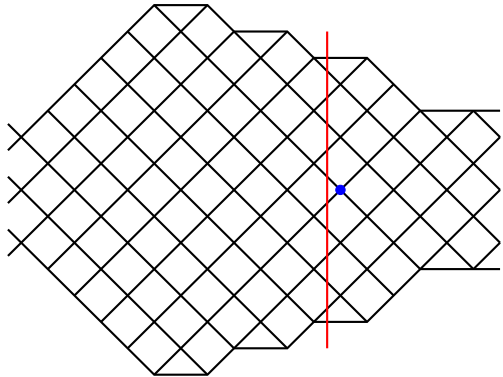

\vspace{0.2cm}
\noindent \textbf{Value of $\delta$:} To understand what values of $\delta$ we can achieve in the holographic context, let's return to the tensor network model discussed in section \ref{sec:PLreview}.
This model provides a suggestion of how small we are allowed to take $\delta$. 
To see this, suppose we wish to recover the state of a quantum system located at a site $p$ inside the lunch. 
Consider any cut $\gamma$ through the network which is homologous to $\R$ and encloses site $p$. 
Then let $\mathbf{V}_{TN}$ be the unitary defined by the tensor network that maps from $\gamma$ to $\R$, so we have
\begin{align}
    \rho_{\hat{\mathcal{R}}} = \frac{1}{Z} \bra{0}^{\otimes m_R'} \mathbf{V}_{TN}\left(\ketbra{\alpha}{\alpha}_Q\otimes \rho_b \right)\mathbf{V}_{TN}\ket{0}^{\otimes m_R'}
\end{align}
Here $Z$ is a normalization factor, $Q$ is the bulk subsystem we are trying to learn about, and $\rho_{b}$ describes the remaining bulk degrees of freedom, including the cut $\gamma$ and any remaining bulk legs.  
See figure \ref{fig:tikztensor}. 

Note that we don't expect to be able to fully invert the map and recover $\ket{\alpha}$ with high fidelity for a general cut $\gamma$.  
However, we can still learn something about $\ket{\alpha}$.
For simplicity, suppose $\ket{\alpha} = \ket{i_*}$ is uniformly drawn from some orthonormal basis $\{\ket{i}\}_i$; this is not general but will suffice for our purposes. 
We can then try to measure $\rho_{\R}$ in a suitable basis and determine $i_*$. 
One (low-complexity) strategy we have for doing this is the following. 
We introduce an ancilla in the state $\ket{0}^{\otimes m_R'}$, apply $\mathbf{V}_{TN}^\dagger$, and then measure $\{\ket{i}\}_i$. 
Measuring each value of $i$ occurs with probability
\begin{align}
    p(i|i_{*}) = \bra{i}_Q\text{tr}_{b}\left(  \mathbf{V}_{TN}(\ketbra{0}{0}^{\otimes m_R'}\otimes \rho_{\hat{\mathcal{R}}})\mathbf{V}_{TN}^\dagger \right)\ket{i}_Q .
\end{align}
Computing this for a Haar random $\mathbf{V}_{TN}$, we find at leading order that
\begin{align}
    p(i|i_{*}) = \frac{1}{d_{Q}} + \frac{1}{2^{m_{R}'}} \delta_{i i_{*}} \: .
\end{align}
This allows us to produce a guess of the secret $i$ which is biased towards being correct by $1/{2^{m_{R}'}}$. 
Recalling the discussion after definition \ref{def:CDQS}, we note that this means the security parameter appearing in the CDQS definition is not smaller than $\delta=1/2^{m_R'}$.

In the gravity picture we would like to identify this number of projections with an appropriate area difference. 
To do this we follow a similar procedure as appears in the initial argument for the python's lunch conjecture.\footnote{We note that there are some ambiguities in choosing a covariant geometrical version of the the number of projections $m_{R}'$ encountered in the tensor network model; for example, rather than maximizing over $\Sigma_{\text{PL}}$, one could fix the slice $\Sigma_{\text{PL}}$ to be that singled out by the maximinimax prescripition for the usual bulge surface. We leave exploring these alternatives to future work.} 
Let $\mathcal{W}_{\text{PL}}$ denote the domain of dependence of any partial Cauchy slice whose boundary is $\gamma_{\text{RT}} \cup \gamma_{ \text{app}}$, and let $\Sigma_{\text{PL}}$ denote such a partial Cauchy slice. For spacetime region $Z$, let $\Sigma_{\text{PL}}^{Z} = \Sigma_{\text{PL}} \cap Z$; we will generally consider $Z$ such that the intersection of $Z$ with every $\Sigma_{\text{PL}}$ is non-empty. 
Moreover, let $\gamma^{Z}$ be a curve in $\Sigma_{\text{PL}}$ which is homologous to $\gamma_{\text{app}}$ and such that the partial Cauchy slice $\Sigma_{O}[\gamma^{Z}] \equiv \Sigma_{\text{PL}} \cap W_{O}[\gamma^{Z}]$ contains $\Sigma_{\text{PL}}^{Z}$ (recall from section \ref{sec:PLreview} that $W_{O}[\gamma]$ is the outer wedge of $\gamma$). In other words, we require that all of $\Sigma_{\text{PL}}^{Z}$ is outside of $\gamma^{Z}$. 
Then let $f_{\gamma^{Z}} : \Sigma_{O}[\gamma^{Z}] \rightarrow [0, 1]$ be a smooth function with $f_{\gamma^{Z}}(\gamma^{Z}) = 0$ and $f_{\gamma^{Z}}(\gamma_{\text{app}}) = 1$. 
Finally, we consider the ``maximinimax'' procedure
\begin{equation} \label{eq:restricted_bulge}
    \max_{\{\Sigma_{\text{PL}}\}} \min_{\{\gamma^{Z}, f_{\gamma^{Z}} \}} \max_{0 \leq \eta \leq 1} S_{\text{gen}}(f_{\gamma^{Z}}^{-1}(\eta)) \: .
\end{equation}
We define $\gamma_{\text{bulge}}^{Z}$ to be the surface on which this maximinimax takes its value. 
In other words, we are performing a procedure similar to that for the usual bulge, but where the sweep-outs only need to cover the part of the lunch intersecting with $Z$. 

We claim that, taking $Z = X$ in the above definition with $X$ defined in \eqref{eq:X_def}, the area difference between the ``restricted bulge'' surface $\gamma^{X}$ and the appetizer should be identified with the number of projections $m_{R}'$ appearing in the corresponding network.
In this case, $\Sigma_{\text{PL}}^{X}$ is the subset of $\Sigma_{\text{PL}}$ into which the secret could possibly be hidden, and we expect $m_{R}'$ to correspond to a ``geometric bottleneck'', i.e. the area difference between the appetizer surface and a minimax surface with respect to sweep-outs covering $\Sigma_{\text{PL}}^{X}$ on an appropriate slice. 
With this identification, 
\begin{align}
    \delta \sim \exp \left( -\frac{A[\gamma^{X}_{\text{bulge}}]-A[\gamma_{\text{app}}]}{4G_N} \right) \: .
\end{align}
An observation is that the minimization includes cases for which $\gamma^{X}$ is the RT surface and $f_{\gamma^{X}}$ is therefore maximized on the true bulge $\gamma_{\text{bulge}}$, implying that $A[\gamma_{\text{bulge}}^{X}] \leq A[\gamma_{\text{bulge}}]$. 
This means
\begin{align}
    \delta \geq \exp \left( -\frac{A[\gamma_{\text{bulge}}]-A[\gamma_{\text{app}}]}{4G_N} \right) = \frac{1}{\mathcal{C}} \: .
\end{align}

\vspace{0.2cm}
\noindent \textbf{Value of $d_Q$:}
Let the number of qubits in the secret be $n_Q$ so that we have a $d_Q=2^{n_Q}$-dimensional secret. 
We should choose the size of the secret such that sending it into the bulk does not back-react on the geometry, so we want to have $n_Q$ grow strictly slower than $1/G_N$.  
Thus we set
\begin{align}
    n_Q=h(-\ln \delta),\qquad h(x)=o(x).
\end{align}
Since $-\ln \delta$ is $\Theta(1/G_N)$, this ensures $n_Q=o(1/G_N)$ so we do not induce a back-reaction on the geometry. 

\vspace{0.2cm}
\noindent \textbf{Value of $\epsilon$:} Naively, in the semi-classical bulk picture, the referee can recover the secret perfectly when it is placed into the easy wedge. 
However, this is corrected by sub-dominant saddles in the path integral, so we expect errors of order $\epsilon=2^{-\Theta(1/G_N)}$. 
We will assume the errors can indeed be taken this small. 
This means they are sub-leading compared to the $1/2^{n_Q/2}$ term and so can be dropped. 

With these choices, our lower bound is
\begin{align}
    I_{\text{min}} \geq - \ln \left(\frac{1}{2^{h(-\ln \delta)}} + \delta \right).
\end{align}
Notice that if $I_{\text{min}}$ grew slower than $-\ln \delta$ as $\delta \rightarrow 0$, we could choose an $h(\cdot)$ growing close enough to linearly such that our lower bound would grow faster than $I_{\text{min}}$. 
To avoid this contradiction then we must have
\begin{align}
    I_{\text{min}} \geq -\alpha_0\ln \delta
\end{align}
for a constant strictly positive $\alpha_0$. 
Importantly, we needed to lower bound $I_{\text{min}}$, rather than $I(\V_1:\V_2)_\Psi$ for any fixed choice of $\Psi$. 
Had we applied the above argument to $I(\V_1:\V_2)_\Psi$, the `constant' $\alpha_0$ appearing could depend on $\Psi$, and so wouldn't be universal. 
Because we bound $I_{\text{min}}$, however, which allows for any choice of $\Psi_{\V_1\cup \V_2}$, we get a universal $\alpha_0$.  
We can now use that $I(\V_1:\V_2)_\Psi \geq I_{\text{min}}$ to find the same lower bound on the information in any concrete setting we wish to study. 

Finally, we record for later use our main claim from this section, which is that
\begin{align}\label{eq:mainlowerbound}
    \boxed{I(\V_1:\V_2) \geq \alpha_0 \left( \frac{A[\gamma_{\text{bulge}}^{X}] - A[\gamma_{\text{app}}]}{4G_N} \right) \: .}
\end{align}
Note that the factor of $4$ in the denominator is conventional.
We also note that this bound trivializes when the scattering region is empty; in that case, $X$ is empty, so the appetizer surface itself is a legitimate choice of $\gamma^{X}$, and we obtain $A[\gamma_{\text{bulge}}^{X}] = A[\gamma_{\text{app}}]$. 

\vspace{0.2cm}
\noindent \textbf{Comments on assumptions:} 
An assumption we have made can be found in our matching of the protocol structure of CDQS, given in figure \ref{fig:CDSprotocolstructure}, to the holographic setting. 
In the CDQS protocol structure, we assume the referee does not hold any quantum systems correlated with Alice and Bob before receiving their messages.
In the holographic setting, the referee's spacetime region $\R$ has $\V_1$ and $\V_2$ in its past (where we think of the messages as originating), but additionally some degrees of freedom in $\R$ do not come from the $\V_i$, and those degrees of freedom are correlated with the $\V_i$. 
A more complete model then would allow the referee in CDS to have some additional quantum systems potentially correlated with the resource system held by Alice and Bob.
Unfortunately in that case we do not know of a lower bound that suffices for our purposes.  

Nonetheless, we argue \eqref{eq:mainlowerbound} should hold in holographic systems. 
Partly this is because this same assumption regarding extra regions was used in \cite{may2019quantum,may2020holographic,may2021holographic,may2021quantum,may2022connected}, where it was used to give an information-theoretic argument for a result known as the connected wedge theorem. 
The connected wedge theorem could then be proven independently, which suggests the validity of the assumption. 
More broadly the intuition is that correlation between Alice and Bob is necessary, because the task they are trying to achieve is dependent on how Alice and Bob's inputs are correlated. 
Nonetheless, we ideally would remove this assumption, perhaps using a setting similar to the one discussed in \cite{may2024non}. 

A minimal interpretation of the python's lunch conjecture is that recovery from a lunch with fidelity near $1$ requires complexity of $\text{poly}(1/G_N) \exp \left( \frac{1}{2} \frac{\Delta A_{\text{PL}}}{4G_N} \right)$. 
To derive \eqref{eq:mainlowerbound}, we had to use instead that the number of projections appearing in the tensor network is controlling $\delta$. 
This assumption however seems closely tied to the validity of the python's lunch conjecture. 
Indeed, the argument for the lunch conjecture within the tensor network model is very strong. 
The crucial idea however is the claim that the tensor network model accurately describes the geometry and it's computational properties. 
Here, we are performing a check on the validity of this model.
Ideally, we would improve our argument to take as input only the python's lunch conjecture itself, rather than the model it is argued from, since in principle the conjecture could be true even while the model is inaccurate in some ways. 
This is what motivates consideration of the stronger geometrical bound \eqref{eq:conjlowerbound}, which directly relates the complexity as predicted by the PL conjecture to the mutual information $I(\V_{1}:\V_{2})$; we elaborate on this bound in the next subsection.  

\subsection{A lower bound from complexity?} \label{sec:lb_comp}

To directly relate the python's lunch conjecture and boundary entanglement, we should look for a lower bound on correlation in CDQS where the complexity of recovering with high probability appears.  
A natural such statement is given by the following conjecture. 
\begin{conjecture}\label{conj:CDS}
    There exists a polynomial complexity family of functions $f_n:X\times Y\rightarrow Z$ such that performing computationally secure CDQS on this family of functions $f_n$, with $\epsilon$, $\delta$ sufficiently small constants, requires the resource system $\Psi_{AB}$ to have mutual information $I(A:B)_{\Psi}$ lower bounded according to
    \begin{align}
        I(A:B)_{\Psi} \geq \min\{I_{IT}(f_n), \alpha \ln t\}
    \end{align}
    where $I_{IT}(f_n)$ is the mutual information needed to perform information-theoretic CDS for $f_n$, and $t$ is the complexity below which the CDS protocol is $\delta$ secure.
\end{conjecture}
This conjecture claims a lower bound in terms of the complexity at fixed $\epsilon$, $\delta$.
Notice also that the minimization in the lower bound ensures it is never larger than the information-theoretic lower bound (which is secure with access to arbitrary complexity).

One line of thought that suggests this conjecture is that we have been able to prove an analogous statement with the mutual information replaced by the number of qubits held by each of Alice and Bob.\footnote{This is straightforward to show. Assume the protocol is information-theoretically insecure, so that there exists a map acting on the messages received by the referee that recovers the secret. At worst this map is exponential complexity in the number of qubits of message, so that $t \lesssim 2^{n_M}$ and $\ln t \lesssim n_m$. Then we use that the message is not bigger than the resource system (counting all ancillas) used by Alice and Bob, so $n_m\leq n_{AB}$ and so $\ln t \lesssim n_{AB}$.} 
Thus, we've shown that Alice and Bob's resource system size must grow as we scale $t$ (until saturating at the information-theoretic value), but not that the amount of correlation they share must grow. 
In practice, we know of no protocols that use a local system much larger than the shared correlation, so we are led to expect the above lower bound.

Assuming this bound and the python's lunch conjecture, we can reason as follows. 
In the holographic setting, so long as we can scatter to somewhere in the lunch, the discussion in section \ref{sec:AdSlowerbound} gives that $\epsilon$, $\delta\rightarrow 0$ as $G_N\rightarrow 0$.
This means that we can always take $G_N$ small enough so that the bound in conjecture \ref{conj:CDS} would hold.
But then according to the PL conjecture, we should identify $\ln t=\frac{A[\gamma_{\text{bulge}}]-A[\gamma_{\text{app}}]}{4G_N}$, so we are led to the claim
\begin{align}\label{eq:conjlowerboundmain}
    I(\V_1:\V_2) \stackrel{?}{\geq} \alpha_0 \left( \frac{ A[\gamma_{\text{bulge}}]-A[\gamma_{\text{app}}]}{4G_N} \right) \: .
\end{align}
which should hold for $A[\gamma_{\text{bulge}}]-A[\gamma_{\text{app}}]$ small enough, and in particular smaller than $I_{IT}$.
Since $I_{IT}$ is expected to be very large, this would imply \eqref{eq:conjlowerboundmain} would hold in many settings.
Geometrically, the simplest way to enforce this would just be to have \eqref{eq:conjlowerboundmain} be true generally, so it seems worth exploring if this could be the case. 

Intriguingly, we find geometrically that \eqref{eq:conjlowerboundmain} is false in general, though the violations occur in a setting where we expect $I_{IT}$ is becoming small.
This allows for consistency with conjecture \ref{conj:CDS}. 
This is discussed in section \ref{sec:vacuumexample}. 
In other examples we find consistency with \eqref{eq:conjlowerboundmain} with the largest consistent $\alpha_0$ being $1$. 

\section{Geometry of the lunch and correlation}\label{sec:geometryproof}

In the last section, we arrived at a prediction for the behaviour of the mutual information by assuming the python's lunch conjecture. 
We now begin studying geometrically if this prediction holds.

\subsection{Proof of a \texorpdfstring{$\Theta(1/G_N)$}{TEXT} lower bound}

In this section we verify a generic aspect of the lower bound: we show that whenever the bulk thought experiment realizing CDQS is possible, the boundary has $\Theta(1/G_N)$ mutual information. 
We show this in asymptotically AdS$_{2+1}$ dimensional spacetimes satisfying the null energy condition (NEC). 
This verifies a prediction of the tensor network model; in particular it is consistent with having a parameter $\delta$ capturing the hardness of reconstructing within the lunch with low complexity operations where $\delta\sim e^{-\alpha \ell/G_N}$ for $\ell$ a quantity with dimensions of length and $\alpha$ a dimensionless constant, rather than $\ell=\Delta A_{\text{PL}}$ in particular. 
We study the more detailed version of our claim in the next section, though we view this check as already a significant test of the projective tensor network model.  

The result we prove in this section is the following. 
\begin{theorem}\label{thm:crudetheorem}
Consider an asymptotically global AdS$_{2+1}$ spacetime satisfying the null energy condition. 
Choose boundary points $c_1$ and $c_2$, and a boundary region $\R$ that is the domain of dependence of a single interval. 
Then define
\begin{align}
    \V_1 &= J^+(c_1) \cap J^-(\R)\cap J^-(\hat{\mathcal{R}'}), \nonumber \\
    \V_2 &= J^+(c_2) \cap J^-(\R)\cap J^-(\hat{\mathcal{R}'}) \: .
\end{align}
Define the scattering region, 
\begin{align}
    J_{2 \rightarrow 2}\equiv J^+(E_{\V_1})\cap J^+(E_{\V_2}^{\text{easy}}) \cap J^-(E^{\text{easy}}_{\hat{\mathcal{R}}}) \cap J^-((E^{\text{easy}}_{\R})') \: .
\end{align}
Then, $J_{2 \rightarrow 2} \neq \emptyset \implies I(\V_1:\V_2)=\Theta(1/G_N)$.
\end{theorem}
Notice this is a strictly weaker claim than our lower bound in \eqref{eq:mainlowerbound}, which in particular implies a $\Theta(1/G_N)$ lower bound for the mutual information whenever the scattering region $J_{2 \rightarrow 2}$ is non-empty. 

To prove this theorem, we observe that a more general claim can already be extracted from the literature. 
\begin{theorem}\label{thm:connectedwedge}
    Consider an asymptotically global AdS$_{2+1}$ spacetime satisfying the null energy condition. 
    Choose points $c_1$, $c_2$ and boundary domains of dependence $\R,\S$. From these objects, define boundary regions 
    \begin{align}
        \V_1 = \hat{J}^+(c_1)\cap \hat{J}^-(\hat{\mathcal{R}}) \cap \hat{J}^-(\hat{\mathcal{S}}), \nonumber \\
        \V_2 = \hat{J}^+(c_2)\cap \hat{J}^-(\hat{\mathcal{R}}) \cap \hat{J}^-(\hat{\mathcal{S}}) \: .
    \end{align}
    and let $\mathcal{R}$ meet the conformal boundary at $\R$, and $\mathcal{S}$ meet the conformal boundary at $\S$.
    Then whenever 
    \begin{align}
        J(\mathcal{R},\mathcal{S}) &\equiv J^+(E_{\V_1}) \cap J^+(E_{\V_2}) \cap J^-(\mathcal{R}) \cap J^-(\mathcal{S})
    \end{align}
    is non-empty and $\partial \mathcal{R}$, $\partial \mathcal{S}$ are extremal surfaces, we have
    \begin{align}
        \frac{1}{2}I(\V_1:\V_2) \geq \frac{A[r]}{4G_N}
    \end{align}
    where $r$ is known as the ridge and is defined by
    \begin{align}
        r=\partial J^+(E_{\V_1})\cap \partial J^+(E_{\V_2}) \cap J^-(\mathcal{R})\cap J^-(\mathcal{S}) \: .
    \end{align}
    In particular we have $I(\V_1:\V_2)=\Theta(1/G_N)$.
\end{theorem}
At a geometrical level, this is similar to a statement proven in \cite{may2020holographic, may2021holographic,may2022connected}, which discusses the case where $\mathcal{S}=E_{\S}$, $\mathcal{R}=E_{\R}$. 
That case has an information-theoretic interpretation in terms of entanglement requirements on non-local quantum computation. 
Inspecting their geometrical proof, we can notice that the somewhat more general statement given in the theorem above also holds. 

Theorem \ref{thm:crudetheorem} then follows by applying this generalized connected wedge theorem to the choices
\begin{align}
    \S &= \R' \: , \nonumber \\
    \mathcal{R} &= E^{\text{easy}}_{\R} \: , \nonumber \\
    \mathcal{S} &= (E^{\text{easy}}_{\R})' \: .
\end{align}
Note that with this choice we have that $\partial \mathcal{R}=\gamma_{ \text{app}}=\partial \mathcal{S}$, so in particular these surfaces are extremal as needed for the connected wedge theorem. 
Further, we notice that 
\begin{align}
    J_{2 \rightarrow 2}\equiv J^+(E_{\V_1})\cap J^+(E_{\V_2}^{\text{easy}}) \cap J^-(E^{\text{easy}}_{\hat{\mathcal{R}}}) \cap J^-((E^{\text{easy}}_{\R})') \: .
\end{align}
being non-empty implies that 
\begin{align}
    J_{2 \rightarrow 2}'\equiv J^+(E_{\V_1})\cap J^+(E_{\V_2}) \cap J^-(E^{\text{easy}}_{\hat{\mathcal{R}}}) \cap J^-((E^{\text{easy}}_{\R})') \: .
\end{align}
is non-empty, since $E_{\V_2}^{\text{easy}} \subseteq E_{\V_2}$. 

\subsection{Proof of a weakened lower bound} \label{sec:weak_bound}

In this section we prove a weakened version of our claimed lower bound \eqref{eq:lowerboundintro} with the region $X$ replaced by a subset $Y\subseteq X$. 

\begin{theorem}\label{thm:Ylowerbound}
    Consider an asymptotically global AdS$_{2+1}$ spacetime satisfying the null energy condition. 
    Choose points $c_1$, $c_2$ and a boundary domain of dependence $\R$. From these objects, define boundary regions 
    \begin{align}
        \V_1 &= \hat{J}^+(c_1)\cap \hat{J}^-(\R) \cap \hat{J}^-(\R'), \nonumber \\
        \V_2 &= \hat{J}^+(c_2)\cap \hat{J}^-(\R) \cap \hat{J}^-(\R') \: .
    \end{align}
    Consider the ridge surface $r \subseteq J_{2 \rightarrow 2}$ as defined in theorem \ref{thm:connectedwedge}, and define 
    \begin{align}\label{eq:Ydef}
        Y = J^+(r) \cap \partial J^+(E_{\V_1}) \cap \partial J^+(E_{\V_2}^{\text{easy}}) \cap (E^{\text{easy}}_{\R})' \: .
    \end{align}
    Then we have that
    \begin{align}
        I(\V_1:\V_2) \geq \frac{1}{2}\left(\frac{A[\gamma_{\text{bulge}}^Y]-A[\gamma_{\text{app}}]}{4G_N}\right) \: ,
    \end{align}
    where the definition of $\gamma_{\text{bulge}}^{Z}$ for spacetime region $Z$ appears around \eqref{eq:restricted_bulge}. 
\end{theorem}

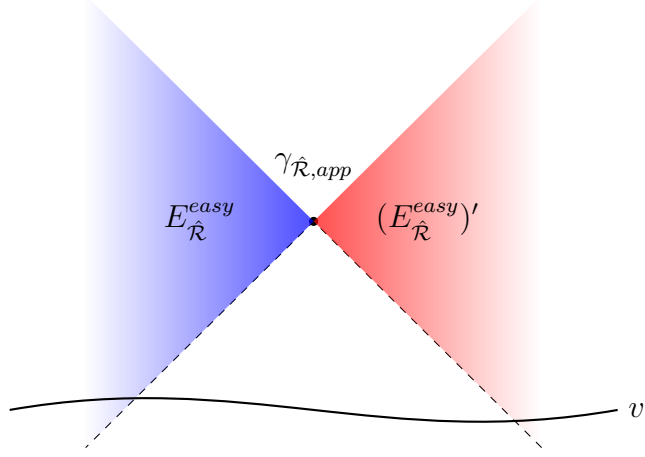
\begin{figure}
    \centering
    \begin{tikzpicture}

    \draw[black] plot [mark=*, mark size=1.5] coordinates{(3,2.5)};
    \node[above] at (3,2.9) {$\gamma_{\R, app}$};
    \draw[dashed] (6,-0.5) -- (3,2.5) -- (0,-0.5);

    \fill[smooth,path fading=fade left, blue] (0,5.5) -- (3,2.5) -- (0,-0.5);
    \fill[smooth,path fading=fade right, red] (6,-0.5) -- (3,2.5) -- (6,5.5);
    
    \node at (1.5,2.5) {$E^{easy}_{\R}$};
    \node at (4.5,2.5) {$(E^{easy}_{\R})'$};
    
    \draw[thick] (-1,0) to [out=10,in=-170] (7,0);
    \node[right] at (7,0) {$v$};
        
    \end{tikzpicture}
    \caption{The past null sheets of the appetizer surface and the ridge. The surface $v$ shown is defined by $v=\partial J^+(E_{\V_1})\cap \partial J^+(E_{\V_2}^{\text{easy}})$. The portion in the past of the appetizer surface is the ridge, $r$. The proof of theorem \ref{thm:Ylowerbound} involves focusing the ridge in two directions, along the light sheets $\partial J^+(E_{\V_1})$ and $\partial J^+(E_{\V_2}^{\text{easy}})$. In each case, one end-point of the resulting surface sits inside of $E_{\R}^{\text{easy}}$ and the other sits inside of $(E_{\R}^{\text{easy}})'$.}
    \label{fig:ridge}
\end{figure}

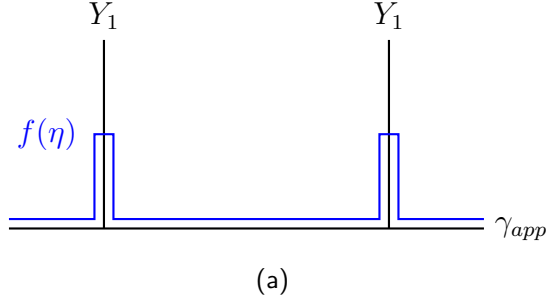
\begin{figure}
    \centering
    \subfloat[]{
    \begin{tikzpicture}[scale=1.25]
    
    \draw[thick] (0,0) -- (5,0);
    \node[right] at (5,0) {$\gamma_{app}$}; 
    \draw[thick] (1,0) -- (1,2);
    \node[above] at (1,2) {$Y_1$}; 
    \draw[thick] (4,0) -- (4,2);
    \node[above] at (4,2) {$Y_2$}; 

    \draw[blue,thick] (0,0.1) -- (0.9,0.1) -- (0.9,1) -- (1.1,1) -- (1.1,0.1) -- (3.9,0.1) -- (3.9,1) -- (4.1,1) -- (4.1,0.1) -- (5,0.1);

    \node[blue] at (0.4,1) {$f(\eta)$};
    
    \end{tikzpicture}
    }
    \caption{The region $Y=Y_1\cup Y_2$. The two segments $Y_1$ and $Y_2$ begin on the appetizer and extend into the hard region. We construct a sweep-out that covers $Y$. The surface $f(\eta)$ in the sweep-out follows the appetizer, and extends into the hard wedge a fraction $\eta$ of the distance along the $Y_i$ surfaces.}
    \label{fig:Yregion}
\end{figure}

\begin{proof}\,
    When $J_{2 \rightarrow 2}$ is empty, we have that $r$ is empty and hence $Y$ is empty, so the lower bound is trivial and we are done. 
    Otherwise, when $Y$ is non-empty, we consider defining the bulge surface $\gamma_{\text{bulge}}^Y$ as indicated around \eqref{eq:restricted_bulge}. 
    Let $\Sigma$ be any of the co-dimension 1 spacelike surfaces picked out by the outer maximization step in this procedure; accordingly, $\gamma_{\text{bulge}}^Y$ is the smallest maximum occurring among sweep-outs in $\Sigma$ that enclose $Y$. 

    By assumption $J_{2 \rightarrow 2} \neq \emptyset$, so we can use the connected wedge theorem (theorem \ref{thm:connectedwedge}) to obtain
    \begin{align}\label{eq:Iandr}
        \frac{1}{2} I(\V_1:\V_2) \geq \frac{A[r]}{4G_N} \: .
    \end{align}
    The ridge $r$ is related to the appetizer surface as shown in figure \ref{fig:ridge}. 
    Recall that the ridge is defined as 
    \begin{align}
        r=\partial J^+(E_{\V_1})\cap \partial J^+(E_{\V_2}^{\text{easy}}) \cap J^-(E_{\R}^{\text{easy}})\cap J^-((E_{\R}^{\text{easy}})') \: ,
    \end{align}
    so that in particular it sits on the null sheets $\partial J^+(E_{\V_1})$ and $\partial J^+(E_{\V_2}^{\text{easy}})$. 
    We consider continuing null rays starting on $r$ along one of these two sheets, say $\partial J^+(E_{\V_1})$, until reaching $\Sigma$. 
    Notice that since one endpoint of $r$ is in the domain of dependence of $E_{\R}^{\text{easy}}$ and the other endpoint is in $(E_{\R}^{\text{easy}})'$, the resulting surface $Y_1'$ will intersect the appetizer surface.
    We remove the portion that sits inside of $E_{\R}^{\text{easy}}$ and denote the remaining segment as $Y_1$. 
    We repeat this procedure with $\partial J^+(E_{\V_2}^{\text{easy}})$, producing $Y_2$. 
    The region $Y$ defined in equation \eqref{eq:Ydef} is also equal to $Y_1\cup Y_2$.
    By the focusing theorem, and because we only removed a portion of the focused forward surfaces, $Y_1$ and $Y_2$ both have area less than $r$, so 
    \begin{align}\label{eq:Yandr}
        A[Y_1]+A[Y_2] \leq 2 A[r] \: .
    \end{align}
    
    Now we construct a sweep-out $f$ with bulge surface of area at least $2 A[Y_1] + 2 A[Y_2] + A[\gamma_{\text{app}}]$ in the surface $\Sigma$ that encloses $Y=Y_1 \cup Y_2$. 
    To do this, we begin with $f(\eta=0)=\gamma_{\text{app}}$, and then consider adding `spurs' to this surface that follow $Y_i$ for distance $\eta \cdot A[Y_i]$, then double back to $\gamma_{\text{app}}$. 
    See figure \ref{fig:Yregion}. 
    At $\eta=1$ this encloses $Y$, so defines a candidate sweep-out. 
    But the maximal surface in this sweep out is just the $f(\eta=1)$ surface, which has area 
    \begin{align}
        A[f(\eta=1)] = A[\gamma_{\text{app}}]+ 2 A[Y_1]+ 2 A[Y_2] \: .
    \end{align}
    Then by \eqref{eq:Yandr},
    \begin{align}
        4 A[r] \geq A[f(\eta=1)] - A[\gamma_{\text{app}}] \: .
    \end{align}
    As well, since $\gamma_{\text{bulge}}^{Y}$ is the smallest maximum surface among sweep-outs in $\Sigma$ which cover $Y$, we must have
    \begin{align}\label{eq:Aandr}
        A[f(\eta=1)] - A[\gamma_{\text{app}}] \geq A[\gamma_{\text{bulge}}^{Y}] - A[\gamma_{\text{app}}] \: .
    \end{align}
    Combining this with \eqref{eq:Iandr} and \eqref{eq:Yandr}, we obtain
    \begin{align}
         I(\V_1:\V_2) \geq \frac{1}{2} \left( \frac{A[\gamma_{\text{bulge}}^Y] -A[\gamma_{\text{app}}]}{4G_N} \right) \: .
    \end{align}
    as needed. 
\end{proof}

\section{Example geometries} \label{sec:examples}

In this section, we will test the speculative lower bound \eqref{eq:conjlowerbound}.
Specifically, we will investigate this bound in the context of vacuum AdS$_{2+1}$ with $\R$ consisting of two disjoint intervals, and in the context of AdS$_{2+1}$ with a conical defect, non-rotating BTZ, and vacuum AdS$_{2+1}$ with a static end-of-the-world (ETW) brane. 
While most examples are consistent with the bound, we find a counter-example in the case that $\R$ consists of two intervals in the vacuum; nevertheless, we find that the weaker geometrical bound \eqref{eq:lowerboundintro}, following from our theorem \ref{thm:CDSlowerbound}, remains satisfied. 

\subsection{Two intervals in pure AdS\texorpdfstring{$_{2+1}$}{TEXT}}\label{sec:vacuumexample}

\begin{figure}
\begin{subfigure}{0.5\textwidth}
\begin{center}
\begin{tikzpicture}[scale=0.4]

        \draw[thick] (-10,0) -- (10,0) -- (10,10) -- (-10,10) -- (-10,0);
        \draw[fill=blue,opacity=0.3] (3,5) -- (0,8) -- (-3,5) -- (0,2) -- (3,5);
        \draw[thick,black] (3,5) -- (0,8) -- (-3,5) -- (0,2) -- (3,5);
        \draw[thick] (-10,2) -- (-7,5) -- (-10,8);
        \draw[fill=blue,opacity=0.3] (-10,2) -- (-7,5) -- (-10,8);
        \draw[thick] (10,2) -- (7,5) -- (10,8);
        \draw[fill=blue,opacity=0.3] (10,2) -- (7,5) -- (10,8);

        \node[above] at (0,5) {$\R_1$};
        \node[above right] at (-10,5) {$\R_2$};
        \node[above left] at (10,5) {$\R_2$};

        \draw[green, ultra thick] (-3,5) -- (3,5);
        \draw[green, ultra thick] (-7,5) -- (-10,5);
        \draw[green, ultra thick] (7,5) -- (10,5);
        \node[below] at (0,5) {$\mu$};

        \draw[red, ultra thick] (-7,5) -- (-3,5);
        \node[below] at (-5,5) {$\nu$};

        \draw[red, ultra thick] (3,5) -- (7,5); 

        \draw[thick] (-7,5) -- (-5,7) -- (-3,5) -- (-5,3) -- (-7,5);
        \draw[thick] (7,5) -- (5,7) -- (3,5) -- (5,3) -- (7,5);

        \draw[thick] (-4.25,3.75) -- (-3,2.5) -- (-1.75,3.75) -- (-3,5);
        \draw[fill=gray, opacity=0.3] (-4.25,3.75) -- (-3,2.5) -- (-1.75,3.75) -- (-3,5);

        \draw[thick] (4.25,3.75) -- (3,2.5) -- (1.75,3.75) -- (3,5);
        \draw[fill=gray, opacity=0.3] (4.25,3.75) -- (3,2.5) -- (1.75,3.75) -- (3,5);

        \draw[black] plot [mark=*, mark size=3] coordinates{(-3,2.5)};
        \node[left] at (-3,2.5) {$c_1$};

        \draw[black] plot [mark=*, mark size=3] coordinates{(3,2.5)};
        \node[left] at (3,2.5) {$c_2$};
        
    \end{tikzpicture}
\end{center}
\caption{}
\label{fig:twointervalboundary}
\end{subfigure}
\begin{subfigure}{0.5\textwidth}
\begin{center}
\tdplotsetmaincoords{15}{0}
    \begin{tikzpicture}[scale=1.6,tdplot_main_coords]
    \tdplotsetrotatedcoords{0}{90}{0}
    
    \draw[gray] (-2,0,0) -- (-2,3,0);
    \draw[gray] (2,0,0) -- (2,3,0);
    
    \begin{scope}[tdplot_rotated_coords]

    \draw plot [mark=*, mark size=1.1] coordinates{({-2*sin(135)}, {0.8889}, {2*cos(135)})};

    \draw[domain=0:25,variable=\x,smooth,thick] plot ({-2*sin(\x+135)}, {2-\x/45}, {2*cos(\x+135)});
    \draw[domain=0:25,variable=\x,smooth,thick] plot ({-2*sin(\x+135)}, {0.8889+\x/45}, {2*cos(\x+135)});
    \draw[domain=0:25,variable=\x,smooth,thick] plot ({-2*sin(-\x+135)}, {0.8889+\x/45}, {2*cos(-\x+135)});
    \draw[domain=0:25,variable=\x,smooth,thick] plot ({-2*sin(-\x+135)}, {2-\x/45}, {2*cos(-\x+135)});

    \foreach \x in {1,...,50}
    {
    \draw[gray,opacity=0.3] ({-2*sin(\x/2+135)}, {2-\x/90}, {2*cos(\x/2+135)}) -- ({-2*sin(\x/2+135)}, {0.8889+\x/90}, {2*cos(\x/2+135)});
    \draw[gray,opacity=0.3] ({-2*sin(-\x/2+135)}, {2-\x/90}, {2*cos(-\x/2+135)}) -- ({-2*sin(-\x/2+135)}, {0.8889+\x/90}, {2*cos(-\x/2+135)});
    }
    \draw[gray,opacity=0.3] ({-2*sin(135)}, {2}, {2*cos(135)}) -- ({-2*sin(135)}, {0.8889}, {2*cos(135)});
    
    \draw[domain=0:45,variable=\x,smooth,thick] plot ({-2*sin(\x)}, {1+\x/45}, {2*cos(\x)});
    \draw[domain=0:45,variable=\x,smooth,thick] plot ({2*sin(\x)}, {1+\x/45}, {2*cos(\x)});
    \draw[domain=0:45,variable=\x,smooth,thick] plot ({-2*sin(\x)}, {3-\x/45}, {2*cos(\x)});
    \draw[domain=0:45,variable=\x,smooth,thick] plot ({2*sin(\x)}, {3-\x/45}, {2*cos(\x)});
    
    \begin{scope}[canvas is xz plane at y=0]
    \draw[gray] (0,0) circle[radius=2] ;
    \end{scope}
    
    \begin{scope}[canvas is xz plane at y=3]
    \draw[gray] (0,0) circle[radius=2] ;
    \end{scope}
    
    \foreach \x in {0,...,90}
    {
    \draw[blue,opacity=0.3] ({-2*sin(\x/2+180)}, {1+\x/90}, {2*cos(\x/2+180)}) -- ({-2*sin(\x/2+180)}, {3-\x/90}, {2*cos(\x/2+180)});
    \draw[blue,opacity=0.3] ({2*sin(\x/2+180)}, {1+\x/90}, {2*cos(\x/2+180)}) -- ({2*sin(\x/2+180)}, {3-\x/90}, {2*cos(\x/2+180)});

    \draw[blue,opacity=0.3] ({-2*sin(\x/2+180)}, {1+\x/90}, {-2*cos(\x/2+180)}) -- ({-2*sin(\x/2+180)}, {3-\x/90}, {-2*cos(\x/2+180)});
    \draw[blue,opacity=0.3] ({2*sin(\x/2+180)}, {1+\x/90}, {-2*cos(\x/2+180)}) -- ({2*sin(\x/2+180)}, {3-\x/90}, {-2*cos(\x/2+180)});
    }
    
    \draw[domain=0:45,variable=\x,smooth,thick] plot ({-2*sin(\x+180)}, {1+\x/45}, {2*cos(\x+180)});
    \draw[domain=0:45,variable=\x,smooth,thick] plot ({2*sin(\x+180)}, {1+\x/45}, {2*cos(\x+180)});
    \draw[domain=0:45,variable=\x,smooth,thick] plot ({-2*sin(\x+180)}, {3-\x/45}, {2*cos(\x+180)});
    \draw[domain=0:45,variable=\x,smooth,thick] plot ({2*sin(\x+180)}, {3-\x/45}, {2*cos(\x+180)});

    \begin{scope}[canvas is xz plane at y=2]
    \draw [green,ultra thick,domain=-45:45] plot ({2*cos(\x-90)}, {2*sin(\x-90)});
    \draw [red,ultra thick,domain=45:135] plot ({2*cos(\x-90)}, {2*sin(\x-90)});
    \draw [red,ultra thick,domain=-45:-135] plot ({2*cos(\x-90)}, {2*sin(\x-90)});
    \end{scope}

    \begin{scope}[canvas is xz plane at y=2]
    \draw [green,ultra thick,domain=-45:45] plot ({2*cos(\x+90)}, {2*sin(\x+90)});
    
    \draw[blue,thick] (1.41,1.41) to [out=-135,in=+135] (1.41,-1.41);
    \draw[blue,thick] (-1.41,1.41) to [out=-45,in=45] (-1.41,-1.41);
    \draw[blue, dashed] (1.41,1.41) to [out=-135,in=-45] (-1.41,1.41);
    \draw[blue, dashed] (1.41,-1.41) to [out=135,in=45] (-1.41,-1.41);
    \end{scope}
    
    \end{scope}

    \draw[domain=0:25,variable=\x,smooth,thick] plot ({-2*sin(\x+135)}, {2-\x/45}, {2*cos(\x+135)});
    \draw[domain=0:25,variable=\x,smooth,thick] plot ({-2*sin(\x+135)}, {0.8889+\x/45}, {2*cos(\x+135)});
    \draw[domain=0:25,variable=\x,smooth,thick] plot ({-2*sin(-\x+135)}, {0.8889+\x/45}, {2*cos(-\x+135)});

    \foreach \x in {1,...,50}
    {
    \draw[gray,opacity=0.3] ({-2*sin(\x/2+135)}, {2-\x/90}, {2*cos(\x/2+135)}) -- ({-2*sin(\x/2+135)}, {0.8889+\x/90}, {2*cos(\x/2+135)});
    \draw[gray,opacity=0.3] ({-2*sin(-\x/2+135)}, {2-\x/90}, {2*cos(-\x/2+135)}) -- ({-2*sin(-\x/2+135)}, {0.8889+\x/90}, {2*cos(-\x/2+135)});
        }
    \draw[gray,opacity=0.3] ({-2*sin(135)}, {2}, {2*cos(135)}) -- ({-2*sin(135)}, {0.8889}, {2*cos(135)});

    \draw plot [mark=*, mark size=1.1] coordinates{({-2*sin(135)}, {0.8889}, {2*cos(135)})};

    \node[left] at (-2,2,0) {$\R_1$};
    \node[right] at (2,2,0) {$\R_2$};
    
    \end{tikzpicture}
\end{center}
\caption{}
\label{fig:twointervalsetup}
\end{subfigure}
\caption{(a) Boundary view of the two interval set-up. $\R$ consists of the two green intervals taken together, $\R=\R_1\cup \R_2$. Regions $\V_1$ and $\V_2$ are the grey diamonds shown, which are associated with inputs points $c_1$ and $c_2$ shown as black dots. (b) The bulk view. The appetizer surface is the pair of dashed blue lines, the RT surface for $\R$ is the pair of solid blue lines.
}
\end{figure}
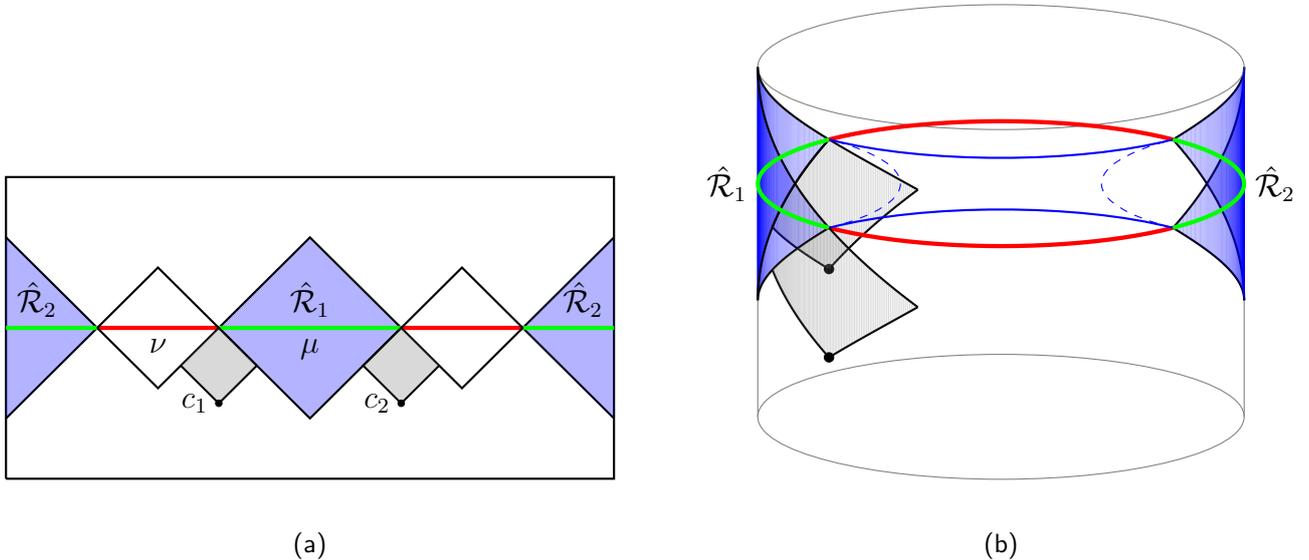

In this section we describe our first explicit setting where we study the lower bound \eqref{eq:mainlowerbound}, wherein the referee's region consist of two intervals in pure AdS$_{2+1}$. 
More details can be found in appendix~\ref{app:two_intervals}. 
Consider pure AdS$_{2+1}$, described by the metric 
\begin{align}
    ds^2 = -\cosh^2\rho \,dt^2 + d\rho^2 + \sinh^2 \rho \,d\phi^2
\end{align}
where here we will set $\lads=1$. 
We choose the referee's region $\R$ to consist of two intervals $\R_1\cup \R_2$ which have endpoints $(t,\phi)$ given by
\begin{align}
\p \R_1 :&~~ \left(0,-\frac{\mu}{2}\right),~~~\left(0,\frac{\mu}{2}\right)  \\
\p \R_2 :&~~ \left(0,\frac{2\pi-\mu}{2}\right),~~~\left(0,-\frac{2\pi-\mu}{2}\right)
\end{align}
and choose the input points to be
\begin{align}
    c_1: \left(-\tau, -\frac{\mu}{2}\right),~~~c_2: \left(-\tau, \frac{\mu}{2}\right) \: .
\end{align}
The set-up is shown in figure \ref{fig:twointervalboundary}.

Notice that for $\V_1$ and $\V_2$ to be domains of dependence we need to take 
\begin{align}\label{eq:cond1}
    0\leq \tau \leq \pi-\mu \: .
\end{align}
With this choice, we obtain intervals $\V_1$, $\V_2$ with endpoints
\begin{align}
\p \V_1 :&~~ \left(-\frac{\tau}{2},-\frac{\mu+\tau}{2}\right),~~~\left(-\frac{\tau}{2},-\frac{\mu-\tau}{2}\right)  \\
\p \V_2 :&~~ \left(-\frac{\tau}{2},\frac{\mu-\tau}{2}\right),~~~\left(-\frac{\tau}{2},\frac{\mu+\tau}{2}\right) \: .  
\end{align}
Further, for $\R_1$ and $\R_2$ to not overlap but still have a connected wedge (and hence a lunch we could scatter into), we need
\begin{align}\label{eq:cond2}
    \pi/2 \leq \mu \leq \pi \: .
\end{align}
Within this domain, we are interested in understanding when we have a non-empty scattering region $J(E_{\R}^{\text{easy}},(E_{\R}^{\text{easy}})')$, and hence when we should satisfy the lower bound \eqref{eq:mainlowerbound}. 
In appendix \ref{app:two_intervals}, we study the bulk light rays and find this occurs when 
\begin{align}\label{eq:intervalscatteringcondition}
    \cos \tau \leq \cos^2\left( \frac{\mu}{2}\right) \: .
\end{align}
We note that there indeed are choices of $\mu, \tau$ allowed by the above constraints and such that this scattering condition is satisfied, for example $\mu=\pi/2$, $\tau=\pi/2$.

Next, we study the value of the mutual information and $\Delta A_{\text{PL}}$ within this range. 
For the mutual information we find
\begin{align}
    \Delta A_{\text{MI}} = \ln\left( \frac{\sin^2(\frac{\tau}{2})}{\sin(\frac{\mu-\tau}{2})\sin(\frac{\mu+\tau}{2})} \right)
\end{align}
where $\Delta A_{\text{MI}}=4G_NI(\V_1:\V_2)$, and $\Delta A_{\text{PL}}$ is
\begin{align}
    \Delta A_{\text{PL}} = \ln \left( \csc^2\left( \frac{\mu}{2} \right) \right)\: .
\end{align}
Within the parameter region allowed by the scattering condition \eqref{eq:intervalscatteringcondition} and the conditions \eqref{eq:cond1} and \eqref{eq:cond2}, we then should have the lower bound \eqref{eq:mainlowerbound} which becomes
\begin{align}\label{eq:cond3}
    \csc^{2\alpha_0}\left( \frac{\mu}{2} \right) \leq \frac{\sin^2(\frac{\tau}{2})}{\sin(\frac{\mu-\tau}{2})\sin(\frac{\mu+\tau}{2})} \: .
\end{align}
Observe that for any finite $\alpha_0$, this will be violated by a point inside the allowed range for $\tau, \mu$. 
This is because if we take $\tau$ to its minimal value allowed by the scattering condition, 
\begin{align}
    \tau \rightarrow \tau_* = \arccos \left(\cos^2\left(\frac{\mu}{2}\right) \right)
\end{align}
then the right hand side of \eqref{eq:cond3} limits to $1$, but the left hand side is larger than $1$ whenever $\mu<\pi$. 

Note that, in this example, how deeply into the lunch the scattering can reach is becoming small as $\tau \rightarrow \tau_*$. 
Thus the lower bound appearing in \eqref{eq:lowerboundintro} is vanishing in the same limit that the mutual information is vanishing.
More specifically, if we take $\tau = \tau_{*} + \epsilon$ for small $\epsilon$, we see that the mutual information is $O(\epsilon)$. On the other hand, we expect that the region $X$ can be covered by a sweep-out $\{\gamma_{\eta}\}$ with $\gamma_{\eta}$ of the form
\begin{equation}
    (t(s), r(s), \phi(s)) = (t_{\text{app}}(s), r_{\text{app}}(s), \phi_{\text{app}}(s)) + \epsilon (t_{\eta}(s), r_{\eta}(s), \phi_{\eta}(s)) \: ,
\end{equation}
with $t_{\eta}(s), r_{\eta}(s), \phi_{\eta}(s)$ all uniformly compactly supported and bounded with respect to $\eta$. 
Since the appetizer surface is locally minimal, it follows that the area difference on the righthand side of \eqref{eq:lowerboundintro} is at most $O(\epsilon^{2})$, so there is no apparent contradiction with this bound at leading order. 

We can also compare our result with the conjectured lower bound \ref{eq:conjlowerbound}.
Let's for a moment assume this bound is true, as well as assume the python's lunch conjecture is true.
Since in this example the mutual information $I$ is going to zero as $\tau\rightarrow \tau_*$, but, according to the lunch conjecture $t$ is remaining fixed, consistency with the python's lunch conjecture requires that 
\begin{align}
    \min\{I_{IT}(f_n), \alpha \ln t \} = I_{IT}(f_n) \rightarrow 0
\end{align}
as $\tau\rightarrow \tau_*$. 
More concretely, the upper bound is like $r(\tau)/G_N$ with $r$ a length that goes to zero as $\tau\rightarrow \tau_*$. 
Naively, this seems surprising as for many functions we expect that $I_{IT}(f_n)$ can be a quickly growing function of $n$\footnote{For instance, the most efficient information-theoretic protocols known so far for CDS for an arbitrary function use $2^{\Theta(\sqrt{n\log n})}$ randomness.}, thus if we can take $n=g(n)$ for $g(n)$ any function which is $o(1/G_N)$, there would seem to be a contradiction. 
The situation here is the same as occurs in the context of holographic scattering and non-local quantum computation \cite{may2020holographic,may2022complexity}: there again we have a $1/G_N$ upper bound on the available mutual information, and an expectation that the entanglement cost of implementing NLQC is much larger if we take inputs of size nearly $1/G_N$. 
The resolution in that case was argued to be that we can't in fact take the inputs to be nearly size $1/G_N$ for most functions, and we expect the same resolution applies here. 
In particular, functions that have a large CDQS (or NLQC) cost must also be hard in some appropriate sense to implement in the bulk, so that they can only be computed in the bulk region for much smaller inputs. 
Thus we can have $I_{IT}(f_n)\leq r/G_N$ and $I_{IT}$ a quickly growing function of $n$, if $n$ is restricted by bulk physics to be a very slowly growing function of $1/G_N$.\footnote{Note that in the argument in section \ref{sec:AdSlowerbound} we took $n=g(n)$ for $g(n)=o(1/G_N)$, so we need in that argument to be using a sufficiently simple function that we don't expect further restrictions (aside from the input size) to appear on computing it in the bulk. Fortunately, the lower bound \ref{thm:CDSlowerbound} we used there applies already for very simple functions.} 
A further discussion of this issue appears in \cite{may2022complexity}. 

\subsection{ETW brane geometry}\label{sec:ETWbrane}

\begin{figure}
    \centering
    \includegraphics[height=8cm]{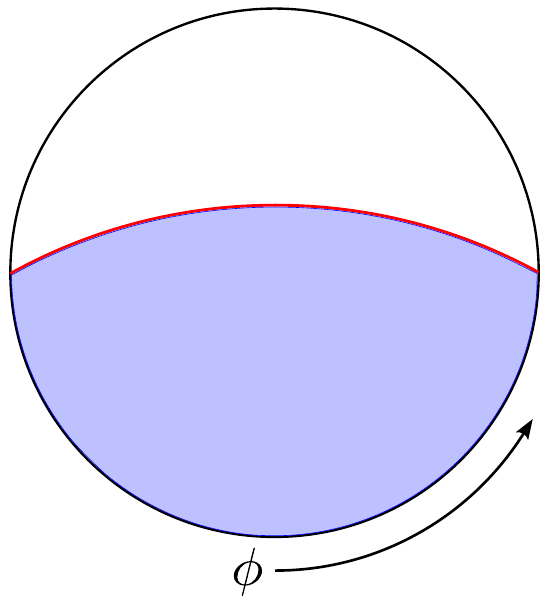}
    \caption{Constant $t$ slice of the AdS$_{2+1}$ geometry (light blue) ending on an ETW brane (red). }
    \label{fig:AdS_Slice}
\end{figure}

In this subsection, we consider a global AdS spacetime terminating on a static ETW brane. 
The bulk metric is
given in global coordinates by
\begin{equation} \label{eq:pure_AdS_metric}
    ds^{2} = - \left( \frac{r^{2}}{\lads^{2}} + 1 \right) dt^{2} + \frac{dr^{2}}{\left( \frac{r^{2}}{\lads^{2}} + 1 \right)} + r^{2} d \phi^{2} \: , \qquad \phi \in [- \pi, \pi)
\end{equation}
while the ETW brane has trajectory
\begin{equation} \label{eq:ETW_traj}
    r(\phi) = - \frac{\lads T}{\sqrt{1 - T^{2}}} \sec \phi \: , \qquad \phi \in \left( - \frac{3 \pi}{2}, -\frac{\pi}{2} \right) \: ,
\end{equation}
with $0 < T < 1$ a dimensionless tension parameter (see figure \ref{fig:AdS_Slice} for the depiction of a constant $t$ slice). 
The asymptotic boundary is no longer the full cylinder but rather a strip with $\phi \in (- \frac{\pi}{2}, \frac{\pi}{2})$ and $\hat{t} \equiv \frac{t}{\lads} \in (- \infty, \infty)$, with some CFT boundary condition $\mathcal{B}$ at the spatial endpoints.

\begin{figure}
    \centering
    \includegraphics[height=8cm]{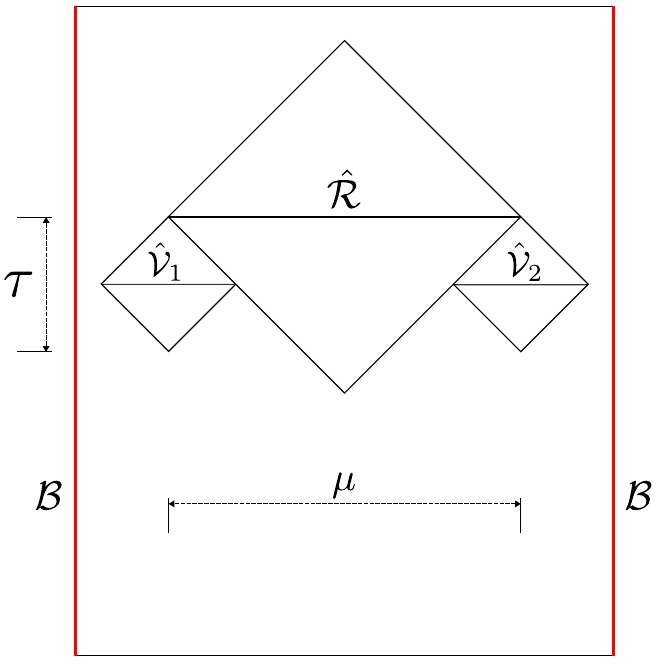} \quad
    \includegraphics[height=8cm]{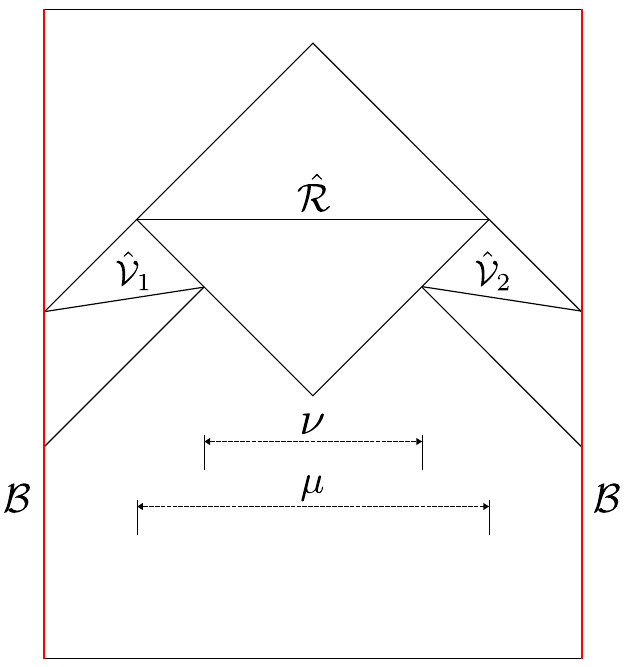}
    \caption{Two different choices of placement of the input regions $\V_1, \V_2$ and output region $\R$ on the boundary of the ETW brane spacetime.}
    \label{fig:ETW_V1V2R_maintext}
\end{figure}

We would like to specify the geometric configuration of the input regions $\V_1, \V_2$ and output region $\R$ on the boundary. 
We consider two possible choices of configurations for these regions, illustrated in figure \ref{fig:ETW_V1V2R_maintext}. The first choice is analogous to that described in the preceding section on vacuum AdS: $\R$ subtends angle $\mu \in (0, \pi)$, while $\V_{1}, \V_{2}$ subtend angle $\tau \in (0, \min \{ \mu, \pi - \mu \})$. The second choice has $\V_{1}$ and $\V_{2}$ causal developments of intervals anchored on the boundary; we parametrize it by an angle $\mu \in (0, \pi)$ subtended by $\R$, and an angle $\nu \in (0, \mu)$ between the interior endpoints of $\V_1$ and $\V_2$ (see figure \ref{fig:ETW_V1V2R_maintext}). 

We analyze this set-up in detail in appendix \ref{app:ETW}. Restricting to configurations for which scattering is possible, we find the range of $\Delta A_{\text{PL}}$ obtained for the configurations of $\V_{1}, \V_{2}$ parametrized by $(\mu, \tau)$ to be a subset of that parametrized by $(\mu, \nu)$, and the minimum value of $\Delta A_{\text{MI}}$ at a given $\Delta A_{\text{PL}}$ is always smaller in the latter case. Consequently, we will review only the latter case in this section, since it more strongly constrains a putative lower bound. 
The information relevant for our purposes is as follows:
\begin{itemize}
    \item A python's lunch exists provided $\tan^{2} \left( \frac{\mu}{2} \right) \geq \frac{1+T}{1-T}$. When this is the case, the area difference appearing in the python's lunch conjecture is given by
    \begin{equation}
        \Delta A_{\text{PL}} = \begin{cases} 
        2 \lads \ln \left( \frac{1}{\sin(\mu/2)} \frac{B \left( B + \sqrt{B^{2} - 1 + T^{2}} \right)}{\sqrt{1 - T^{2}} \left( B(B-A) + AT \right)} \right)  & A < T \\
        2 \lads \ln \left( \frac{1}{\sin(\mu/2)} \frac{B \sqrt{1 - T^{2}}}{ \left( B (B-A) + A T \right) \left( B + \sqrt{B^{2} - 1 + T^{2}} \right)} \right) & A > T 
        \end{cases} \: ,
    \end{equation}
    where
    \begin{equation}
        A = \tan(\mu/2) \left( \frac{1 - \sin(\mu/2)}{1 - \sin(\mu/2) + T \cos(\mu/2)} \right) \: , \quad B = \sqrt{1 - 2T A + A^{2}} \: .
    \end{equation}
    \item When a python's lunch exists, the condition under which scattering is possible is
    \begin{equation}
        \nu \leq \frac{\mu}{2} \: .
    \end{equation}
    \item When scattering is possible, the area difference $\Delta A_{\text{MI}}$ appearing in the mutual information $I(\V_{1} : \V_{2})$ is
    \begin{equation}
        \Delta A_{\text{MI}} = 2 \lads \ln \left[ \cot \left( \frac{\nu}{2} \right) \right] + \lads \ln \left( \frac{1+T}{1-T} \right) \: .
    \end{equation}
\end{itemize}

\begin{figure}
    \centering
    \includegraphics[height=8cm]{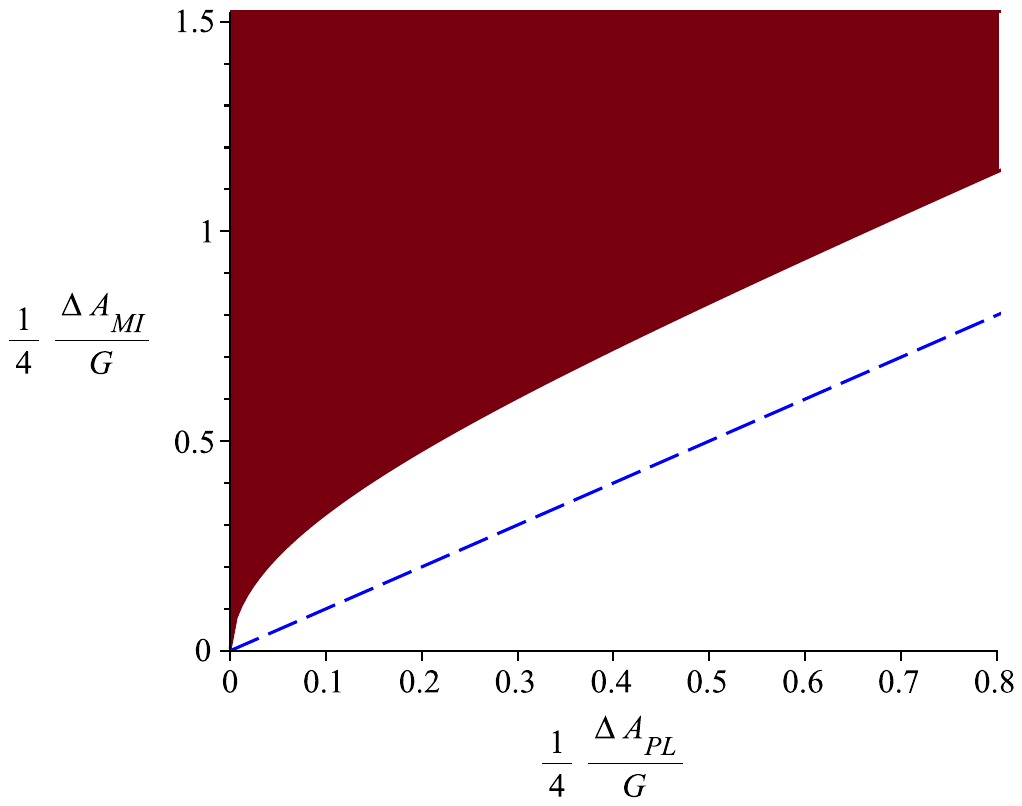}
    \caption{Allowed values of $\frac{\Delta A_{\text{MI}}}{4 G}$ as a function of $\frac{\Delta A_{\text{PL}}}{4 G}$ (solid red), as well as the linear relationship $\Delta A_{\text{MI}} = \Delta A_{\text{PL}}$ (dashed blue), for the ETW brane geometry. Here we are using units $\frac{\lads}{G}$ for both axes, which is roughly the number of local degrees of freedom in the dual CFT (recall that the Brown-Henneaux central charge is $c = \frac{3 \lads}{2 G}$). }
    \label{fig:LowerBound_ETW_2}
\end{figure}

With this information, we can plot the allowed values of $\Delta A_{\text{MI}}$ as a function of $\Delta A_{\text{PL}}$ whenever scattering is possible. We show this plot in figure \ref{fig:LowerBound_ETW_2}, obtained by varying $T, \mu$, and $\nu$ subject to the relevant constraints and computing $\Delta A_{\text{PL}}, \Delta A_{\text{MI}}$. 
We find that the largest lower bound of the form $\Delta A_{\text{MI}} \geq \alpha_0 \Delta A_{\text{PL}}$ consistent with this plot has $\alpha_0 = 1$. 
In particular, this example is consistent with the lower bound from \eqref{eq:lowerboundintro}.

\subsection{AdS\texorpdfstring{$_{2+1}$}{TEXT} defect and BTZ black hole}\label{sec:defectandBTZ}

The metric of the defect and BTZ spacetimes in global coordinates is given by
\begin{equation} \label{eq:defect_metric}
    ds^{2} = - \left( \frac{r^{2}}{\lads^{2}} - M \right) dt^{2} + \frac{dr^{2}}{\left( \frac{r^{2} }{\lads^{2}} - M \right)} + r^{2} d \phi^{2} \: , \qquad \phi \in [0, 2 \pi) \: .
\end{equation}
The parameter $M$ is related to the ADM mass by $M = 8 G_N M_{\text{ADM}}$. This metric encapsulates the following geometries:
\begin{itemize}
    \item $M = -1$: Pure AdS$_{3}$
    \item $-1 < M < 0$: Conical defect with a particle of mass $m = \frac{1}{4 G} \left( 1 - \sqrt{|M|} \right)$
    \item $M > 0$: Non-rotating BTZ black hole.
\end{itemize}

\begin{figure}
    \centering
    \includegraphics[height=8cm]{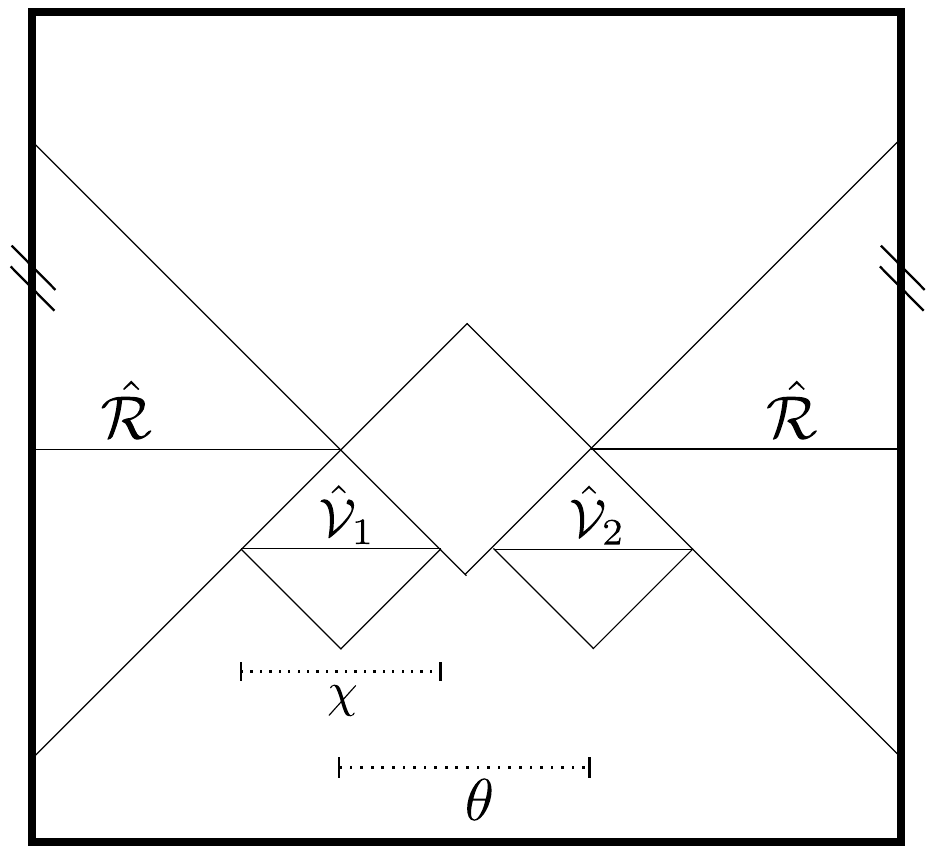}
    \caption{Placement of input regions $\V_1, \V_2$ and output region $\R$ on the boundary of the defect/BTZ spacetime. We take $\chi$ to be the angular size of the regions $\V_1$ and $\V_2$, and $\theta$ to be their angular separation.}
    \label{fig:V1V2R_defect}
\end{figure}

The conformal boundary of the defect spacetime is the Lorentzian cylinder, which we can describe with angular coordinate $\phi$ and time coordinate $\hat{t} = t / \lads$. The configuration of input regions $\V_1, \V_2$ and output region $\R$ on the cylinder will be taken as shown in figure \ref{fig:V1V2R_defect}. 
For simplicity, we are restricting to the case that $\V_1, \V_2$ lie in a slice with fixed $\hat{t} = \hat{t}_{\text{i}}$ and have equal angular size $\chi$. 
We will assume that the centres of these intervals are separated by angle $\theta$ satisfying $0 < \chi < \theta < \pi$, so that the intervals don't overlap. 

We analyze this set-up in detail in appendices \ref{app:defect} and \ref{app:BTZ}. Restricting to configurations for which scattering is possible, we find the range of $\Delta A_{\text{PL}}$ obtained for the BTZ black hole to be a subset of that obtained for the defect, and the minimum value of $\Delta A_{\text{MI}}$ at a given $\Delta A_{\text{PL}}$ is always smaller for the defect than for the black hole. Consequently, we will review only the defect case in this section, since it more strongly constrains a putative lower bound. 
The information relevant for our purposes is as follows:
\begin{itemize}
    \item A python's lunch exists provided either $M < -\frac{1}{4}$ and $(2 \pi - \theta) < \frac{\pi}{\sqrt{|M|}}$, or $M \geq - \frac{1}{4}$. When this is the case, the area difference appearing in the python's lunch conjecture is given by
    \begin{equation}
        \Delta A_{\text{PL}} = \begin{cases}
            2 \lads \ln \left[ \csc \left( \frac{\sqrt{|M|}}{2} (2 \pi - \theta) \right) \right] & M < -\frac{1}{9} \quad \text{and} \quad \theta > \pi \big| 2 - \frac{1}{\sqrt{|M|}} \big|  \\
            2 \lads \ln \left[ \frac{\sin \left( \frac{\sqrt{|M|}}{2} (2 \pi + \theta) \right)}{\sin \left( \frac{\sqrt{|M|}}{2} (2 \pi - \theta) \right)}\right] & M > - \frac{1}{4} \quad \text{and} \quad \theta < \pi \big| 2 - \frac{1}{\sqrt{|M|}} \big|
        \end{cases} \: .
    \end{equation}
    \item When a python's lunch exists, the condition under which scattering is possible is
    \begin{equation}
        \cos^{2}\left( \frac{\sqrt{|M|}}{2} (2 \pi - \theta) \right) - \cos(\sqrt{|M|} \chi) \geq 0 \: .
    \end{equation}
    \item When scattering is possible, the area difference $\Delta A_{\text{MI}}$ appearing in the mutual information $I(\V_{1} : \V_{2})$ is
    \begin{equation}
        \Delta A_{\text{MI}} = 2 \lads \ln \left[ \frac{\sin^{2} \left( \frac{\sqrt{|M|}}{2} \chi \right)}{\sin^{2} \left( \frac{\sqrt{|M|}}{2} ( \pi - \chi) \right) - \sin^{2} \left( \frac{\sqrt{|M|}}{2} ( \pi - \theta) \right)} \right] \: .
    \end{equation}
\end{itemize}

\begin{figure}
    \centering
    \includegraphics[height=10cm]{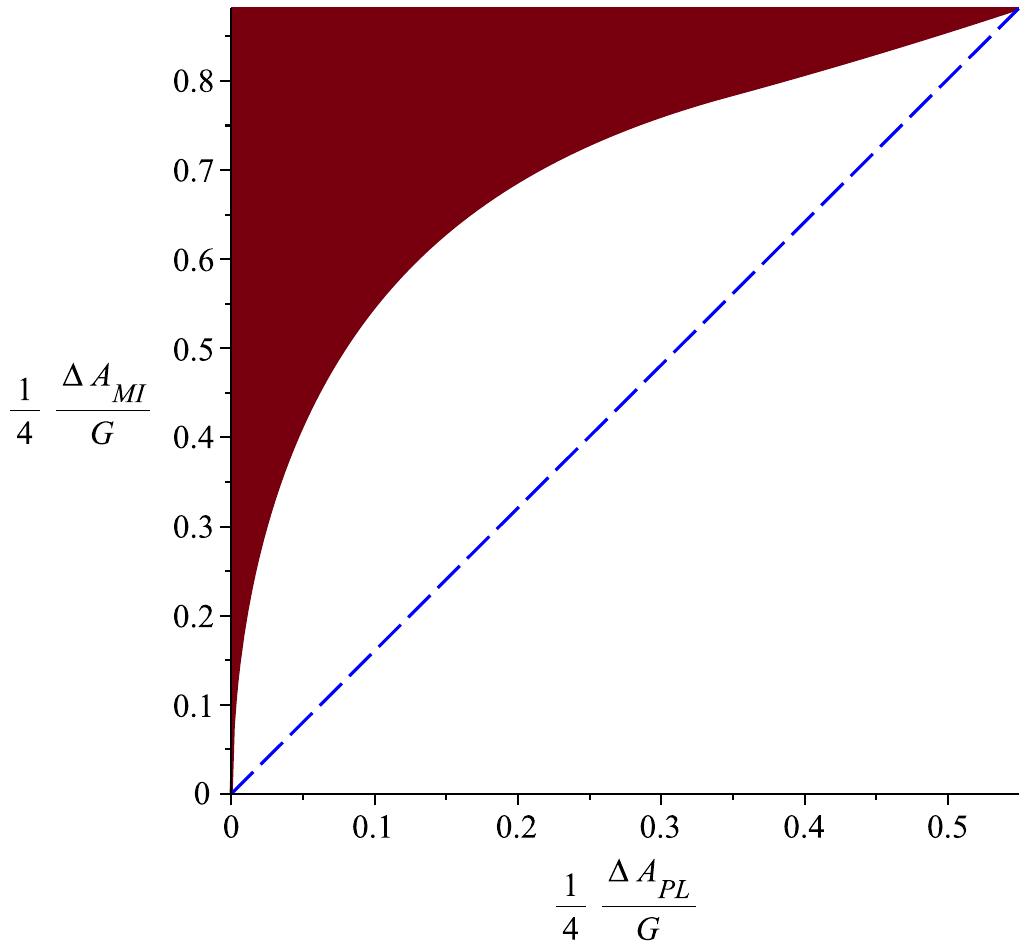}
    \caption{Allowed values of $\frac{\Delta A_{\text{MI}}}{4 G}$ as a function of $\frac{\Delta A_{\text{PL}}}{4 G}$ (solid red), as well as the linear relationship $\Delta A_{\text{MI}} = 1.6 \Delta A_{\text{PL}}$ (dashed blue), for the AdS defect. Here we are using units $\frac{\lads}{G}$ for both axes, which is roughly the number of local degrees of freedom in the dual CFT (recall that the Brown-Henneaux central charge is $c = \frac{3 \lads}{2 G}$). Note that $0 \leq \frac{\Delta A_{\text{PL}}}{4 G} \leq \frac{1}{2} \ln(3) \frac{\lads}{G}$ given the constraints on parameters. }
    \label{fig:LowerBoundAMI_defect_maintext}
\end{figure}

With this information, we can plot the allowed values of $\Delta A_{\text{MI}}$ as a function of $\Delta A_{\text{PL}}$ whenever scattering is possible. We show this plot in figure \ref{fig:LowerBoundAMI_defect_maintext}, obtained by varying $M, \chi,$ and $ \theta$ subject to the relevant constraints and computing $\Delta A_{\text{PL}}, \Delta A_{\text{MI}}$. 
We find that the largest lower bound of the form $\Delta A_{\text{MI}} \geq \alpha_0 \Delta A_{\text{PL}}$ consistent with this plot has $\alpha_0 \approx 1.6$. 
In particular, this example is consistent with the lower bound from \eqref{eq:lowerboundintro} and less constraining for the value of $\alpha_0$ than the ETW brane example. 

\section{Discussion}\label{sec:discussion}

In this work, we argued for boundary correlation implications of the projective tensor network model description of bulk AdS geometry. 
Our key claim is the lower bound \eqref{eq:lowerboundintro}. 
Interpreting this statement geometrically, we obtain a precise implication of this model for bulk geometry which can be checked by direct computations. 
We verify this implication partially and hence provide evidence supporting the tensor network model by proving two weakened versions of this geometrical statement. 
Further investigating this setting opens up a number of opportunities to better understand the tensor network model, the pythons lunch conjecture, and aspects of cryptography.

\vspace{0.2cm}
\noindent \textbf{Resolving $\alpha_0$:} Currently, our argument for the lower bound \eqref{eq:lowerboundintro} requires a delicate choice of secret size $d_Q$, and an assumption that $\epsilon$ (the correctness error) is very small. 
Because of this, the undetermined constant $\alpha_0$ appears in our eventual lower bound. 

This situation is reminiscent of initial quantum information arguments for the connected wedge theorem \cite{may2020holographic}, which required similar arguments, while later improvements to the lower bound removed this need.
Similarly improvements to theorem \ref{thm:CDSlowerbound} may be able to determine $\alpha_0$. 
In particular, a lower bound where $-\ln \delta$ appears directly, rather than $-\ln(1/d_Q^{1/2}+\delta+\epsilon)$ may be possible. 
For instance $d_Q$ and $\epsilon$ could appear in a term multiplying $\ln \delta$, or in other terms. 
We leave investigating this to future work. 
Determining $\alpha_0$ would allow a more precise comparison to be carried out in studying the lower bound \eqref{eq:lowerboundintro}, and hence a more precise check of the projective tensor network model. 

\vspace{0.2cm}
\noindent \textbf{Conjecture \ref{conj:CDS} and the python's lunch:} Recall that our current argument tests the projective tensor network model by extracting consequences of these projections for boundary correlation. 
Since these projections also lead to the large complexities appearing in the lunch conjecture, this indirectly tests the conjecture. 
It would be an improvement to our work to be able to relate the complexity appearing in the python's lunch conjecture and boundary correlation directly. 
Towards doing this we introduced conjecture \ref{eq:conjlowerboundmain}, which relates complexity and mutual information directly. 
Unfortunately, we couldn't prove this conjecture, but doing so would open up more direct tests of the lunch conjecture. 

\vspace{0.2cm}
\noindent \textbf{Other implications of the python's lunch conjecture:} We can view our set-up as an example sitting within a broader framework. 
The work \cite{may2019quantum} advocated for taking an operational perspective on AdS/CFT: quantum information processing tasks with inputs and outputs at the conformal boundary could be understood and reasoned about from both a bulk and boundary perspective. 
Requiring consistency of the bulk and boundary perspectives on these information processing tasks implies constraints on the CFT that can be checked and confirmed explicitly. 

This was extended in \cite{may2021holographic} to allow for inputs and outputs beginning and ending at bulk points. 
To do so, we assume entanglement wedge reconstruction, which tells us which boundary regions can access information at a given bulk point and lets us define the corresponding boundary task. 
Our work here presents a further extension of this framework: Assuming the python's lunch conjecture, we can characterize when reconstructing a bulk point is computationally hard, and thereby reason about information processing tasks which have computational constraints. 
Our work derives an initial consequence of this perspective, but it seems likely we can find others. 
For example in the information-theoretic context, \cite{may2021bulk} used the quantum tasks perspective to derive boundary entanglement consequences of a bulk geometric feature they labelled a `private curve'. 
It may be possible to revisit this in the computational setting, or develop novel settings that don't have information-theoretic analogues.

\subsection*{Acknowledgements} We thank Lana Bozanic, Aidan Herderschee, Juan Maldacena, Geoff Penington, and Jon Sorce for insightful discussions.
Geoff Penington pointed out an oversight in the original version of this work which led to substantial improvements. 
The research of SP is supported by the Celestial Holography Initiative at the Perimeter Institute for Theoretical Physics and by the Simons Collaboration on Celestial Holography. 
MX is supported by the US National Science Foundation GRFP and a Stanford Physics Departmental fellowship.
Research at the Perimeter Institute is supported by the Government of Canada through the Department of Innovation, Science and Industry Canada and by the Province of Ontario through the Ministry of Colleges and Universities.
AM benefited from the hospitality of the Institute for Advanced Study, where a portion of this work was completed. 

\appendix

\section{Complexity-correlation relationship in examples}\label{app:examples}

In this appendix we provide details for the analysis of the examples appearing in section \ref{sec:examples}. Our goal is to verify whether the lower bound \eqref{eq:conjlowerbound} on the mutual information $I(\V_1 : \V_2)$ is violated in these examples. Of course, this lower bound comes with an undetermined universal coefficient $\alpha_{0}$, which represents an obstacle for direct comparison of our results with this bound. Nevertheless, if we can identify a sequence of geometries for which $I(\V_1 : \V_2)$ approaches zero while $\Delta A_{\text{PL}}$ is bounded from below, we can demonstrate a violation of the bound. Moreover, given a family of geometries where the bound is expected to hold, studying the complexity-entanglement relationship within the family can provide non-trivial upper bounds on the coefficient $\alpha_{0}$. 

We will consider a two-interval example in vacuum AdS$_{2+1}$, as well as two classes of locally AdS$_{2+1}$ geometries: 
\begin{itemize}
    \item Conical defects/BTZ black holes 
    \item Global AdS$_{2+1}$ with a static end-of-the-world (ETW) brane. 
\end{itemize}
Each of these two classes is a one-parameter family of geometries, with the parameter corresponding to the mass of the defect/black hole and the tension of the ETW brane respectively. We will not consider the most general input regions $\V_{1}, \V_{2}$, but rather assume that they are intervals of equal size to simplify calculations.  

\subsection{Two intervals in pure AdS\texorpdfstring{$_{2+1}$}{2+1}}\label{app:two_intervals}

We start with the details of the two-interval example.

\subsubsection*{Bulk geometry}

The bulk geometry is global AdS$_{2+1}$ with the metric
\begin{equation} 
    ds^{2} = - \left( \frac{r^{2}}{\lads^{2}} +1\right) dt^{2} + \frac{dr^{2}}{\left( \frac{r^{2} }{\lads^{2}} +1 \right)} + r^{2} d \phi^{2} \: , \qquad \phi \in [0, 2 \pi) \: .
\end{equation}
Since we are doing computations in vacuum AdS, we can take advantage of the embedding space formalism. Recall that we can write AdS$_{2+1}$ as the unit hyperboloid in $\mathbb{R}^{2,2}$ defined by the locus $X^2=-L_{AdS}^2$. Namely
\begin{align}\label{xemb}
X^a=L_{AdS}(\cosh\rho\cos \hat t, \cosh\rho\sin \hat t, \sinh\rho \cos\phi,\sinh\rho\sin\phi)
\end{align}
where 
\be
r=L_{AdS}\sinh \rho,~~\hat t=t / \lads.
\ee

\subsubsection*{Boundary geometry}

The conformal boundary is the Lorentzian cylinder, which we can describe with angular coordinate $\phi$ and time coordinate $\hat{t}$. 
The configuration of input regions $\V_1, \V_2$ and output region $\R$ is illustrated in figure~\ref{fig:twointervalsetup}. Here $\R$ has two disconnected components, each of angular width $\mu$. 
The two complementary intervals have size $\nu$
\be
\nu=\pi-\mu.
\ee
We place them at $\hat t=0$. 
We will assume that $\mu>\nu$ to ensure that $\R$ gives a connected wedge, and so that this configuration has a lunch.  
In terms of $(\hat t,\phi)$ coordinates we have
\begin{align}
\p \R_1 :&~~ \left(0,-\frac{\mu}{2}\right),~~~\left(0,~\frac{\mu}{2}\right)  \\
\p \R_2 :&~~ \left(0,\frac{2\pi-\mu}{2}\right),~~~\left(0,-\frac{2\pi-\mu}{2}\right).
\end{align}
We will now define the input regions $\V_1$ and $\V_2$ in terms of two input points $c_1$ and $c_2$. For convenience, we will place them at fixed $\hat{t} = -\tau$ and symmetrically beneath the end points of $\R_1$
\begin{align}\label{ci}
c_1: \left(-\tau, -\frac{\mu}{2}\right),~~~c_2: \left(-\tau, ~\frac{\mu}{2}\right).
\end{align}
 This makes  $\V_1, \V_2$ intervals of size $\tau$ with endpoints 
 \begin{align}
\p \V_1 :&~~ \left(-\frac{\tau}{2},-\frac{\mu+\tau}{2}\right),~~~\left(-\frac{\tau}{2},-\frac{\mu-\tau}{2}\right)  \\
\p \V_2 :&~~ \left(-\frac{\tau}{2},\frac{\mu-\tau}{2}\right),~~~\left(-\frac{\tau}{2},\frac{\mu+\tau}{2}\right)  .
\end{align}
To make sure the $\V_i$ are domains of dependence we need to take
\begin{align}\label{nu}
\tau\in(0,\nu)
\end{align}
where again we've assumed $\nu<\mu$.  

\subsubsection*{Spacelike geodesics}

For our purposes, we will only need spacelike geodesics on equal time slices. These take the form
\begin{equation} \label{eq:r_phi}
   {r(\phi) =  \lads \sqrt{ \frac{\sec^{2}(\phi-\phi_0)}{ \tan^{2} \left( \frac{1}{2} \Delta \phi \right) - \tan^{2} (\phi-\phi_0)} }} \: ,
\end{equation}
where $\phi_0$ is the center and $\Delta \phi$ is the angular width of the interval. These have regularized lengths 
\begin{equation} \label{eq:l_general}
   {\ell = 2 \lads \ln \left[ \frac{2}{\epsilon } \sin \left(  \frac{1}{2} \Delta \phi \right) \right]} \: ,
\end{equation}
where we regulate by cutting of the bulk at $r = \frac{\lads}{\epsilon}$. 

Since it will be useful to us later, note that we can also describe these spacelike geodesics in terms of intersections of hyperplanes. As we limit to the boundary, the leading term in~\eqref{xemb} is proportional to the following null vector
\begin{align}\label{qemb}
X^a\simeq L_{AdS}\frac{e^\rho}{2}(\cos \hat t,\sin \hat t,\cos\phi,\sin\phi)=L_{AdS} e^\rho q^a(\hat t,\phi).  
\end{align}
The Rindler horizons null separated from a point on the boundary is described by the locus
\be
q^a(\hat t, \phi )X_a=0
\ee
and each spacelike geodesic can be recast as the intersection of two such Rindler horizons.

\begin{figure}
    \centering
    \begin{tikzpicture}[scale=0.9]
    
    \draw[thick] (0,0) circle (3);
    
    \draw[blue,dashed, thick] (1.5, 2.58) to [out=-120,in=120] (1.5, -2.58);
    \draw[blue,dashed, thick] (-1.5, 2.58) to [out=-60,in=60] (-1.5, -2.58);
    
    \draw[blue, thick] (-1.5, 2.58) to [out=-30,in=-150] (1.5, 2.58);
    \draw[blue, thick] (-1.5, -2.58) to [out=30,in=150] (1.5, -2.58);
    
    \node[right] at (3,0) {$\R_2$};
    \node[left] at (-3,0) {$\R_1$};

    \draw[red] (-1.5, 2.58) -- (1.5, -2.58);
    \draw[red] (-1.5, -2.58) -- (1.5, 2.58);
    
    \end{tikzpicture}
    \caption{The RT surface $\gamma_{\R_1\cup \R_2}$ defining the entanglement wedge of $\R_1\cup \R_2$ is shown in solid blue; the appetizer surface $\gamma_{\R_1}\cup \gamma_{\R_2}$ is in dashed blue. The bulge is depicted in red. The lunch region sits between the appetizer and RT surfaces.}
    \label{fig:counterexample2}
\end{figure}
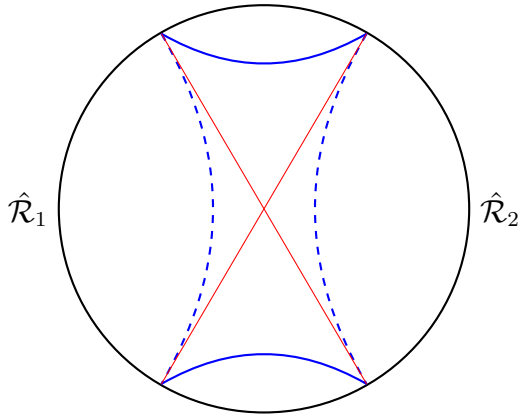

Now the particular RT surfaces relevant for our lunch configuration are illustrated in figure~\ref{fig:counterexample2}. The ``appetizer'' surface, the true RT surface, and the bulge surface are as follows:
\begin{itemize}
    \item The appetizer surface corresponds to the union of ${ \Delta \phi_{\text{app}} = \mu,~~ \phi_0=0,\pi}$ shown in dashed blue.
    \item The true RT surface corresponds to the union of  ${ \Delta \phi_{\text{RT}} = \nu,~~ \phi_0=\pm\frac{\pi}{2}}$ shown in solid blue.
    \item The bulge surface corresponds to the union of  ${ \Delta \phi_{\text{bulge}} = \pi,~~ \phi_0=\pm\frac{\nu}{2}}$ shown in solid red.
\end{itemize}
Technically the bulge can be viewed as the union of two kink surfaces, each homologous to one of the connected intervals.  The length of the kink is manifestly larger than either candidate RT surface and our choice $\mu>\nu$ makes the wedge for $\R$ connected.

\subsubsection*{Null geodesics}
We can write the null geodesics in global AdS$_{2+1}$ as \cite{may2019quantum} 
\be
\begin{aligned}
\rho(\lambda) & =\operatorname{arccosh}\left(\sqrt{\lambda^2\left(1-\ell^2\right)+\frac{1}{1-\ell^2}}\right) \\
\hat t(\lambda) & = \hat t_0+\frac{\pi}{2}+\arctan \left(\lambda\left(1-\ell^2\right)\right)\\
\phi(\lambda) & =\phi_0+\frac{\pi}{2}+\arctan \left(\frac{\lambda\left(1-\ell^2\right)}{\ell}\right) 
\end{aligned}
\ee
where $\lambda \in (-\infty,\infty) $ is an affine parameter, $\ell\in (-1,1)$ is the angular momentum, and $(t_0, \phi_0)$ is the starting points of the null geodesic on the conformal boundary.

\subsubsection{Scattering}

We will now set up the scattering problem relevant to our lunch configuration. See also appendix C of \cite{may2019quantum} for equations relevant to scattering in global AdS$_{2+1}$. Our goal is to determine for what ranges of $\tau$ and $\mu$ scattering is possible. Our scattering problem is determined by the input points in~\eqref{ci},
\begin{align}
c_1: \left(-\tau, -\frac{\mu}{2}\right),~~~c_2: \left(-\tau, \frac{\mu}{2}\right)
\end{align}
and the output point lying on the appetizer surface associated with $\R$. This will correspond to being in the past of both $\R_1$ and $\R_1'$. 
In vacuum AdS, we can phrase this in terms of output points
\begin{align}\label{ri}
r_1: \left(\frac{\mu}{2},0\right),~~~r_2: \left(\frac{2\pi-\mu}{2},\pi\right).
\end{align}
The past lightcones of these points will intersect at the RT surface for $\R_1$ in figure~\ref{fig:counterexample2}. 

\subsubsection*{Ingoing process}
 
Lightlike geodesics from $c_1$ and $c_2$ will intersect along a ridge defined by the intersection of the following null hyperplanes
\begin{align}
q^a\left(-\tau,-\frac{\mu}{2}\right)X_a=0,~~q^a\left(-\tau,\frac{\mu}{2}\right)X_a=0,
\end{align} 
or
\begin{align}\label{rhotau}
\cosh\rho \cos(\hat t+\tau)=\sinh\rho \cos(\phi+\frac{\mu}{2})=\sinh\rho \cos(-\phi+\frac{\mu}{2}).
\end{align}
One can see straightforwardly that $\phi=0,\pi$, which is also clear by symmetry. Equation~\eqref{rhotau} then reduces to
\begin{align}
\cos(\hat t+\tau) =\pm \tanh \rho \cos(\frac{\mu}{2}) 
\end{align}
which implicitly defines the profile $\hat t(\rho)$. 

\subsubsection*{Outgoing process}

We next want to ask when this ridge lies in the past of $r_1$ and $r_2$ in~\eqref{ri} so that easy-hard scattering is allowed. 
The past light cones of $r_1$ and $r_2$ are defined in embedding space by the hyperplanes, 
\begin{align}\label{r12}
q^a\left(\frac{\mu}{2},0\right)X_a=0,~~q^a\left(\frac{2\pi-\mu}{2},\pi\right)X_a=0
\end{align} 
Substituting in~\eqref{xemb} and~\eqref{qemb} 
we see that the lightcone of $r_1$ satisfies
\begin{align}
\cosh \rho \cos(\hat t-\frac{\mu}{2}) = \sinh\rho \cos\phi,
\end{align}
while the lightcone of $r_2$ satisfies
\begin{align}
\cosh \rho \cos(\hat t+\frac{\mu}{2}) = \sinh\rho \cos\phi .
\end{align}
Let us label the intersections of the ridge~\eqref{rhotau} with these loci $p_1$ and $p_2$. 
In order for the ridge to have a nonzero length, we want the intersections with the past lightcones to overlap. 

\subsubsection*{Scattering condition}

Our goal is now to see for what values of $\tau$ this occurs. 
The easiest way to do this is to identify the marginal case where the ridge gets cut off to a point. 
All together we want
\begin{align}
\tanh\rho=\cos(\hat t-\frac{\mu}{2})=\cos(\hat t+\frac{\mu}{2})=\pm \frac{\cos(\hat t+\tau)}{\cos(\frac{\mu}{2})}
\end{align}
The middle equality tells us
\begin{align}
\hat t+\frac{\mu}{2}=\pm\left(\hat t-\frac{\mu}{2}\right)+2n\pi.
\end{align}
Since we don't want to restrict $\mu$ we should have the minus sign. Then we obtain
\begin{align}
\hat t=n\pi .
\end{align}
Now we also need the scattering point to be in the future of the $c_i$ and past of the $r_i$. 
The fact that $\tau$ is upper bounded by $\pi-\mu$ tells us
\begin{align}
\hat t\in \left(\mu-\pi,\pi-\frac{\mu}{2}\right)\in (-\pi,\pi),
\end{align}
so we should have $\hat t=0$.  For $\phi=0,\pi$ we have 
\begin{align}
\cos\tau=\pm\cos^2\frac{\mu}{2}.
\end{align}
Now we've assumed $\mu> \frac{\pi}{2}$ so that $\nu<\frac{\pi}{2}$ and by~\eqref{nu}
\begin{align}
\tau\in \left(0,\frac{\pi}{2}\right).
\end{align}
We thus need to pick the $+$ signed solution corresponding to $\phi=0$. In the end, the marginal scattering point is at 
\begin{align}
\hat t=0,~~\phi=0,~~\cos\tau=\cos^2\frac{\mu}{2},~~\tanh\rho=\cos\frac{\mu}{2}
\end{align}
in agreement with~\cite{may2019quantum} and we get bulk-only scattering for $ \cos^{-1}[\cos^2(\frac{\mu}{2})]\le \tau \le \pi-\mu$. 

\subsubsection{Complexity and correlation}

Now we would like to see how this configuration leads to a counterexample where $\tau$ can be tuned so that $\Delta A_{\text{MI}}$ appearing in the RT formula for the mutual information $I(\V_1 : \V_2)$ vanishes while $\Delta A_{\text{PL}}$ remains fixed and non-zero. We will start by computing each of these quantities in turn.

\subsubsection*{Mutual information}

Recall that in order for the boundary input regions to be domains of dependence we needed to restrict to $\tau <\nu$ and, further, we have $\nu<\frac{\pi}{2}$ by assumption. As such, we know each of the input intervals is less than $\frac{\pi}{2}$ in size. Meanwhile, their centers are separated by $\pi$. Comparing to~\eqref{eq:l_general} we thus see that the entanglement wedge will be connected if the right hand side of
\begin{equation} \label{eq:AMI_2int}
   { \Delta A_{\text{MI}} 
    = \lads \ln \left[ \frac{ \sin^2 \left(  \frac{1}{2} \tau \right)}{\sin \left(  \frac{1}{2} (\mu-\tau) \right)  \sin \left(  \frac{1}{2} (2\pi-\mu-\tau) \right)  } \right]  } \: 
\end{equation}
is positive, which will be the case when
\be
\cos\tau\le \cos^2\frac{\mu}{2}
\ee
which was exactly our scattering condition above once we note $\cos\tau$ is monotonically decreasing over the range $\tau\in(0,\frac{\pi}{2})$. When $\tau<\cos^{-1}[\cos^2(\frac{\mu}{2})]$, we have $\Delta A_{\text{MI}} =0$.

\subsubsection*{Python's lunch}

Now the existence of a python's lunch in our set up is independent of $\tau$ and was guaranteed by our choice that $\mu>\nu$ so that the entanglement wedge of $\R$ was connected. We then find that the python's lunch area difference is given by 
\begin{equation}
    { \Delta A_{\text{PL}} 
    = \lads \ln \left[ \csc^2 \left(  \frac{\mu}{2}  \right) \right]  } \: 
\end{equation}
which is the difference of the bulge surface built with two $\Delta\phi=\pi$ segments and the appetizer surface with $\Delta\phi=\mu$.

\subsubsection*{Connected wedge theorem}

The connected wedge theorem tells us that scattering implies a connected wedge; however, above we saw that the conditions for scattering and positive $\Delta A_{\text{MI}} $ were precisely equal. 
This arises from the fact that in global AdS$_{2+1}$ the connected wedge theorem actually has a converse. 
While one might be worried that our two-interval set up is different than the one considered in~\cite{may2019quantum}, for our discussion of easy-hard scattering we are effectively studying the scattering problem with input regions $\V_i$ and output regions $\R_1$ and $\R_1'$, which reduces to the points based configuration defined by the $c_i$ and $r_i$ above. 
Then, the lift and slope surfaces in the proof of \cite{may2020holographic} have no caustics, and the length of the ridge is precisely equal to twice the mutual information, rather than giving a lower bound. 

\subsubsection*{Complexity versus correlation}

Our above discussions show that we can tune $\tau$ to be small enough that the entanglement wedge for the $\V_i$ becomes disconnected and $\Delta A_{\text{MI}}$ vanishes. However,  $\Delta A_{\text{PL}}$ is independent of $\tau$. We thus have a counterexample where $\Delta A_{\text{MI}} \ngeq \alpha_{0} \Delta A_{\text{PL}}$ for any $\alpha_{0}$.

\subsection{\texorpdfstring{AdS$_{2+1}$}{} with conical defect} \label{app:defect}

We now consider the case of a single conical defect in the centre of AdS$_{2+1}$. 
Note that some of our calculations overlap with those of \cite{Caminiti:2024ctd}.

\subsubsection*{Bulk geometry}

The metric in global coordinates $(t, r, \phi)$ can be found in \eqref{eq:defect_metric}; recall that the case $-1 < M < 0$ corresponds to the defect. Note that, in the case of conical defects, the special values $M = - \frac{1}{N^{2}}$ with $N \in \mathbb{N}$ correspond to orbifolds AdS$_{2+1} / \mathbb{Z}_{N}$, but we can consider $M$ as a continuous parameter more generally. In this more general case, we will isolate the integer part of $1 / \sqrt{|M|}$ by writing
\begin{equation}
    \frac{1}{\sqrt{|M|}} = N + \alpha \: , \qquad N = \left\lfloor \frac{1}{\sqrt{|M|}} \right\rfloor \: , \qquad 0 < \alpha < 1 \: . 
\end{equation}

\subsubsection*{Boundary geometry}

The conformal boundary of the defect spacetime with the relevant configuration of input regions $\V_1, \V_2$ and output region $\R$ is shown in figure \ref{fig:V1V2R_defect}. 
Recall that, using angular coordinate $\phi$ and time coordinate $\hat{t} = t / \lads$, we choose $\V_1, \V_2$ to lie in a slice with fixed $\hat{t} = \hat{t}_{\text{i}}$, to have equal angular size $\chi$, and to be placed symmetrically about $\phi = 0$ and separated by angle $\theta$ satisfying
\begin{equation}
    0 < \chi < \theta < \pi \: ,
\end{equation}
so that the intervals don't overlap. 
The region $\R$ is then in the slice $\hat{t} = \hat{t}_{i} + \frac{\chi}{2}$, has angular size $2 \pi - \theta$, and is centred at $\phi = \pi$.

\subsubsection*{Spacelike geodesics}

Consider the geodesics whose endpoints are at a fixed coordinate time $t$ and separated by an angle $\Delta \phi$ (which may for example be larger than $2 \pi$ if the geodesic self-intersects). Such geodesics centred at $\phi=0$ have trajectory (see e.g. \cite{Caminiti:2024ctd})
\begin{equation} \label{eq:r_phi_general_defect}
    r(\phi) = \sqrt{|M|} \lads \sqrt{ \frac{\sec^{2}(\sqrt{|M|} \phi)}{ \tan^{2} \left( \frac{\sqrt{|M|}}{2} \Delta \phi \right) - \tan^{2}(\sqrt{|M|} \phi)} } \: ,
\end{equation}
and they have regularized lengths
\begin{equation} \label{eq:l_general_defect}
    \ell = 2 \lads \ln \left[ \frac{2}{\epsilon \sqrt{|M|}} \sin \left(  \frac{\sqrt{|M|}}{2} \Delta \phi \right) \right] \: ,
\end{equation}
where we regulate by cutting off the bulk at $r = \frac{\lads}{\epsilon}$. 

\begin{figure}
    \centering
    \includegraphics[height=8cm]{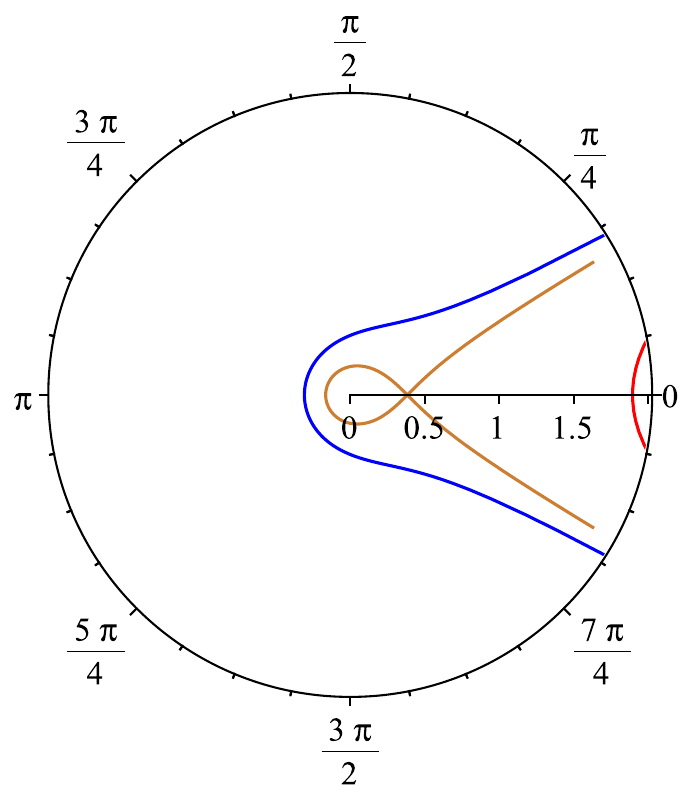}
    \includegraphics[height=8cm]{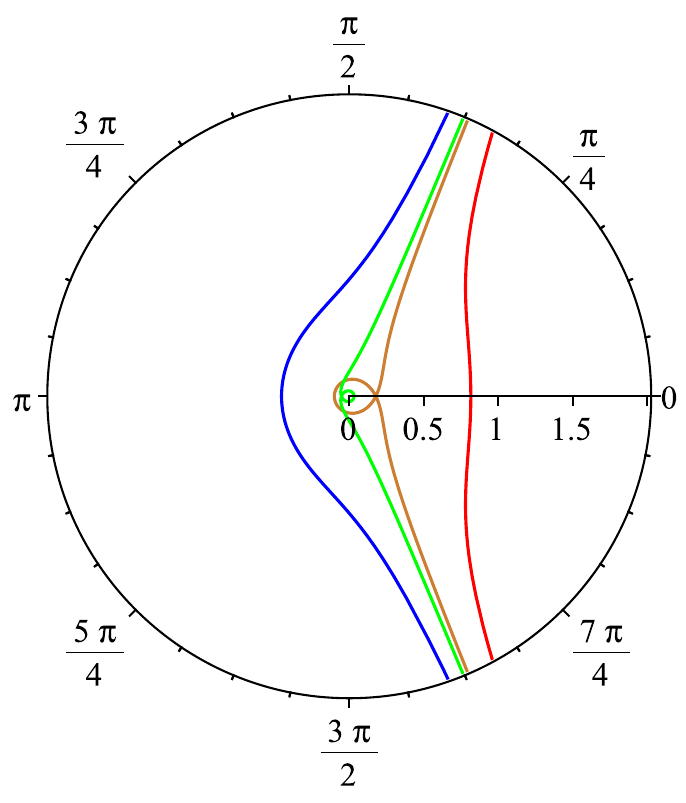}
    \caption{Spacelike geodesics in the case of defect with $\frac{1}{\sqrt{|M|}} = 3.5$, i.e. $N = 3$ and $\alpha = 0.5$. In the left figure, we take $\Theta = \frac{\pi}{3}$ less than $(1 - \alpha) \pi = \frac{\pi}{2}$, and therefore obtain only three geodesics. In the right figure, we take $\Theta = \frac{3 \pi}{4}$ greater than $\frac{\pi}{2}$, and therefore obtain four geodesics.}
    \label{fig:DefectGeods_NonInt}
\end{figure}

We would like to fix $0 < \Theta < \pi$ and determine how many distinct geodesics have endpoints subtending angle $\Theta$.
In \eqref{eq:r_phi_general_defect} and \eqref{eq:l_general_defect}, we may take $\Delta \phi \in (0, \pi/ \sqrt{|M|})$ without loss of generality due to the periodicity of the trigonometric functions. 
If $M = - \frac{1}{N^{2}}$ as in the previous case, then this would imply that, for any angle $0 < \Theta < \pi$, we have $N$ distinct geodesics subtending angle $\Theta$, corresponding to
\begin{equation}
    \Delta \phi \in \begin{cases}
        \{ 2 \pi k + \Theta | k \in \{0, \ldots, \frac{N-2}{2} \} \} \cup \{2 \pi (k+1) - \Theta | k \in \{ 0, \ldots, \frac{N-2}{2} \} \} & 2 \mid N \\
        \{ 2 \pi k + \Theta | k \in \{0, \ldots, \frac{N-1}{2} \} \} \cup \{2 \pi (k+1) - \Theta | k \in \{ 0, \ldots, \frac{N-3}{2} \} \}  & 2 \nmid N
    \end{cases} \: .
\end{equation}
However, more generally, the number of geodesics will vary discontinuously with $\Theta$: when $\Delta \phi \in (0, (N + \alpha) \pi)$, then compared to the above accounting, there is an extra geodesic for $0 < \Theta < \alpha \pi$ when $2 \mid N$, and an extra geodesic for $(1 - \alpha) \pi < \Theta < \pi$ when $2 \nmid N$. 
See figure \ref{fig:DefectGeods_NonInt} for an example. 

\begin{figure}
    \centering
    \includegraphics[height=8cm]{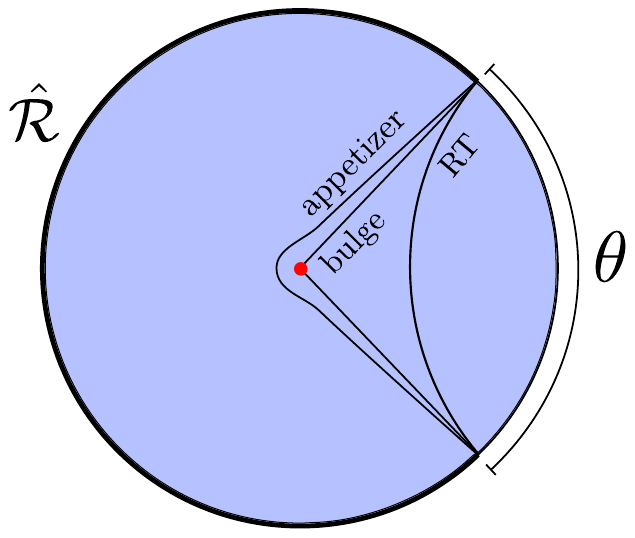}
    \caption{Constant $t$ slice of the AdS defect, and the appetizer, bulge, and RT surfaces associated with boundary region $\R$. Here, the bulge is taken to be the ``kink'' surface.}
    \label{fig:kink}
\end{figure}

In the analysis of the python's lunch conjecture, we will also consider a candidate path which consists of two radial segments joined at a kink where the defect is located, as shown in figure \ref{fig:kink}. The length of such a path is
\begin{equation} \label{eq:l_kink}
    \ell_{\text{kink}} = 2 \int_{0}^{\lads/\epsilon} \frac{dr}{\sqrt{r^{2} / \lads^{2} - M}} = 2 \lads \ln \left[ \frac{2}{\epsilon \sqrt{|M|}} \right] \: .
\end{equation}
This is manifestly larger than the length \eqref{eq:l_general_defect} of any of the extremal surfaces discussed above, and will thus only be the bulge surface when no other candidate bulge surface is available, i.e. in the case $N=1$ or in the case $N=2$ and $\alpha \pi < \Theta < \pi$. This is a consequence of the ``maximinimax prescription'' (see Appendix B of \cite{brown2020python}, or Section 5 of \cite{Arora:2024edk} where the AdS$_{2+1}/\mathbb{Z}_{n}$ orbifolds are explicitly considered). 

We will take a moment to clarify which geodesics are the ``appetizer'' surface, the true RT surface, and the bulge surface. 
For any $-1 < M < 0$, assuming that we take $0 < \theta < \pi$ and are interested in a boundary region $\R$ of size $2 \pi - \theta$, the values of $\Delta \phi$ appearing in \eqref{eq:r_phi_general_defect} and \eqref{eq:l_general_defect} are as follows:
\begin{itemize}
    \item The appetizer surface corresponds to $\Delta \phi_{\text{app}} = 2 \pi - \theta$.
    \item The true RT surface corresponds to $\Delta \phi_{\text{RT}} = \theta$. 
\end{itemize}
For the bulge surface, we have a few cases:
\begin{itemize}
    \item Suppose that $N = 1$.
    \begin{itemize}
        \item For $0 < \theta < (1 - \alpha) \pi$, there is no candidate appetizer surface, and therefore no python's lunch.
        \item For $(1 - \alpha) \pi < \theta < \pi$, the bulge surface is the ``kink'' surface discussed around \eqref{eq:l_kink}. 
    \end{itemize}
    \item Suppose that $N=2$. \begin{itemize}
        \item For $0 < \theta < \alpha \pi$, the bulge surface corresponds to $\Delta \phi_{\text{bulge}} = 2 \pi + \theta$.
        \item For $\alpha \pi < \theta < \pi$, the bulge surface is the kink surface discussed around \eqref{eq:l_kink}. 
    \end{itemize}
    \item Suppose that $N>2$. Then the bulge surface corresponds to $\Delta \phi_{\text{bulge}} = 2 \pi + \theta$. 
\end{itemize}
We'll call the surface with $\Delta \phi = 2 \pi + \theta$ the non-kink bulge surface. 

\subsubsection*{Null geodesics}

Up to a choice of affine parameter, the general solution to the null geodesic equation originating from the conformal boundary at $(t_{-\infty}, \phi_{-\infty})$ is (see e.g. \cite{Caminiti:2024ctd})
\begin{equation} \label{eq:null_geod}
    \begin{split}
        r(s) & = \lads \sqrt{(1 - b^{2}) s^{2} + \frac{b^{2}}{(1 - b^{2})} |M| } \\
        t(s) & = \frac{\lads}{\sqrt{|M|}} \tan^{-1} \left[ \frac{(1 - b^{2})}{\sqrt{|M|}} s \right] + \frac{\pi}{2 \sqrt{|M|}} \lads + t_{-\infty} \\
        \phi(s) & = \frac{1}{\sqrt{|M|}} \tan^{-1} \left[ \frac{(1 - b^{2})}{b\sqrt{|M|}} s \right] + \frac{\pi}{2 \sqrt{|M|}} \text{sgn}(b) + \phi_{-\infty} \: ,
    \end{split}
\end{equation}
where $|b| < 1$. 

A degenerate special case of interest occurs when the angle $\phi$ is constant. In this case, the null equation is $\dot{t} = \pm \frac{\dot{r}}{f(r)}$, with solution
\begin{equation} \label{eq:geod_radial}
    t = \mp \frac{\lads}{\sqrt{|M|}} \left[ \tan^{-1} \left( \frac{r}{\sqrt{|M|} \lads} \right) - \frac{\pi}{2} \right] + t_{\mp \infty} \: ,
\end{equation}
with negative and positive signs for $t$ corresponding to ingoing and outgoing geodesics respectively. 

\subsubsection{Scattering} \label{app:scattering_defect}

We would like to determine under what conditions the ``easy-hard'' two-to-two scattering region $J(E_{\R}^{\text{easy}}, (E_{\R}^{\text{easy}})')$ is non-empty. 
By continuity, and since the entanglement wedges of $\V_{1}, \V_{2}$ coincide with the causal wedges of these regions, it will suffice to determine when we have a non-empty two-to-one scattering region with input points on the conformal boundary at
\begin{equation}
    (\hat{t}, \phi) = \left( \hat{t}_{i} - \frac{\chi}{2}, - \frac{\theta}{2}\right) \qquad \text{and} \qquad (\hat{t}, \phi) = \left( \hat{t}_{i} - \frac{\chi}{2}, \frac{\theta}{2} \right) \: ,
\end{equation}
and the output point lying on the appetizer surface associated with $\hat{\mathcal{R}}$, at some 
\begin{equation}
    (t, r, \phi) = (\hat{t}_{i} \lads + \frac{\chi}{2} \lads, r_{\text{app}}, \phi_{\text{app}}) \: .
\end{equation}

Suppose that we have a bulk scattering point at $(r_{*}, \phi_{*})$. We first determine the elapsed coordinate times $\Delta t_{\text{in}}^{(+)}(r_{*}, \phi_{*}), \Delta t_{\text{in}}^{(-)}(r_{*}, \phi_{*})$ for the ingoing parts of a scattering process where null geodesics originate at the input points and terminate at $(t, r, \phi) = (\hat{t}_{i} \lads - \frac{\chi}{2} \lads + \Delta t_{\text{in}}^{(\pm)}, r_{*}, \phi_{*})$. We then determine the elapsed time $\Delta t_{\text{out}}(r_{*}, \phi_{*})$ for the outgoing part of the scattering process where a null geodesic originates at a point $(t, r, \phi) = (\hat{t}_{i} \lads + \frac{\chi}{2} \lads - \Delta t_{\text{out}}, r_{*}, \phi_{*})$ and terminates at the output point. 
It is evident that the desired scattering process is possible if and only if the time $t$ at which the ingoing part of the process ends is smaller than the time $t$ at which the outgoing part of the process begins, namely
\begin{equation} \label{eq:max_tot_Delta_t}
    \min_{(r_{*}, \phi_{*})} \Big\{ \max \{ \Delta t_{\text{in}}^{(+)}(r_{*}, \phi_{*}), \Delta t_{\text{in}}^{(-)}(r_{*}, \phi_{*}) \} + \Delta t_{\text{out}}(r_{*}, \phi_{*}) \Big\} \leq \chi \lads \: .
\end{equation}
We note that the elapsed time at $(r, \phi)$ for a null geodesic originating at the conformal boundary is given by \cite{Caminiti:2024ctd}
\begin{equation}
    \Delta t = \frac{\lads}{\sqrt{|M|}}  \left[ \frac{\pi}{2} - \tan^{-1} \left( \frac{\cot( \sqrt{|M|} \Delta \phi)}{\sqrt{1 + \frac{|M| \lads^{2}}{r^{2}} \csc^{2}(\sqrt{|M|} \Delta \phi)}} \right) \right] \: ,
\end{equation}
where $\Delta \phi \in [0, \pi]$ is the angular distance between the input point and $\phi$. 

Taking $\phi_{*} \in [-\pi, 0]$ without loss of generality, we find that the viable local minima for $\Delta t_{\text{tot}}(r_{*}, \phi_{*})$ occur 
at $\phi_{*} = -\pi$. 
We therefore find
\begin{equation} \label{eq:Delta_t_in}
    \frac{\Delta t_{\text{in}}(r_{*})}{\lads} = \frac{1}{\sqrt{|M|}} \left[ \frac{\pi}{2} - \tan^{-1} \left( \frac{\cot \left( \frac{\sqrt{|M|}}{2} (2 \pi - \theta) \right) }{\sqrt{1 + \left( \frac{\sqrt{|M|} \lads}{r_{*}} \right)^{2} \csc^{2} \left( \frac{\sqrt{|M|}}{2} (2 \pi - \theta)) \right)}} \right)  \right] \: .
\end{equation}
Moreover, if we take an outgoing radial geodesic on the $\phi = -\pi$ axis, we have by \eqref{eq:geod_radial},
\begin{equation} \label{eq:Delta_t_out_app}
    \frac{\Delta t_{\text{out}}(r_{*})}{\lads} = \Bigg| \frac{\theta - 2 \pi}{2} + \frac{\pi}{2 \sqrt{|M|}} - \frac{1}{\sqrt{|M|}} \tan^{-1} \left( \frac{r_{*}}{\sqrt{|M|} \lads} \right) \Bigg| \: .
\end{equation}

\subsubsection*{Scattering condition}

We can now determine $r_{*}$ by minimizing the total elapsed time $\Delta t_{\text{tot}} = \Delta t_{\text{in}} + \Delta t_{\text{out}}$, and then determine whether this exceeds the upper bound $\chi \lads$ from \eqref{eq:max_tot_Delta_t}. 

\begin{figure}
    \centering
    \includegraphics[height=5cm]{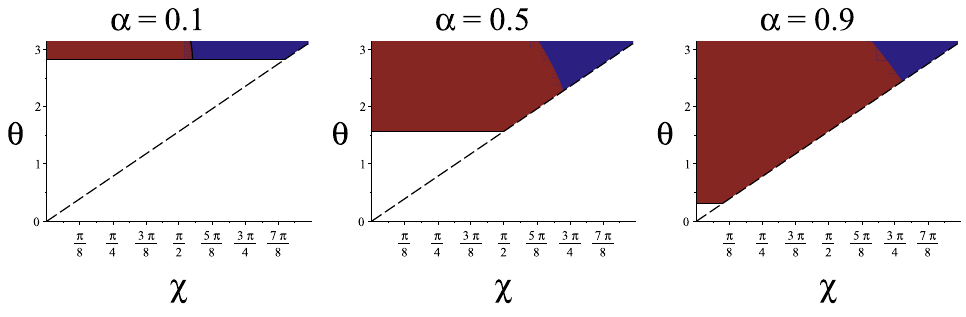}
    \caption{Phase diagrams for $N=1$ and various values of $\alpha$. The blue region is where scattering is allowed, the red region is where a python's lunch exists but scattering is not allowed, and the white region with $\chi < \theta$ (above the dashed line) is where a python's lunch does not exist. }
    \label{fig:PhaseDiagrams}
\end{figure}

We observe that $\Delta t_{\text{in}}(r_{*})$ is decreasing as a function of $r_{*}$, whereas $\Delta t_{\text{out}}(r_{*})$ is decreasing until $r_{*} = r_{\text{app}}$, then increasing. It follows that one should have $r_{*} \geq r_{\text{app}}$. In fact, we find that $\Delta t_{\text{tot}}(r_{*})$ does not have any local extrema on $r_{*} \in (r_{\text{app}}, \infty )$, so we should take $r_{*} = r_{\text{app}}$. Whether scattering is possible or not therefore reduces to whether $\Delta t_{\text{in}}(r_{\text{app}}) \leq \chi \lads$. 
Explicitly, we find the condition
\begin{equation} \label{eq:scatter_defect}
    f_{\text{scatter}}(\chi, \theta) \equiv \cos^{2}\left( \frac{\sqrt{|M|}}{2} (2 \pi - \theta) \right) - \cos(\sqrt{|M|} \chi) \geq 0 \: .
\end{equation}
In figure \ref{fig:PhaseDiagrams}, we plot some examples of phase diagrams depicting when scattering is possible and when a python's lunch exists.  

\subsubsection{Complexity and correlation}

We would like to see whether the area difference $\Delta A_{\text{MI}}$ appearing in the RT formula for the mutual information $I(\V_1 : \V_2)$ and the area difference $\Delta A_{\text{PL}}$ appearing in the python's lunch proposal for the complexity are related in the case where bulk scattering is possible. We begin by calculating these quantities in our setup. 

\subsubsection*{Mutual information}

The entanglement wedge of $\V_1 \cup \V_2$ can either be disconnected or connected, implying vanishing or non-vanishing mutual information respectively. From \eqref{eq:l_general_defect}, the disconnected candidate RT surface has regularized length
\begin{equation}
    \ell_{\text{dis}}
    = 4 \lads \ln \left[ \frac{2 \lads}{\epsilon} \sin \left( \frac{\sqrt{|M|}}{2} \chi \right) \right] \: .
\end{equation}
The candidate connected entanglement wedge includes the defect if and only if $\theta + \chi > \pi$, and therefore the corresponding RT surface has regularized length
\begin{equation}
    \ell_{\text{conn}}
    = \begin{cases}
    2 \lads \ln \left[ \left(\frac{2 \lads}{\epsilon} \right)^{2} \sin \left( \frac{\sqrt{|M|}(\theta - \chi)}{2} \right) \sin \left( \frac{\sqrt{|M|}(\theta + \chi)}{2} \right) \right] & \theta + \chi < \pi \\
    2 \lads \ln \left[ \left( \frac{2 \lads}{\epsilon} \right)^{2} \sin \left( \frac{\sqrt{|M|} (\theta - \chi)}{2} \right) \sin \left( \frac{\sqrt{|M|} (2 \pi - \theta - \chi)}{2} \right) \right]  & \theta + \chi > \pi
    \end{cases} \: .
\end{equation}

For $\theta + \chi < \pi$, the condition for $\ell_{\text{conn}} < \ell_{\text{dis}}$, and therefore to have a connected entanglement wedge, is
\begin{equation} \label{eq:conn_con_1}
    \theta < \frac{2}{\sqrt{|M|}} \sin^{-1} \left[ \sqrt{2} \sin \left( \frac{\sqrt{|M|}}{2} \chi \right) \right] \: ,
\end{equation}
whereas for $\theta + \chi > \pi$, it is that either $\chi > \frac{\pi}{2}$ or 
\begin{equation} \label{eq:conn_con_2}
    \theta < \pi - \frac{2}{\sqrt{|M|}} \sin^{-1} \left( \sqrt{\sin^{2} \left( \frac{\sqrt{|M|}}{2} (\pi - \chi) \right) - \sin^{2} \left( \frac{\sqrt{|M|}}{2}\chi \right)} \right) \: .
\end{equation}

For any situation $0 < \chi < \theta < \pi$ with $\chi > \frac{\pi}{2}$, one necessarily has $\theta + \chi > \pi$, so the entanglement wedge is connected if and only if one of the following mutually disjoint conditions holds:
\begin{enumerate}
    \item $\chi < \frac{\pi}{2}$ and $\chi + \theta < \pi$ and \eqref{eq:conn_con_1}
    \item $\chi < \frac{\pi}{2}$ and $\chi + \theta > \pi$ and \eqref{eq:conn_con_2}
    \item $\chi > \frac{\pi}{2}$. 
\end{enumerate}

The area difference appearing in the mutual information $I(\V_1 : \V_2)$ is, in the case of a connected entanglement wedge,
\begin{equation} \label{eq:AMI_defect}
    \Delta A_{\text{MI}} = \begin{cases}
        2 \lads \ln \left[ \frac{\sin^{2}\left( \frac{\sqrt{|M|}}{2} \chi \right)}{\sin^{2}\left( \frac{\sqrt{|M|}}{2} \theta \right) - \sin^{2}\left( \frac{\sqrt{|M|}}{2} \chi \right)} \right] & \chi + \theta < \pi \\
        2 \lads \ln \left[ \frac{\sin^{2}\left( \frac{\sqrt{|M|}}{2} \chi \right)}{\sin^{2}\left( \frac{\sqrt{|M|}}{2} (\pi - \chi) \right) - \sin^{2}\left( \frac{\sqrt{|M|}}{2} (\pi - \theta) \right)} \right] & \chi + \theta > \pi
    \end{cases} \: .
\end{equation}

\subsubsection*{Python's lunch}

The area difference appearing in the python's lunch conjecture is the difference between the area of the bulge surface and the appetizer surface.
We find for cases where the bulge is the kink surface, i.e. both $N=1$, $(1 - \alpha) \pi < \theta$ and $N=2$, $\alpha \pi < \theta$, that
\begin{equation}
    \Delta A_{\text{PL}} = 2 \lads \ln \left[ \csc \left( \frac{\sqrt{|M|}}{2} (2 \pi - \theta) \right) \right] \: ,
\end{equation}
whereas for cases where the bulge is a non-kink surface, i.e. $N=2$, $\theta < \alpha \pi$ and $N > 2$,
\begin{equation}
    \Delta A_{\text{PL}} = 2 \lads \ln \left[ \frac{\sin \left( \frac{\sqrt{|M|}}{2} (2 \pi + \theta) \right)}{\sin \left( \frac{\sqrt{|M|}}{2} (2 \pi - \theta) \right)} \right] \: .
\end{equation}
We note that the inequalities imply
\begin{equation}
    0 \leq \frac{1}{\lads} \Delta A_{\text{PL}} \leq 2 \ln (3) \: ,
\end{equation}
with the upper bound corresponding to the limit $M \rightarrow 0$ and $\theta \rightarrow \pi$. 

\subsubsection*{Connected wedge theorem}

We can verify whether scattering implies a connected entanglement wedge in the current setup. We can consider the contrapositive of this statement: suppose that the entanglement wedge is not connected, such that $\chi < \frac{\pi}{2}$, and either one is in the case where $\chi + \theta < \pi$ and
\begin{equation} \label{eq:cw_inequalities1}
     \theta > \frac{2}{\sqrt{|M|}} \sin^{-1} \left[ \sqrt{2} \sin \left( \frac{\sqrt{|M|}}{2} \chi \right) \right] \: ,
\end{equation}
or one is in the case where $\chi + \theta > \pi$ and
\begin{equation} \label{eq:cw_inequalities2}
    \theta > \pi - \frac{2}{\sqrt{|M|}} \sin^{-1} \left( \sqrt{\sin^{2} \left( \frac{\sqrt{|M|}}{2} (\pi - \chi) \right) - \sin^{2} \left( \frac{\sqrt{|M|}}{2}\chi \right)} \right) \: .   
\end{equation}
We recall from \eqref{eq:scatter_defect} that the condition for scattering is the positivity of 
\begin{equation}
    f_{\text{scatter}}(\chi, \theta) = \cos^{2} \left( \frac{\sqrt{|M|}}{2} (2 \pi - \theta) \right) - \cos(\sqrt{|M|} \chi) \: .
\end{equation}
We can maximize this quantity with respect to either of the above constraints in \eqref{eq:cw_inequalities1} and \eqref{eq:cw_inequalities2}, along with the constraints $0 \leq \chi \leq \theta \leq \pi$, and the additional constraints 
\begin{equation} \label{eq:add_const}
    0 < \alpha < 1 \: , \qquad (1 - \alpha) \pi \leq \theta \: ,
\end{equation}
in the case $N=1$, which ensure that a python's lunch exists. 
In both cases, we find that the maximum value subject to the constraints is non-positive. 
The largest value of $f_{\text{scatter}}(\chi, \theta)$ occurs in the limit $(\chi, \theta) \rightarrow (\frac{\pi}{2}, \pi)$ and in the limit $M \rightarrow 0$, in which cases $f_{\text{scatter}}(\chi, \theta)$ approaches zero. 
Thus, scattering is not possible. This establishes that the connected wedge theorem holds in this setting.

Furthermore, similar to the result of \cite{Caminiti:2024ctd}, we find that holographic scattering requires that we are in the phase where the entanglement wedge includes the defect. Indeed, we can do a similar maximization, now subject to the constraints
\begin{equation} 
      \chi \leq \frac{\pi}{2} \: , \quad \chi + \theta \leq \pi \: , \quad \theta \leq \frac{2}{\sqrt{|M|}} \sin^{-1} \left[ \sqrt{2} \sin \left( \frac{\sqrt{|M|}}{2} \chi \right) \right] \: ,
\end{equation}
corresponding to a connected wedge which does not include the defect, 
along with $0 \leq \chi \leq \theta \leq \pi$, and the additional constraints \eqref{eq:add_const}
in the case $N=1$. We find that the maximum is again negative, implying that the situation in which the entanglement wedge does not include the defect is inconsistent with scattering. 
In particular, this means that we can always restrict to the second case in \eqref{eq:AMI_defect} when evaluating $\Delta A_{\text{MI}}$ for scattering configurations. 

\subsubsection*{Complexity versus correlation}

We can at last investigate whether there is any relationship between $\Delta A_{\text{MI}}$ and $\Delta A_{\text{PL}}$ whenever scattering is possible.  
We observe that, for fixed $M$ and $\theta$, we have $\Delta A_{\text{PL}}$ constant with respect to $\chi$, while $\Delta A_{\text{MI}}$ is increasing with respect to $\chi$. The minimum value of $\Delta A_{\text{MI}}$ for a given $\Delta A_{\text{PL}}$ therefore corresponds to the minimum value of $\chi$ consistent with scattering, i.e. where $f_{\text{scatter}}(\chi, \theta) = 0$ in \eqref{eq:scatter_defect}. Letting $\chi_{\text{min}}(\theta)$ denote this value of $\chi$, we find that the lower bound on $\Delta A_{\text{MI}}$ can be found by fixing $\chi = \chi_{\text{min}}(\theta)$, setting $\theta = \pi$, and varying $-1 < M < 0$. From these considerations, we can obtain an implicit lower bound which we plot in figure \ref{fig:LowerBoundAMI_defect_maintext}, given by
\begin{equation}
\begin{split}
    \Delta A_{\text{MI}}^{\text{min}}(x) & = 2 \lads \ln \left[ \frac{1}{1 + 2 \cos^{2}(x) - 2 \cos(x) \sqrt{1 + \cos^{2}(x)}} \right] \\
    \Delta A_{\text{PL}}(x) & = \begin{cases}
        2 \lads \ln \left( \frac{\sin (3x)}{\sin(x)} \right) & 0 < x < \frac{\pi}{6} \\
        2 \lads \ln \left( \csc (x) \right) & \frac{\pi}{6} < x < \frac{\pi}{2}
    \end{cases} \: .
\end{split}
\end{equation}

As mentioned in the main text, we find that the largest lower bound of the form $\Delta A_{\text{MI}} \geq \alpha_0 \Delta A_{\text{PL}}$ consistent with this phase boundary is
\begin{equation}
    \alpha_0 = - \frac{2 \ln\left( \sqrt{2} - 1 \right)}{\ln(3)} \approx 1.6 \: ,
\end{equation}
obtained from the limit $M \rightarrow 0$. 

\subsection{BTZ black hole} \label{app:BTZ}

We next consider the case of a non-rotating BTZ black hole. 

\subsubsection*{Bulk geometry}

The metric is the same as the defect metric in \eqref{eq:defect_metric}, though now with $M > 0$. 

\subsubsection*{Boundary geometry}

We specify the boundary geometry exactly as in the analysis of the conical defect; see figure \ref{fig:V1V2R_defect} and the surrounding discussion for details. 

\subsubsection*{Spacelike geodesics}

Consider the geodesics whose endpoints are at a fixed coordinate time $t$ and separated by an angle $\Delta \phi$. Such geodesics have trajectory \cite{Caminiti:2024ctd}
\begin{equation}
    r(\phi) = \sqrt{M} \lads \sqrt{ \frac{\sech^{2}(\sqrt{M} \phi)}{ \tanh^{2} \left( \frac{\sqrt{M}}{2} \Delta \phi \right) - \tanh^{2}(\sqrt{M} \phi)} } \: ,
\end{equation}
and they have regularized lengths
\begin{equation}
    \ell = 2 \lads \ln \left[ \frac{2}{\epsilon \sqrt{M}} \sinh \left(  \frac{\sqrt{M}}{2} \Delta \phi \right) \right] \: ,
\end{equation}
where we regulate by cutting off the bulk at $r = \frac{\lads}{\epsilon}$. 

\begin{figure}
    \centering
    \includegraphics[height=8cm]{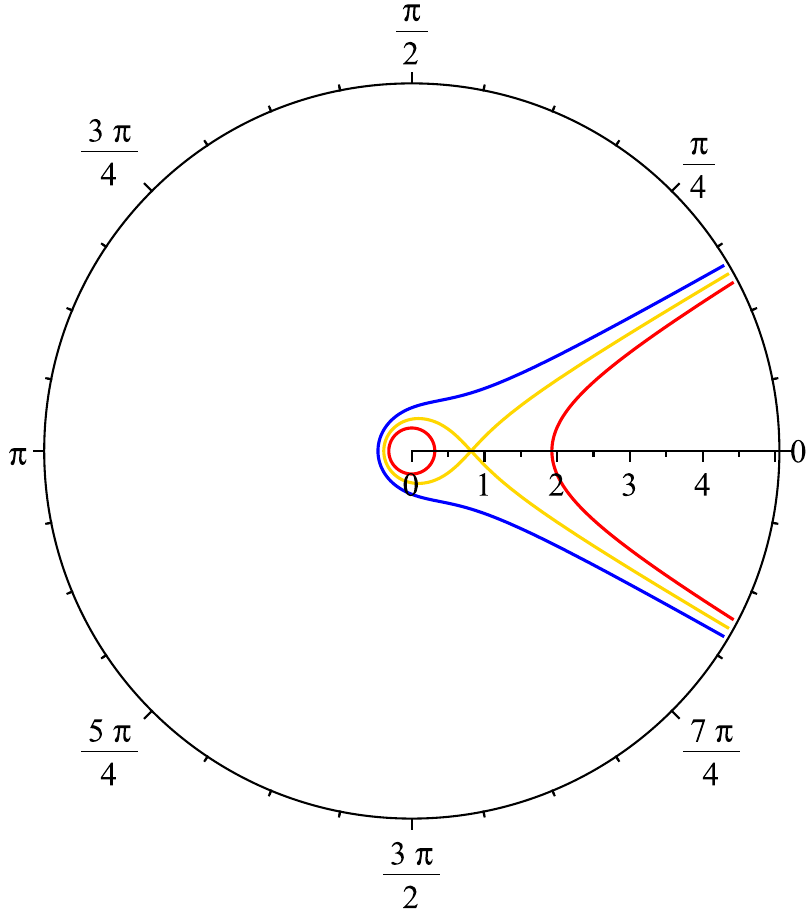}
    \caption{Spacelike geodesics in the case of the BTZ black hole with $M = 0.1$ and $\theta = \frac{\pi}{3}$. The solid red, gold, and blue curves denote the RT surface, bulge surface, and appetizer surface respectively.}
    \label{fig:BTZ_geods}
\end{figure}

Unlike in the case of the defect, we now have infinitely many geodesics with fixed endpoints, which can have arbitrarily many self-intersections. The bulge surface will presumably always be the shortest self-intersecting surface, with a single self-intersection. 
Thus, for any $M > 0$, assuming that we take $0 < \theta < \pi$ and are interested in a boundary region $\R$ of size $2 \pi - \theta$, the values of $\Delta \phi$ appearing in the expressions for the trajectory of the geodesic are as follows:
\begin{itemize}
    \item The appetizer surface corresponds to $\Delta \phi_{\text{app}} = 2 \pi - \theta$.
    \item The true RT surface corresponds to $\Delta \phi_{\text{RT}} = \theta$. Additionally, there is a separate connected piece of the RT surface at the horizon $r = \sqrt{M} \lads$. 
    \item The bulge surface corresponds to $\Delta \phi_{\text{bulge}} = 2 \pi + \theta$. 
\end{itemize}

We note that the exchange of dominance between the candidate RT surfaces, required for the existence of a python's lunch for $\hat{\mathcal{R}}$, occurs for $\theta = \theta_{\text{crit}}(M)$ with 
\begin{equation}
    \theta_{\text{crit}}(M) = \pi - \frac{1}{\sqrt{M}} \ln \cosh \left( \pi \sqrt{M} \right) \: .
\end{equation}
This is a decreasing function of $M$, with $\theta_{\text{crit}}(0) = \pi$ and $\theta_{\text{crit}}(\infty) = 0$. 

\subsubsection*{Null geodesics}

Up to a choice of affine parameter, the general solution to the null geodesic equation originating from the conformal boundary at $(t_{-\infty}, \phi_{-\infty})$ is \cite{Caminiti:2024ctd}
\begin{equation}
    \begin{split}
        r(s) & = \lads \sqrt{(1 - b^{2}) s^{2} - \frac{b^{2}}{(1 - b^{2})} M } \\
        t(s) & = - \frac{\lads}{\sqrt{M}} \tanh^{-1} \left[ \frac{\sqrt{M}}{(1 - b^{2})} \frac{1}{s} \right] + t_{-\infty} \\
        \phi(s) & = - \frac{1}{\sqrt{M}} \tanh^{-1} \left[ \frac{b\sqrt{M}}{(1 - b^{2})} \frac{1}{s} \right] + \phi_{-\infty} \: ,
    \end{split}
\end{equation}
where $|b| < 1$. 

A degenerate special case of interest occurs when the angle $\phi$ is constant. In this case, the solution becomes
\begin{equation} \label{eq:geod_radial_btz}
    t = \pm \frac{\lads}{\sqrt{M}} \tanh^{-1} \left( \frac{\sqrt{M} \lads}{r} \right) + t_{\infty} \: ,
\end{equation}
with negative and positive signs for $t$ corresponding to ingoing and outgoing geodesics respectively. 

\subsubsection{Scattering} \label{sec:scattering_btz}

Again, we will consider the two-to-one scattering region with input points on the conformal boundary at
\begin{equation}
    (\hat{t}, \phi) = \left( \hat{t}_{i} - \frac{\chi}{2}, - \frac{\theta}{2}\right) \qquad \text{and} \qquad (\hat{t}, \phi) = \left( \hat{t}_{i} - \frac{\chi}{2}, \frac{\theta}{2} \right) \: ,
\end{equation}
and the output point lying on the appetizer surface associated with $\hat{\mathcal{R}}$, at some 
\begin{equation}
    (t, r, \phi) = (\hat{t}_{i} \lads + \frac{\chi}{2} \lads, r_{\text{app}}, \phi_{\text{app}}) \: .
\end{equation} 

Suppose that we have a bulk scattering point at $(r_{*}, \phi_{*})$. As in the analysis of the defect in section \ref{app:scattering_defect}, the condition under which scattering is allowed is
\begin{equation}
    \min_{(r_{*}, \phi_{*})} \Bigg\{ \max\{ \Delta t_{\text{in}}^{+}(r_{*}, \phi_{*}), \Delta t_{\text{in}}^{-}(r_{*}, \phi_{*}) \} + \Delta t_{\text{out}}(r_{*}, \phi_{*}) \Bigg\} \leq \chi \lads  \: .
\end{equation}
We note that the elapsed time at $(r, \phi)$ for a null geodesic originating at the conformal boundary is given by \cite{Caminiti:2024ctd}
\begin{equation}
    \Delta t = \frac{\lads}{\sqrt{M}}  \tanh^{-1} \left( \frac{\sqrt{ \frac{M \lads^{2}}{r^{2}} + \sinh^{2}(\sqrt{M} \Delta \phi)}}{\cosh( \sqrt{M} \Delta \phi)} \right) \: ,
\end{equation}
where $\Delta \phi \in [0, \pi]$ is the angular distance between the input point and $\phi$. 

Taking $\phi_{*} \in [-\pi, 0]$ without loss of generality, we appear to find that the viable local minima for $\Delta t_{\text{tot}}(r_{*}, \phi_{*})$ occur at $\phi_{*} = -\pi$. 
We therefore find
\begin{equation} \label{eq:Delta_t_in_BTZ}
    \frac{\Delta t_{\text{in}}(r_{*})}{\lads} = \frac{1}{\sqrt{M}}  \tanh^{-1} \left( \frac{\sqrt{ \frac{M \lads^{2}}{r_{*}^{2}} + \sinh^{2}\left( \frac{\sqrt{M}}{2} (2 \pi - \theta) \right)}}{\cosh\left( \frac{\sqrt{M}}{2} (2 \pi - \theta) \right)} \right) \: .
\end{equation}
Moreover, if we take an outgoing radial geodesic on the $\phi = -\pi$ axis, we have by \eqref{eq:geod_radial_btz},
\begin{equation} \label{eq:Delta_t_out_app_BTZ}
    \frac{\Delta t_{\text{out}}(r_{*})}{\lads} = \Bigg| \frac{2 \pi - \theta}{2} - \frac{1}{\sqrt{M}} \tanh^{-1} \left( \frac{\sqrt{M} \lads}{r_{*}} \right) \Bigg| \: .
\end{equation}

\subsubsection*{Scattering condition}

We can now determine $r_{*}$ by minimizing the total elapsed time $\Delta t_{\text{tot}} = \Delta t_{\text{in}} + \Delta t_{\text{out}}$, and then determine whether this exceeds the upper bound $\chi \lads$ from \eqref{eq:max_tot_Delta_t}. 

We observe that $\Delta t_{\text{in}}(r_{*})$ is decreasing as a function of $r_{*}$, whereas $\Delta t_{\text{out}}(r_{*})$ is decreasing until $r_{*} = r_{\text{app}}$, then increasing. It follows that one should have $r_{*} \geq r_{\text{app}}$. In fact, we find that $\Delta t_{\text{tot}}(r_{*})$ does not have any local extrema on $r_{*} \in (r_{\text{app}}, \infty )$, so we should take $r_{*} = r_{\text{app}}$. Whether scattering is possible or not therefore reduces to whether $\Delta t_{\text{in}}(r_{\text{app}}) \leq \chi \lads$. 
Explicitly, we find the condition
\begin{equation} \label{eq:scatter_BTZ}
    f_{\text{scatter}}(\chi, \theta) \equiv \cosh(\sqrt{M} \chi) - \cosh^{2}\left( \frac{\sqrt{M}}{2} (2 \pi - \theta) \right) \geq 0 \: .
\end{equation} 

\begin{figure}
    \centering
    \includegraphics[height=5cm]{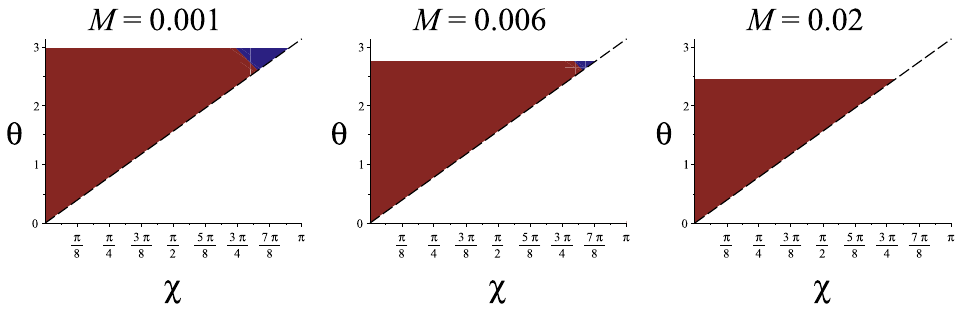}
    \caption{Phase diagrams for various values of $M$. The blue region is where scattering is allowed, the red region is where a python's lunch exists but scattering is not allowed, and the white region with $\chi < \theta$ (above the dashed line) is where a python's lunch does not exist. }
    \label{fig:PhaseDiagrams_BTZ}
\end{figure}

We observe that $f_{\text{scatter}}(\chi, \theta)$ is maximized at $\chi = \theta$, and $f_{\text{scatter}}(\theta, \theta)$ is an increasing function of $0 < \theta < \pi$ for all fixed $M > 0$. 
On the other hand, we recall that there is an upper bound on $\theta$ consistent with $\hat{\mathcal{R}}$ having a python's lunch, given by $\theta < \theta_{\text{crit}}(M)$, which is decreasing with $M$. It turns out that there is a maximum value of $M$ above which this upper bound on $\theta$ is inconsistent with scattering; this maximum value is
\begin{equation}
    M_{\text{max}} \approx 0.01217001701 \: .
\end{equation}
We plot phase diagrams for values of $M$ below and above this threshold in figure \ref{fig:PhaseDiagrams_BTZ}. 

\subsubsection{Complexity and correlation}

We would like to see whether the area difference $\Delta A_{\text{MI}}$ appearing in the RT formula for the mutual information $I(\V_1 : \V_2)$ and the area difference $\Delta A_{\text{PL}}$ appearing in the python's lunch proposal for the reconstruction complexity of $\hat{\mathcal{R}}$ are related in the case where bulk scattering is possible. We proceed to calculate these quantities in our setup. 

\subsubsection*{Mutual information}

We can either have an entanglement wedge for $\V_1 \cup \V_2$ which is disconnected or connected, implying vanishing or non-vanishing mutual information respectively. The disconnected candidate RT surface has regularized length
\begin{equation}
    \ell_{\text{dis}}
    = 4 \lads \ln \left[ \frac{2}{\epsilon \sqrt{M}} \sinh \left( \frac{\sqrt{M}}{2} \chi \right) \right] \: .
\end{equation}
The candidate connected entanglement wedge surrounds the horizon if and only if 
\begin{equation}
     \theta + \chi > 2 \pi - \theta_{\text{crit}}(M) \: ,
\end{equation}
and therefore the corresponding RT surface has regularized length
\begin{equation}
    \ell_{\text{conn}} = 2 \lads \ln \left[ \frac{4}{\epsilon^{2} M}  \sinh \left( \frac{\sqrt{M}(\theta - \chi)}{2} \right) \sinh \left( \frac{\sqrt{M}(\theta + \chi)}{2} \right) \right]
\end{equation}
in the case $\theta + \chi < 2 \pi - \theta_{\text{crit}}(M)$, and
\begin{equation}
    \ell_{\text{conn}}
    = 2 \lads \ln \left[ \frac{4}{\epsilon^{2} M} \sinh \left( \frac{\sqrt{M} (\theta - \chi)}{2} \right) \sinh \left( \frac{\sqrt{M} (2 \pi - \theta - \chi)}{2} \right) \right] + 2 \pi \sqrt{M} \lads 
\end{equation}
in the case $\theta + \chi > 2 \pi - \theta_{\text{crit}}(M)$. 

For $\theta + \chi < 2 \pi - \theta_{\text{crit}}(M)$, the condition for $\ell_{\text{conn}} < \ell_{\text{dis}}$, and therefore to have a connected entanglement wedge, is
\begin{equation} \label{eq:conn_cond_1_defect}
    \theta < \frac{2}{\sqrt{M}} \sinh^{-1} \left[ \sqrt{2} \sinh \left( \frac{\sqrt{M}}{2} \chi \right) \right] \: .
\end{equation}
On the other hand, for $\theta + \chi > 2 \pi - \theta_{\text{crit}}(M)$, the condition to have a connected entanglement wedge is
\begin{equation}
    \sinh^{2} \left( \frac{\sqrt{M} (\pi - \chi)}{2} \right) - e^{-\pi \sqrt{M}} \sinh^{2} \left( \frac{\sqrt{M}}{2} \chi \right) < \sinh^{2} \left( \frac{\sqrt{M} (\pi - \theta)}{2} \right) 
\end{equation}
which means either that 
\begin{equation}
    \chi > \chi_{\text{crit}}(M) \equiv \frac{\pi}{2} + \frac{1}{\sqrt{M}} \ln \left( \cosh \left( \frac{\sqrt{M} \pi}{2} \right) \right) \: ,
\end{equation}
which ensures that the left hand side is negative, or 
\begin{equation}
\label{eq:conn_cond_2_defect}
    \theta < \pi - \frac{2}{\sqrt{M}} \sinh^{-1} \left( \sqrt{\sinh^{2} \left( \frac{\sqrt{M}}{2} (\pi - \chi) \right) - e^{- \pi \sqrt{M}} \sinh^{2} \left( \frac{\sqrt{M}}{2}\chi \right)} \right) \: .
\end{equation}

For $0 < \chi < \theta < \pi$, we see that the entanglement wedge is connected if and only if one of the following mutually disjoint conditions holds:
\begin{enumerate}
    \item $\chi + \theta < 2 \pi - \theta_{\text{crit}}(M)$ and \eqref{eq:conn_cond_1_defect}
    \item $\chi + \theta > 2 \pi - \theta_{\text{crit}}(M)$ and $\chi < \chi_{\text{crit}}(M)$ and \eqref{eq:conn_cond_2_defect}
    \item $\chi + \theta > 2 \pi - \theta_{\text{crit}}(M)$ and $\chi > \chi_{\text{crit}}(M)$.  
\end{enumerate}

The area difference appearing in the mutual information $I(\V_1 : \V_2)$ is, in the case of a connected entanglement wedge,
\begin{equation}
    \Delta A_{\text{MI}} = \begin{cases}
        2 \lads \ln \left[ \frac{\sinh^{2}\left( \frac{\sqrt{M}}{2} \chi \right)}{\sinh^{2}\left( \frac{\sqrt{M}}{2} \theta \right) - \sinh^{2}\left( \frac{\sqrt{M}}{2} \chi \right)} \right] & \chi + \theta < 2 \pi - \theta_{\text{crit}}(M) \\
        2 \lads \ln \left[ \frac{e^{- \pi \sqrt{M}} \sinh^{2}\left( \frac{\sqrt{M}}{2} \chi \right)}{\sinh^{2}\left( \frac{\sqrt{M}}{2} (\pi - \chi) \right) - \sinh^{2}\left( \frac{\sqrt{M}}{2} (\pi - \theta) \right)} \right] & \chi + \theta > 2 \pi - \theta_{\text{crit}}(M)
    \end{cases} \: .
\end{equation}

\subsubsection*{Python's lunch}

The area difference appearing in the python's lunch conjecture is the difference between the area of the bulge surface and the appetizer surface.
When a python's lunch exists, which requires $\theta < \theta_{\text{crit}}(M)$, we have
\begin{equation}
    \Delta A_{\text{PL}} =2 \lads \ln \left[ \frac{\sinh \left( \frac{\sqrt{M}}{2} (2 \pi + \theta) \right)}{\sinh \left( \frac{\sqrt{M}}{2} (2 \pi - \theta) \right)} \right] \: .
\end{equation}
We note that this quantity is bounded from above for fixed $M$ since $\theta < \theta_{\text{crit}}(M)$, with upper bound
\begin{equation}
    \frac{1}{\lads} \Delta A_{\text{PL}} < 2 \ln \left[ \frac{\sinh \left( \frac{\sqrt{M}}{2} (3 \pi  - \frac{1}{\sqrt{M}} \ln \cosh \left( \pi \sqrt{M} \right)) \right)}{\sinh \left( \frac{\sqrt{M}}{2} (\pi + \frac{1}{\sqrt{M}} \ln \cosh \left( \pi \sqrt{M} \right)) \right)} \right]  \: .
\end{equation}
This upper bound is a decreasing function of $M$, equal to $2 \lads \ln(3)$ at $M=0$ and $2 \lads \ln(2)$ at $M = \infty$. We also find that there is a lower bound
\begin{equation}
    - 2 \ln(\sqrt{2}-1) \leq \frac{1}{\lads} \Delta A_{\text{PL}} \: ,
\end{equation}
obtained from the limit $M \rightarrow 0$ with $\theta$ taking its minimum value consistent with \eqref{eq:scatter_BTZ} and $\chi \leq \theta$. 

\subsubsection*{Connected wedge theorem}

As for the defect, we can verify that the possibility of scattering implies a connected entanglement wedge for $\V_{1} \cup \V_{2}$. 
Suppose that the entanglement wedge is not connected. Then we either have $\chi + \theta < 2 \pi - \theta_{\text{crit}}(M)$ and
\begin{equation}
    \theta > \frac{2}{\sqrt{M}} \sinh^{-1} \left( \sqrt{2} \sinh \left( \frac{\sqrt{M}}{2} \chi \right) \right) \: ,
\end{equation}
or $\chi + \theta > 2 \pi - \theta_{\text{crit}}(M)$, $\chi < \chi_{\text{crit}}(M)$, and
\begin{equation}
    \theta > \pi - \frac{2}{\sqrt{M}} \sinh^{-1} \left[ \sqrt{\sinh^{2}\left( \frac{\sqrt{M}}{2} ( \pi - \chi ) \right) - e^{- \pi \sqrt{M}} \sinh^{2} \left( \frac{\sqrt{M}}{2} \chi \right) } \right] \: .
\end{equation}
Maximizing $f_{\text{scatter}}(\chi, \theta)$ from \eqref{eq:scatter_BTZ} subject to these constraints, along with $0 \leq \chi \leq \theta \leq \theta_{\text{crit}}(M)$, 
we find a non-positive result, demonstrating that scattering is not possible whenever the entanglement wedge is not connected. 

Similarly, we also find that whenever scattering is possible, we are in the phase where the entanglement wedge of $\V_{1} \cup \V_{2}$ surrounds the horizon. 

\subsubsection*{Complexity versus correlation}

\begin{figure}
    \centering
    \includegraphics[height=8cm]{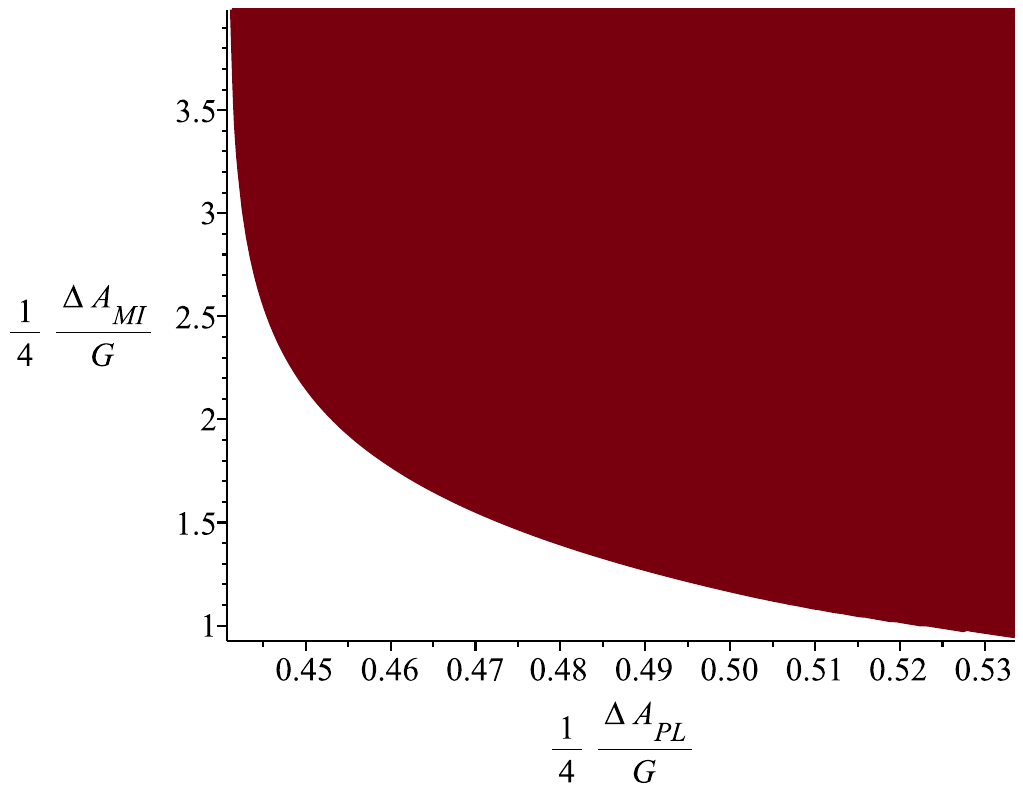}
    \caption{Allowed values of $\frac{\Delta A_{\text{MI}}}{4 G}$ as a function of $\frac{\Delta A_{\text{PL}}}{4 G}$ (solid red) for the BTZ black hole. Here we are using units $\frac{\lads}{G}$ for both axes, which is roughly the number of local degrees of freedom in the dual CFT (recall that the Brown-Henneaux central charge is $c = \frac{3 \lads}{2 G}$). Note that $-\frac{\ln(\sqrt{2}-1)}{2} \frac{\lads}{G} \leq \frac{\Delta A_{\text{PL}}}{4 G} \leq \frac{1}{2} \ln(3) \frac{\lads}{G}$ whenever scattering is possible, so we are plotting the full allowed range on the horizontal axis.   }
    \label{fig:LowerBoundAMI_BTZ}
\end{figure}

We can now investigate the relationship between $\Delta A_{\text{MI}}$ and $\Delta A_{\text{PL}}$.
We observe that, for fixed $M$ and fixed $\theta$, we have $\Delta A_{\text{PL}}$ constant with respect to $\chi$, while $\Delta A_{\text{MI}}$ is increasing with respect to $\chi$. The minimum value of $\Delta A_{\text{MI}}$ for a given $\Delta A_{\text{PL}}$ therefore corresponds to the minimum value of $\chi$ consistent with scattering, i.e. where $f_{\text{scatter}}(\chi, \theta) = 0$. From these considerations, we can obtain a lower bound which we plot in figure \ref{fig:LowerBoundAMI_BTZ}; note that we are plotting the full range of $\Delta A_{\text{PL}}$ consistent with scattering in this figure. 

The largest coefficient $\alpha_0$ in a lower bound of the form $\Delta A_{\text{MI}} \geq \alpha_0 \Delta A_{\text{PL}}$ can be found from taking $M \rightarrow 0$, $\theta = \theta_{\text{crit}}(M)$, and $\chi = \frac{1}{\sqrt{M}} \cosh^{-1} \left[ \cosh^{2} \left( \frac{\sqrt{M}}{2} (2 \pi - \theta) \right) \right]$. 
We thereby obtain
\begin{equation}
    \alpha_0 = \frac{\ln \left( \frac{1}{3 - 2 \sqrt{2}} \right)}{\ln(3)} \approx 1.6 \: ,
\end{equation}
precisely the same coefficient as in the defect case. In fact, the defect case provides stronger constraints on a putative lower bound on $\Delta A_{\text{MI}}$ than the BTZ case for all $\Delta A_{\text{PL}}$. 

\subsection{\texorpdfstring{AdS$_{2+1}$}{} with ETW brane} \label{app:ETW}

We lastly turn to the case of a pure AdS$_{2+1}$ spacetime terminating on a static ETW brane of constant tension. 

\subsubsection*{Bulk geometry}

The metric of pure AdS$_{2+1}$ in global coordinates $(t, r, \phi)$, along with the ETW brane trajectory $r(\phi)$ for a given choice of tension parameter $T \in [0, 1)$, can be found in \eqref{eq:pure_AdS_metric} and \eqref{eq:ETW_traj}. 
It will sometimes be useful to change to Poincar{\'e} coordinates, with the Poincar{\'e} patch centred at $\phi = - \frac{\pi}{2}$ and $t = 0$. One has metric
\begin{equation}
    ds^{2} = \frac{\lads^{2}}{z^{2}} \left( dz^{2} - d\tau^{2} + dx^{2} \right) \: ,
\end{equation}
and the ETW brane has trajectory
\begin{equation}
    \frac{x}{z} = - \frac{T}{\sqrt{1 - T^{2}}} \: .
\end{equation}
In this case, the asymptotic boundary is the half-plane $x > 0$. See figure \ref{fig:AdS_Slice} for the depiction of the $\tau=0$ slice. 

Denoting $y^{2} = - \tau^{2} + x^{2} + z^{2}$ for convenience, the transformation between these two coordinate systems is
\begin{equation}
    r = \lads \sqrt{ \frac{x^{2}}{z^{2}} + \frac{(1 - y^{2})^{2}}{4 z^{2}} } \: , \quad
    t = \lads \tan^{-1} \left[ \frac{2 \tau}{1 + y^{2}} \right] \: , \quad
    \phi = \tan^{-1} \left[ \frac{y^{2} - 1}{2 x} \right] \: .
\end{equation}

\subsubsection*{Boundary geometry}

The conformal boundary of the spacetime with two different choices of regions $\V_{1}, \V_{2}$ and $\R$ can be found in figure \ref{fig:ETW_V1V2R_maintext}. 
Recall that, using angular coordinate $\phi \in (- \frac{\pi}{2}, \frac{\pi}{2})$ and time coordinate $\hat{t} = t / \lads$, we are choosing $\R$ to lie at fixed $\hat{t}$ and to subtend angle $0 < \mu < \pi$. We then consider two cases: either $\V_{1}, \V_{2}$ do not reach the boundary, in which case these intervals have width $\tau \in (0, \min\{\mu, \pi - \mu\})$ and are at fixed $\hat{t}$, or they do reach the boundary, in which case we take $\nu \in (0, \mu)$ to denote the angular separation of their innermost endpoints. 

The total boundary time $\Delta \hat{t}$ elapsed between an input point (the smallest $\hat{t}$ point of the causal development of $\V_1$ or $\V_2$) and the output region $\R$ is $\Delta \hat{t} = \tau$ for the first choice of regions $\V_{1}, \V_{2}$, and
\begin{equation} \label{eq:max_t_ETW}
    \Delta \hat{t}_{\text{tot}} = \frac{\pi + \mu - 2 \nu}{2}
\end{equation}
for the second choice of $\V_{1}, \V_{2}$. 

\subsubsection*{Spacelike geodesics}

\begin{figure}
    \centering
    \includegraphics[height=8cm]{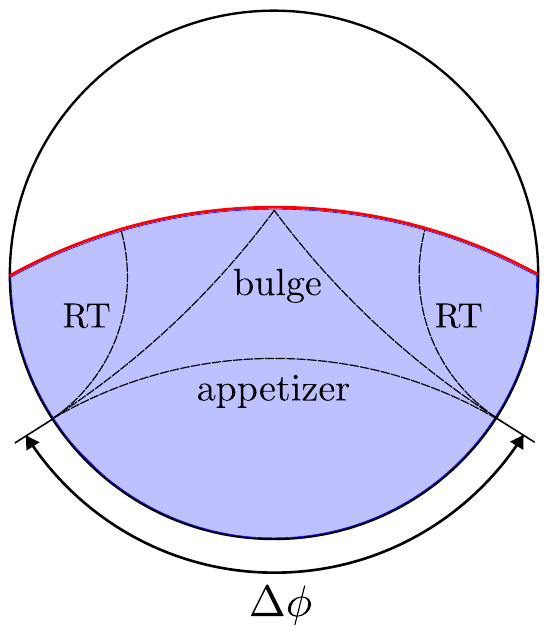}
    \caption{Constant $t$ slice of the AdS$_{3}$ geometry, depicting the appetizer, bulge, and RT surfaces for a boundary interval of size $\Delta \phi$. }
    \label{fig:SES}
\end{figure}

We will only be interested in spacelike extremal surfaces within a constant $t$ slice; for the purposes of calculations, we can assume this is the $t=0$ slice, or equivalently, the $\tau = 0$ slice in Poincar{\'e} coordinates. See figure \ref{fig:SES} for an illustration.

We can first consider the extremal surface, homologous to a boundary interval centred at $\phi=0$, which does not end on the ETW brane. Assuming it subtends angle $\Delta \phi$, we have that the trajectory of the surface is given by 
\begin{equation} \label{eq:traj_RT_ETW}
    r(\phi) = \frac{\lads \sec(\phi)}{\sqrt{\tan^{2}(\Delta \phi / 2) - \tan^{2}(\phi)}} \: ,
\end{equation}
and the regularized length (with regulator at $r = \lads / \epsilon$) is given by 
\begin{equation}
    \ell = 2 \lads \ln \left[ \frac{2}{\epsilon} \sin \left( \frac{\Delta \phi}{2} \right) \right] \: .
\end{equation}

We next consider the locally minimal surface, corresponding to the same boundary interval, which ends on the ETW brane. This will be symmetric about $\phi=0$ and consist of two disconnected pieces at $\phi < 0$ and $\phi > 0$; we restrict to determining the trajectory of the $\phi < 0$ piece without loss of generality. 

It will be most convenient to work in Poincar{\'e} coordinates on the $\tau = 0$ slice. In this case, the relevant extremal surface is a circular arc of a radius we label $R$ centred at the origin in Poincar{\'e} coordinates
\begin{equation}
    x^{2} + z^{2} = R^{2} \: .
\end{equation}
It is readily verified that this is normal to the ETW brane. Reverting to global coordinates, we have trajectory
\begin{equation}
    \frac{1 + \frac{r^{2}}{\lads^{2}} \cos^{2} \phi}{1 + \frac{r^{2}}{\lads^{2}}} = \frac{4 R^{2}}{(1 + R^{2})^{2}} \: , \quad t = 0 \: .
\end{equation}
Note that the boundary value of $\phi$, which should be $- \frac{\Delta \phi}{2}$, is given by $\cos^{-1}\left( \frac{2R}{1 + R^{2}} \right)$, so
\begin{equation}
    R = \sec ( \Delta \phi / 2 ) - \tan( \Delta \phi / 2) \: .
\end{equation}
In particular, the surface has fixed $t$ and
\begin{equation}
   \frac{1 + \frac{r^{2}}{\lads^{2}} \cos^{2} \phi}{1 + \frac{r^{2}}{\lads^{2}}} = \cos^{2}( \Delta \phi / 2) \: .
\end{equation}

We can also find the length of the surface in Poincar{\'e} coordinates. 
We first remark that the regulator $r = \lads / \epsilon$ translates to an $x$-dependent cutoff
\begin{equation} \label{eq:z_cutoff}
    z(x) = \frac{\epsilon}{2}(1 + x^{2}) \: .
\end{equation}
We thus have (including both connected pieces of the surface)
\begin{equation}
\begin{split}
    \ell & = 2 \lads \int_{z_{0}}^{R} \frac{dz}{z} \frac{R}{\sqrt{R^{2} - z^{2}}} + 2 \lads \int_{\frac{\epsilon}{2}(1+R^{2})}^{R} \frac{dz}{z} \frac{R}{\sqrt{R^{2} - z^{2}}} \\
    & = 2 \lads \ln \left( \frac{2 \cos(\Delta \phi / 2)}{\epsilon} \right) + \lads \ln \left( \frac{ 1 +  T}{1 - T } \right) \: , 
\end{split}
\end{equation}
where $z_{0}$ is the solution to $x^{2} + z^{2} = R^{2}$ and $\frac{x}{z} = - \frac{T}{\sqrt{1 - T^{2}}}$, given by
\begin{equation}
    z_{0} = \sqrt{1 - T^{2}} R \: .
\end{equation}
We observe that the two candidate RT surfaces exchange dominance at
\begin{equation} \label{eq:exchange_dom}
     \tan^{2} \left( \frac{\Delta \phi}{2} \right) = \frac{ 1 +  T}{1 - T } \: .
\end{equation}

Lastly, we consider a candidate for the bulge surface. 
We propose that the bulge surface should be analogous to half of the ``X'' surface for global AdS identified in \cite{brown2020python}: it should be symmetric about $\phi = 0$, consisting of two constant $t$ extremal surfaces which meet at the point $\phi=0$ on the ETW brane (see figure \ref{fig:SES}). 

\begin{figure}
    \centering
    \includegraphics[height=8cm]{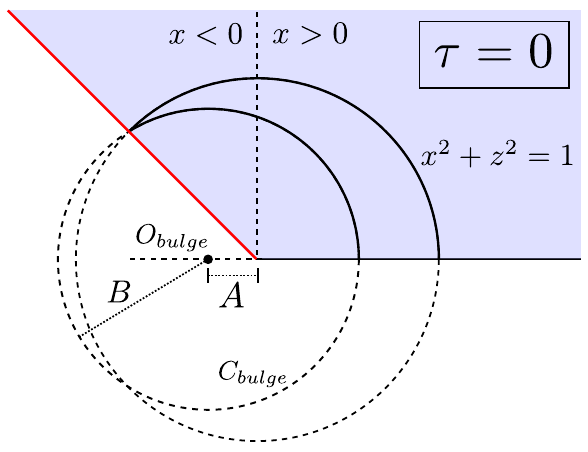}
    \caption{Geometry on the $\tau = 0$ slice. Here, $C_{\text{bulge}}$ is a Poincar{\'e} coordinate circle of radius $B$ centred at $O_{\text{bulge}}$, whose coordinates are $(x, z) = (-A, 0)$; it includes an arc which we identify as half of the bulge surface. The circle $x^{2} + z^{2} = 1$ corresponds to the symmetry plane $\phi = 0$ in global coordinates.}
    \label{fig:geom}
\end{figure}

Again, it will be simplest to work in Poincar{\'e} coordinates at $\tau = 0$. The extremal surface will now be a circular arc of a radius we label $B$ and offset from the origin in the $x$-direction by a shift we label $A$,
\begin{equation}
    (x + A)^{2} + z^{2} = B^{2} \: .
\end{equation}
See figure \ref{fig:geom}. 
We know that the arc has one endpoint at $(x_{1}, z_{1}) = (\frac{1 - \sin(\mu/2)}{\cos(\mu / 2)}, 0)$, and another endpoint at $(x_{2}, z_{2})$ satisfying
\begin{equation}
    x_{2} = - \frac{T}{\sqrt{1 - T^{2}}} z_{2} \: , \qquad x_{2}^{2} + z_{2}^{2} = 1 \: ,
\end{equation}
where we have used that $(x_{2}, z_{2})$ is on the ETW brane and is located at $\phi = -\pi$ in global coordinates. In particular, we have $(x_{2}, z_{2}) = ( - T, \sqrt{1 - T^{2}})$. We therefore find
\begin{equation} \label{eq:AB}
\begin{split}
    A & = \tan(\mu/2) \left( \frac{1 - \sin(\mu/2)}{1 - \sin(\mu/2) + T \cos(\mu/2)} \right) \: , \\
    B^{2} & = 1 - 2 T A + A^{2} = \left( A + \frac{1-\sin(\mu/2)}{\cos(\mu/2)} \right)^{2} \: .
\end{split}
\end{equation}
We note that the inward tangent vector to the bulge surface at the ETW brane, located at $(x_{2}, z_{2})$, is in the positive $z$-direction if and only if $A < T$.

The expression for the regulated length (including both pieces) depends on whether $A<T$ is satisfied; if so, then it is given by (recalling the cutoff \eqref{eq:z_cutoff})
\begin{equation}
    \begin{split}
        \ell & = 2 B \lads \int_{\sqrt{1 - T^{2}}}^{B} \frac{dz}{z} \frac{1}{\sqrt{B^{2} - z^{2}}} + 2 B \lads \int_{\epsilon [ B (B-A) + A T]}^{B} \frac{dz}{z} \frac{1}{\sqrt{B^{2} - z^{2}}} \\
        & = 2 \lads \ln \left( \frac{2}{\epsilon} \frac{B \left( B + \sqrt{B^{2} - 1 + T^{2}} \right)}{\sqrt{1 - T^{2}} \left( B(B-A) + AT \right)} \right)  \qquad (A < T) \: ,
    \end{split}
\end{equation}
while if not, it is given by
\begin{equation}
    \begin{split}
        \ell & = 2 B \lads \int_{\epsilon [ B (B-A) + A T]}^{\sqrt{1 - T^{2}}} \frac{dz}{z} \frac{1}{\sqrt{B^{2} - z^{2}}} \\
        & = 2 \lads \ln \left( \frac{2}{\epsilon} \frac{B \sqrt{1 - T^{2}}}{ \left( B (B-A) + A T \right) \left( B + \sqrt{B^{2} - 1 + T^{2}} \right)} \right)  \qquad (A > T) \: .
    \end{split}
\end{equation}
For convenience, we observe that
\begin{equation}
    \begin{split}
        \sqrt{B^{2} - 1 + T^{2}} & = |A - T| \\
        & = \Big| \frac{(1 - \sin(\mu/2)) \left( \sin(\mu/2) - T \cos(\mu/2) - T^{2} \left( 1 + \sin(\mu/2) \right) \right)}{\cos(\mu/2) \left( 1 - \sin(\mu/2) + T \cos(\mu/2) \right)} \Big|
    \end{split}
\end{equation}
and
\begin{equation}
    B(B-A) + AT = \frac{1}{2} \left( 1 + \left( \frac{1 - \sin(\mu/2)}{\cos(\mu/2)} \right)^{2} \right) \: .
\end{equation}
We therefore find that for $A < T$, 
\begin{multline}
    \ell = 2 \lads \ln \Bigg[ \frac{2}{\epsilon} \frac{1}{\sqrt{1-T^{2}}} \frac{(1 - \sin(\mu/2))(1 + T \cos(\mu/2))}{\left( 1 - \sin(\mu/2) + T \cos(\mu/2) \right)^{2}} \\
    \quad \times \left( 1 - \sin(\mu/2) + 2 T \cos(\mu/2) + T^{2} (1 + \sin(\mu/2)) \right) \Bigg] \: ,
\end{multline}
while for $A > T$, one has for $\tan(\mu/2) > \frac{T \sqrt{1-T^{2}}(\sqrt{1-T^{2}} + T)}{1 - 2 T^{2}}$
\begin{equation}
    \ell = 2 \lads \ln \left[ \frac{2}{\epsilon} \frac{1}{\sqrt{1-T^{2}}} (1 + T \cos(\mu/2)) \right] \: ,
\end{equation}
and for 
$\tan(\mu/2) < \frac{T \sqrt{1-T^{2}}(\sqrt{1-T^{2}} + T)}{1 - 2 T^{2}}$
\begin{equation}
    \ell = 2 \lads \ln \left[ \frac{2}{\epsilon} \frac{\sqrt{1-T^{2}} \left( 1 + \sin(\mu/2) \right) \left( 1 + T \cos(\mu/2) \right)}{\left( 1 - \sin(\mu/2) + 2 T \cos(\mu/2) + T^{2}(1 + \sin(\mu/2)) \right)} \right] \: .
\end{equation}

To justify the proposition that this is the correct bulge surface, we can at least verify that it is maximal under a perturbation similar to that for which the ``X'' surface in vacuum AdS is maximal. In one direction, if the endpoints of the two halves of the bulge surface on the ETW brane move apart from one another, so that for example the segment at $\phi < 0$ now has an endpoint at 
\begin{equation}
    x_{2} = - T (1 - \delta) \: , \qquad z_{2} = \sqrt{1 - T^{2}} (1 - \delta) \: ,
\end{equation}
then $A$ and $B$ change and we find that the first order variation in the length is always negative for any allowed values of $T, \mu$. In the other direction, if the endpoints of the two halves of the bulge surface move away from the ETW brane along the $\phi=0$ axis, so that they are now located at
\begin{equation}
    x_{2} = - T(1 - \delta) \: , z_{2} = \sqrt{1 - T^{2}(1-\delta)^{2}} \: ,
\end{equation}
we again find a negative first order variation for the length. 

\subsubsection*{Null geodesics}

We will not require any null geodesics which reflect from the ETW brane in our analysis; consequently, the relevant formulae correspond to the expressions for the defect in \eqref{eq:null_geod} and \eqref{eq:geod_radial}, in the case $M = -1$. 

\subsubsection{Scattering}

We proceed much as in previous appendices to determine for what parameter values scattering is possible. 
For the first choice of configuration for $\V_{1}, \V_{2}, \R$, the scattering condition is precisely the same as in the vacuum case, i.e.
\begin{equation} \label{eq:ETW_scatt_1}
    \cos \tau \leq \cos^{2} \left( \frac{\mu}{2} \right) \: ,
\end{equation}
so we will now consider the second choice of configuration. 

In accord with \eqref{eq:max_t_ETW}, we will require that the scattering process has input points on the conformal boundary at 
\begin{equation}
    (\hat{t}, \phi) = \left( \hat{t}_{i}, - \frac{\pi}{2} \right) \qquad \text{and} \qquad (\hat{t}, \phi) = \left( \hat{t}_{i}, \frac{\pi}{2} \right)
\end{equation}
and output points in the ``easy'' and ``hard'' regions associated with $\R$ at $t / \lads = \hat{t}_{i} + \frac{\pi + \mu - 2 \nu}{2}$ should satisfy
\begin{equation}
    \text{min}_{(r_{*}, \phi_{*})} \Big\{ \max \{ \Delta t_{\text{in}}^{+}(r_{*}, \phi_{*}), \Delta t_{\text{in}}^{-}(r_{*}, \phi_{*})\} + \Delta t_{\text{out}}(r_{*}, \phi_{*}) \Big\} \leq \left( \frac{\pi + \mu - 2 \nu}{2} \right) \lads \: .
\end{equation}

It transpires that we should take $\phi_{*} = 0$. 
In this case, one has
\begin{equation} \label{eq:t_in}
    \frac{1}{\lads} \Delta t_{\text{in}}(r_{*}) = \frac{\pi}{2} \: ,
\end{equation}
independent of $r_{*}$. 
Moreover, the radial geodesic from the scattering point to the appetizer surface \eqref{eq:traj_RT_ETW} has
\begin{equation} \label{eq:t_out_app}
    \frac{1}{\lads} \Delta t_{\text{out}}(r_{*}) = \big| \frac{\pi - \mu}{2} - \tan^{-1}\left( \frac{r_{*}}{\lads} \right) \big| \: .
\end{equation}
This is of course minimized if $r_{*} = r_{\text{app}}$, in which case we obtain $\Delta t_{\text{out}} = 0$. 

\subsubsection*{Scattering condition}

For the first choice of configuration for $\V_{1}, \V_{2}, \R$, the scattering condition is provided in \eqref{eq:ETW_scatt_1}. 
It is worth noting that this inequality can only be consistent with the requirement $\tau < \min\{\mu, \pi - \mu\}$ for $\mu < 2 \cos^{-1}(1 / \sqrt{3})$. 
For the second choice, 
the coordinate $t$ which is allowed to elapse throughout the entire scattering process is $\frac{\Delta t_{\text{in}} + \Delta t_{\text{out}}}{\lads} \leq \frac{\pi + \mu - 2 \nu}{2}$.
Thus, scattering will be allowed if and only if
\begin{equation}
    \nu \leq \frac{\mu}{2}  \: .
\end{equation}

\subsubsection{Complexity and correlation}

Again, we will check whether the area difference $\Delta A_{\text{MI}}$ appearing in the RT formula for the mutual information $I(\V_1 : \V_2)$ and the area difference $\Delta A_{\text{PL}}$ appearing in the python's lunch proposal for the complexity are related in the case where bulk scattering is possible.

\subsubsection*{Mutual information}

In the configuration of $\V_{1}, \V_{2}$ parametrized by $(\mu, \tau)$, we have three possible phases for RT surfaces: a disconnected entanglement wedge, a connected entanglement wedge which includes part of the ETW brane, and a connected entanglement wedge which does not include part of the ETW brane. The areas of these surfaces are given by
\begin{equation}
    \begin{split}
        \ell_{\text{dis}} & = 2 \lads \ln \left[ \frac{4}{\epsilon^{2}} \sin^{2} \left( \frac{\tau}{2} \right) \right] \\
        \ell_{\text{conn, ETW}} & = 2 \lads \ln \left[ \frac{4}{\epsilon^{2}} \sin \left( \frac{\mu - \tau}{2} \right) \cos \left( \frac{\mu + \tau}{2} \right) \right] + \lads \ln \left( \frac{1+T}{1-T} \right) \\
        \ell_{\text{conn, no ETW}} & = 2 \lads \ln \left[ \frac{4}{\epsilon^{2}} \sin \left( \frac{\mu - \tau}{2} \right) \sin \left( \frac{\mu + \tau}{2} \right) \right] \: . 
    \end{split}
\end{equation}
We see that the second of these dominates the third if and only if
\begin{equation}
    \tan^{2} \left( \frac{\mu+\tau}{2} \right) \leq \frac{1+T}{1-T} \: ,
\end{equation}
so when we are in the connected phase we have
\begin{equation}
    \Delta A_{\text{MI}} = \begin{cases}
        2 \lads \ln \left[ \frac{\sin^{2}(\tau/2)}{\cos^{2}(\tau/2) - \cos^{2}(\mu/2) } \right] & \tan^{2} \left( \frac{\mu + \tau}{2} \right) < \frac{1+T}{1-T} \\
        2 \lads \ln \left[ \sqrt{\frac{1-T}{1+T}} \frac{2 \sin^{2}(\tau/2)}{\sin(\mu) - \sin(\tau)} \right] & \tan^{2} \left( \frac{\mu + \tau}{2} \right) > \frac{1+T}{1-T}
    \end{cases} \: .
\end{equation}
The condition that the entanglement wedge is connected is determined numerically by the requirement that the argument of the logarithm is greater than 1. Note that $\Delta A_{\text{MI}}$ is increasing as a function of $\tau$ for fixed $T$ and $\mu$. 

On the other hand, in the configuration of $\V_{1}, \V_{2}$ parametrized by $(\mu, \nu)$, 
we see from \eqref{eq:exchange_dom} that 
$\V_1 \cup \V_2$ will have a connected entanglement wedge if and only if
\begin{equation} \label{eq:nu_crit}
    \nu < \nu_{\text{crit}}(T) \equiv 2 \tan^{-1} \left( \sqrt{\frac{1+T}{1-T}} \right)  \: .
\end{equation}
In this case, the area difference determining the mutual information will be
\begin{equation} \label{eq:AMI_ETW}
    \Delta A_{\text{MI}} = 2 \lads \ln \left[ \cot \left( \frac{\nu}{2} \right) \right] + \lads \ln \left( \frac{1+T}{1-T} \right) \: .
\end{equation}
We note that $\Delta A_{\text{MI}}$ is monotonically decreasing with $0 < \nu < \pi$, and monotonically increasing with $0 < T < 1$. 

\subsubsection*{Python's lunch}

Recall from \eqref{eq:exchange_dom} that the python's lunch will only exist for the region $\R$ if 
\begin{equation} \label{eq:mu_PL}
    \mu > \mu_{\text{crit}}(T) \equiv 2 \tan^{-1} \left( \sqrt{\frac{1+T}{1-T}} \right) \: ,
\end{equation}
so we assume this in the following. In this case, the area difference appearing in the python's lunch conjecture is
\begin{equation}
    \Delta A_{\text{PL}} = \begin{cases} 
        2 \lads \ln \left( \frac{1}{\sin(\mu/2)} \frac{B \left( B + \sqrt{B^{2} - 1 + T^{2}} \right)}{\sqrt{1 - T^{2}} \left( B(B-A) + AT \right)} \right)  & A < T \\
        2 \lads \ln \left( \frac{1}{\sin(\mu/2)} \frac{B \sqrt{1 - T^{2}}}{ \left( B (B-A) + A T \right) \left( B + \sqrt{B^{2} - 1 + T^{2}} \right)} \right) & A > T 
    \end{cases} \: ,
\end{equation}
where $A, B$ are determined by \eqref{eq:AB}.

Using \eqref{eq:AB}, we observe that a consequence of \eqref{eq:mu_PL} is
\begin{equation} \label{eq:A_lb}
    A(\mu, T) > A(\mu_{\text{crit}}(T), T) = \sqrt{\frac{1+T}{1-T}} \frac{\sqrt{2} - \sqrt{1+T}}{\sqrt{2} + T - \sqrt{1+T}} \: .
\end{equation}

\subsubsection*{Connected wedge theorem}

We can readily verify the connected wedge theorem, namely that scattering implies a connected entanglement wedge for $\V_1 \cup \V_2$. For $\V_{1}, \V_{2}$ parametrized by $(\mu, \tau)$, we know that scattering is equivalent to $\cos \tau \leq \cos^{2}(\mu/2)$. In this case, we find that
\begin{equation}
    \frac{\sin^{2}(\tau/2)}{\cos^{2}(\tau/2) - \cos^{2}(\mu/2)} \geq \frac{\sin^{2}(\tau/2)}{\cos^{2}(\tau/2) - \cos(\tau)} = 1
\end{equation}
so the entanglement wedge must be connected. Whether it includes part of the ETW brane or not depends on whether $\tan^{2} \left( \frac{\mu+\tau}{2} \right) \leq \frac{1+T}{1-T}$; we find that this inequality can never be satisfied, so the entanglement wedge always includes part of the ETW brane whenever scattering is possible. Explicitly, we have
\begin{equation}
\begin{split}
    \tan^{2} \left( \frac{\mu + \tau}{2} \right) & \geq \tan^{2} \left( \frac{\mu + \cos^{-1}(\cos^{2}(\mu/2))}{2} \right) \\
    & = \sin^{2} (\mu / 2) \left( \frac{ \sqrt{1 + \cos^{2}(\mu/2)} + \cos(\mu/2)}{\cos(\mu/2) \sqrt{1 + \cos^{2}(\mu/2)} - \sin^{2}(\mu/2)} \right)^{2} \: .
\end{split}
\end{equation}
This quantity is increasing for $\mu \in (0, 2 \cos^{-1}(1/\sqrt{3}))$, so it must be strictly larger than the same quantity with $\mu$ replaced by $\mu_{\text{crit}}(T)$ if a python's lunch exists; this quantity is equal to
\begin{equation} \label{eq:tan2_T}
    - \frac{1+T}{\sqrt{2}} \frac{\sqrt{3-T} + \sqrt{1-T}}{1+T - \sqrt{(3-T)(1-T)}} \: .
\end{equation}
We also observe that, since $2 \cos^{-1}(1/\sqrt{3}) > \mu > \mu_{\text{crit}}(T)$, we must have $0 < T < \frac{1}{3}$. We find that \eqref{eq:tan2_T} is strictly larger than $\frac{1+T}{1-T}$ for $T$ in this range, verifying our claim.  

Now, for $\V_{1}, \V_{2}$ parametrized by $(\mu, \nu)$, we have established above that scattering in particular requires $\nu \leq \frac{\mu}{2} < \frac{\pi}{2}$. However, as per \eqref{eq:nu_crit}, a connected entanglement wedge for a given $T$ is equivalent to the requirement $\nu < \nu_{\text{crit}}(T)$, and since $\nu_{\text{crit}}(T)$ is monotonically increasing with $0 \leq T < 1$ we have
\begin{equation}
    \nu_{\text{crit}}(T) \geq \nu_{\text{crit}}(0) = \frac{\pi}{2} \: .
\end{equation}
Consequently, whenever scattering is possible, $\nu < \frac{\pi}{2} \leq \nu_{\text{crit}}(T)$, and thus one has a connected entanglement wedge. 

\subsubsection*{Complexity versus correlation}

\begin{figure}
    \centering
    \includegraphics[height=8cm]{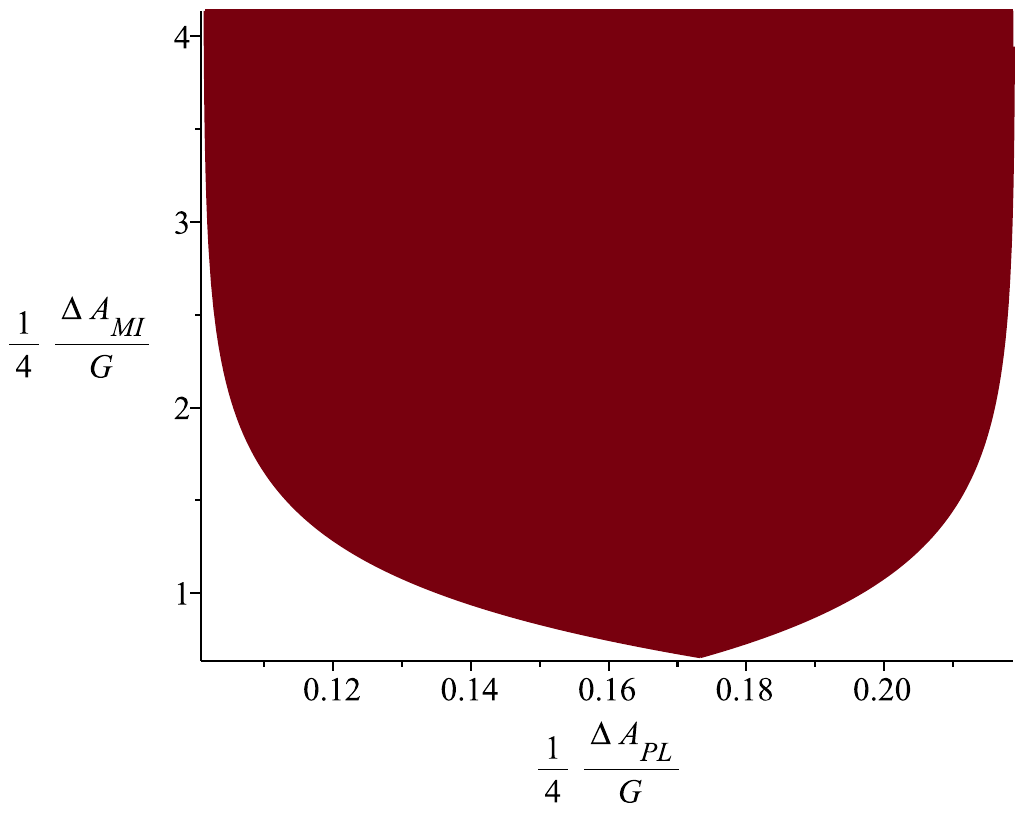}
    \caption{Allowed values of $\frac{\Delta A_{\text{MI}}}{4 G}$ as a function of $\frac{\Delta A_{\text{PL}}}{4 G}$ (solid red) for the ETW brane geometry with boundary regions parametrized by $(\mu, \tau)$. Here we are using units $\frac{\lads}{G}$ for both axes, which is roughly the number of local degrees of freedom in the dual CFT (recall that the Brown-Henneaux central charge is $c = \frac{3 \lads}{2 G}$). Note that $\frac{1}{4} \ln(3/2) \frac{\lads}{G} \leq \frac{\Delta A_{\text{PL}}}{4 G} \lesssim 0.219 \frac{\lads}{G}$ whenever scattering is possible, so we are plotting the full allowed range on the horizontal axis.  }
    \label{fig:LowerBoundAMI_ETW_1}
\end{figure}

We are interested in determining the lower bound on $\Delta A_{\text{MI}}$ for a given value of $\Delta A_{\text{PL}}$. We will begin by considering the configurations of $\V_{1}, \V_{2}$ parametrized by $(\mu, \tau)$. Note that, for fixed $T$, $\Delta A_{\text{PL}}$ depends on $\mu$ and not $\tau$, and recall that $\Delta A_{\text{MI}}$ is an increasing function of $\tau$; thus, for fixed $T$, we obtain a lower bound on $\Delta A_{\text{MI}}$ at fixed $\Delta A_{\text{PL}}$ by evaluating at $\tau_{\text{min}}(\mu)$, the minimum value of $\tau$ consistent with scattering. Explicitly, we have for $0 < \mu < 2 \cos^{-1}(1/ \sqrt{3})$ and $0 < T < \frac{1}{3}$ that
\begin{equation} \label{eq:AMIAPLETW1}
    \begin{split}
        \Delta A_{\text{MI}} & = 2 \lads \ln \left[ \sqrt{\frac{1-T}{1+T}} \frac{\sin(\mu/2)}{2 \cos(\mu/2) - \sqrt{1 + \cos^{2}(\mu/2)}} \right] \\
        \Delta A_{\text{PL}} & = 2 \lads \ln \left( \frac{1}{\sin(\mu/2)} \frac{B \sqrt{1-T^{2}}}{(B(B-A) + AT) (B+\sqrt{B^{2} - 1 + T^{2}})} \right) \: .
    \end{split}
\end{equation}
where we have used that $A>T$ whenever $T$ is in this range, as a consequence of \eqref{eq:A_lb}. We find that the total lower bound is piecewise, coming either from fixing $T = 0$ and varying $\mu$, or from fixing $\mu = 2 \tan^{-1} \left( \sqrt{\frac{1+T}{1-T}} \right)$ and varying $T$. The former applies to $\ln(3/2) < \frac{\Delta A_{\text{PL}}}{\lads} < \ln(2)$ and gives
\begin{equation}
    \Delta A_{\text{MI}} \geq 2 \ln \left( \frac{ e^{- \Delta A_{\text{PL}} / 2 \lads}}{2 \sqrt{1 - e^{- \Delta A_{\text{PL}} / \lads}} - \sqrt{2 - e^{- \Delta A_{\text{PL}} / \lads}}} \right) \: ,
\end{equation}
while the latter applies to $\ln(2) < \frac{\Delta A_{\text{PL}}}{\lads} \lesssim 0.875$ and gives a lower bound which is more straightforward to give implicitly using \eqref{eq:AMIAPLETW1}. 
The largest value of $\alpha_0$ for a putative lower bound $\Delta A_{\text{MI}} \geq \alpha_0 \Delta A_{\text{PL}}$ is therefore $\alpha_0 = - \frac{\ln \left( 2 - \sqrt{3} \right)}{\ln(2)} \approx 1.9$, coming from $T = 0$ and $\mu = \frac{\pi}{2}$. We plot the allowed region for $\Delta A_{\text{PL}}, \Delta A_{\text{MI}}$ in figure \ref{fig:LowerBoundAMI_ETW_1}. 

We now turn to the configurations of $\V_{1}, \V_{2}$ parametrized by $(\mu, \nu)$. 
We can first observe that $\Delta A_{\text{PL}}$ and $\Delta A_{\text{MI}}$ depend only on $\mu$ and $\nu$ respectively (as well as $T$), and that $\Delta A_{\text{MI}}$ is decreasing with $\nu$; consequently, for fixed $\mu$ and $T$, we have fixed $\Delta A_{\text{PL}}$, and the lower bound on $\Delta A_{\text{MI}}$ corresponds to $\nu = \frac{\mu}{2}$, the maximum value consistent with scattering. We will thus always take $\nu = \frac{\mu}{2}$ when investigating the lower bound. 

Fixing various values of $T$ and varying $\mu$ within the allowed range, we obtain the relationship between $\Delta A_{\text{MI}}$ and $\Delta A_{\text{PL}}$ in figure \ref{fig:LowerBound_ETW_2}. 
The strongest constraint on the coefficient $\alpha_{0}$ in a lower bound of the form $\Delta A_{\text{MI}} \geq \alpha_{0} \Delta A_{\text{PL}}$ comes from taking $\mu \rightarrow \pi$ and then $T \rightarrow 1$, which in particular yields large $\Delta A_{\text{PL}}$ and $\Delta A_{\text{MI}}$. 
Setting $\mu \rightarrow \pi$ gives (with $\nu = \frac{\mu}{2}$)
\begin{equation}
    \Delta A_{\text{PL}} = - \lads \ln \left( 4 T^{4} (1 - T^{2}) \right) \: , \quad \Delta A_{\text{MI}} = \lads \ln \left( \frac{1+T}{1-T} \right) \: ,
\end{equation}
and then for $T = 1 - \delta$ with small $\delta$
\begin{equation}
    \Delta A_{\text{PL}} = \lads \ln (1 / \delta) + O(1) \: , \quad \Delta A_{\text{MI}} = \lads \ln (1 / \delta) + O(1) \: .
\end{equation}
It follows that the largest allowed value of $\alpha_0$ for a lower bound of the form $\Delta A_{\text{MI}} \geq \alpha_0 \Delta A_{\text{PL}}$ is $\alpha_0 = 1$. 
We note that the configuration of $\V_{1}, \V_{2}$ parametrized by $(\mu, \nu)$ provides strictly stronger constraints on a putative lower bound on $\Delta A_{\text{MI}}$ than that parametrized by $(\mu, \tau)$. 

\bibliographystyle{unsrtnat}
\bibliography{biblio}

\end{document}